\documentclass[usenatbib]{mn2e}

\usepackage{amsmath}
\usepackage{natbib}
\usepackage{epsfig}
\usepackage{txfonts}

\newcommand{\plotwd}{\columnwidth}

\newcommand{\GHz}{{\rm GHz}}

\newcommand{\expf}[1]{{{\rm e}^{#1}}}
\newcommand{\Jbb}{\mathcal{J}}

\newcommand{\TRJ}{T_{\rm RJ}}
\newcommand{\Tgin}{{\Tg^{\rm in}}}

\newcommand{\zh}{{z_{\rm h}}}

\newcommand{\zmudc}{{z_{\rm dc}}}
\newcommand{\zmubr}{{z_{\rm br}}}

\newcommand{\xmat}{x_{\rm m}}

\newcommand{\xe}{x_{\rm e}}

\newcommand{\xc}{x_{\rm c}}

\newcommand{\id}{{\,\rm d}}

\newcommand{\beq}{\begin{equation}}   %

\newcommand{\eeq}{\end{equation}}   %

\newcommand{\beqa}{\begin{eqnarray}}   %

\newcommand{\eeqa}{\end{eqnarray}}   %

\newcommand{\beal}{\begin{align}}

\newcommand{\enal}{\end{align}}

\newcommand{\bspl}{\begin{split}}

\newcommand{\espl}{\end{split}}

\newcommand{\bsub}{\begin{subequations}}

\newcommand{\esub}{\end{subequations}}

\newcommand{\bmulti}{\begin{multline}}   %

\newcommand{\beqm}{\begin{mathletters}}   %

\newcommand{\eeqm}{\end{mathletters}}   %

\newcommand{\Abst}[1]{\,#1}

\newcommand{\me}{m_{\rm e}}

\newcommand{\Ne}{N_{\rm e}}

\newcommand{\Te}{T_{\rm e}}

\newcommand{\Tg}{T_{\gamma}}

\newcommand{\The}{\theta_{\rm e}}

\newcommand{\Thg}{\theta_{\gamma}}

\newcommand{\sigT}{\sigma_{\rm T}}

\newcommand{\nPl}{n_{\rm Pl}}

\newcommand{\pd}{\partial}

\newcommand{\pAb}[2]{\frac{\displaystyle\pd #1}{\displaystyle\pd #2}}

\newcommand{\PAb}[3]{\frac{\displaystyle\pd^{#3} #1}{\displaystyle\pd {#2}^{#3}}}

\newcommand{\Abl}[2]{\frac{{\rm d} #1}{{\rm d} #2}}

\newcommand{\pot}[2]{#1 \times 10^{#2}}

\newcommand{\Yp}{Y_{\rm p}}

\usepackage{color}

\usepackage{hyperref}
\usepackage{grffile}
\usepackage{graphics}
\usepackage{booktabs}

\title[$\mu$-type distortions]
{Refined approximations for the distortion visibility function and $\mu$-type spectral distortions}

\author[Chluba]{
 J.~Chluba$^{1}$\thanks{E-mail: jchluba@pha.jhu.edu}
 \\
$^{1}$ Department of Physics and Astronomy, Johns Hopkins University, Bloomberg Center 435, 
3400 N. Charles St., Baltimore, MD 21218, USA
}

\date{{Accepted 2014 February 27. Received 2013 December 19}}

\begin{document}

\maketitle

\begin{abstract}
The physical processes affecting the thermalization of cosmic microwave background spectral distortions are very simple and well understood. This allows us to make precise predictions for the distortions signals caused by various energy release scenarios, where the theoretical uncertainty is largely dominated by the physical ingredients that are used for the calculation.
Here, we revisit various approximations for the {\it distortion visibility function} -- defined using the fraction of the released energy that does not thermalize -- and early {\it $\mu$-type distortions}. Our approach is based on a perturbative expansion, which allows us to identify and clarify the origin of different improvements over earlier approximations. It provides a better than $\simeq 0.1\%-1\%$ description of our numerical results over a wide range of parameters. In particular, we are able to capture the high-frequency part of the $\mu$-distortion, which directly depends on the time derivative of the electron temperature. 
We also include lowest order double Compton and Compton scattering relativistic corrections, finding that because of cancelation they increase the thermalization efficiency in the tail of the distortion visibility function by only $\simeq 10\%$ (at $z\simeq \pot{6}{6}$), although individually their effect can reach $\simeq 20\%-40\%$. 
\end{abstract}



\section{Introduction}
\label{sec:intro}
Energy release in the early Universe can cause deviations of the cosmic microwave background (CMB) energy spectrum from a pure blackbody shape \citep[e.g.,][]{Zeldovich1969, Sunyaev1970mu, Illarionov1975, Danese1977}. These spectral distortions can tell us about processes occurring well before photons last scattered at redshift $z\simeq 1100$, allowing us to constrain the thermal history of our Universe looking deep into the pre-recombination plasma. This exciting possibility has recently spurred renewed theoretical interest into how spectral distortions form and evolve, showing that the distortion signals may open a new unexplored window to early-universe and particle physics \citep[see][for broader overview]{Chluba2011therm, Sunyaev2013, Chluba2013fore}.

The physics going into the thermalization problem -- the process that restores the pure blackbody spectrum after some perturbation from thermal equilibrium -- are pretty simple and well understood. For primordial spectral distortions, we are only concerned with the average CMB spectrum, so that spatial perturbations can be neglected and the Universe can be described as uniformly expanding, thermal plasma consisting of free electrons, hydrogen and helium atoms, and their corresponding ions inside a bath of photons from the CMB. 
We shall restrict ourselves to redshifts $z\lesssim \pot{\rm few}{7}$, when electron-positron pairs already completely disappeared. 
With these assumptions, any energy release inevitably causes a momentary distortion of the CMB spectrum. This can be understood with the following simple arguments: a pure blackbody spectrum, $B_\nu(T)$, is fully characterized by one number, its temperature $T$. Changing the energy density, $\rho_\gamma$, of the photon field by some $\Delta \rho_\gamma/\rho_\gamma\ll 1$ (e.g., from some particle decay) means that the photon number also has to be readjusted by $\Delta N_\gamma/N_\gamma\approx (3/4)\Delta \rho_\gamma/\rho_\gamma$. These additional photons, furthermore, have to be distributed according to $\partial B_\nu/\partial T$ in energy, to correctly shift the initial blackbody temperature from $T$ to $T'\approx T+(1/4) \Delta \rho_\gamma/\rho_\gamma$. In the early Universe, the double Compton (DC) and Bremsstrahlung (BR) processes are controlling the number of CMB photons, while Compton scattering (CS) allows photons to diffuse in energy. The exact interplay of these interactions between matter and radiation determines the spectrum of the CMB at any stage of its evolution. When studying different energy release mechanisms, the question thus is whether there was enough time between the energy release event and our observation to produce and redistribute those $\Delta N_\gamma/N_\gamma$ of missing photons, thereby fully digesting the energy injection, rendering the {\it distortion visibility} tiny. 

The thermalization problem has been studied thoroughly both analytically \citep[e.g.,][]{Zeldovich1969, Sunyaev1970mu, Illarionov1975, Danese1977, Burigana1995, Chluba2005, Khatri2012b, Khatri2012mix} and numerically \citep[e.g.,][]{Burigana1991, Hu1993, Burigana2003, Procopio2009, Chluba2011therm, Chluba2013fore}. From these studies, the following simple picture can be drawn: at $z\gtrsim \pot{2}{6}$, the thermalization process is extremely efficient and practically any distortion can be erased until today. 
At lower redshifts, the CMB spectrum becomes vulnerable to disturbances in the thermal history and only small amounts of energy can be ingested without violating the tight experimental bounds from COBE/{\it FIRAS} \citep{Mather1994, Fixsen1996,Fixsen2002} and other distortion measurements \citep{Kogut2006ARCADE, tris1, arcade2}. 

The transition from efficient to inefficient thermalization is encoded by the distortion visibility function, $\Jbb(z, z')$, which  determines by how much the distortion amplitude is suppressed between redshift $z$ and $z'<z$. The shape of the final distortion is close to a superposition of $\mu$- and $y$-type distortion \citep{Zeldovich1969, Sunyaev1970mu}, with a smaller residual (non-$\mu$/non-$y$) which provides additional time-dependent information at $10^4\lesssim z \lesssim \pot{3}{5}$ \citep{Burigana1991, Hu1995PhD, Chluba2011therm, Khatri2012mix, Chluba2013Green, Chluba2013PCA}. 
How far into the {\it cosmic photosphere} \citep{Bond1996} one can view, furthermore depends on the sensitivity of the experiment and how much energy needs to be thermalized.

The distortion visibility function is independent of the precise form of the distortion and just measures how much of the released energy is left in the distorted spectrum. To determine its shape, it is sufficient to study the evolution of distortions in the $\mu$-era ($\pot{3}{5}\lesssim z$), since later the visibility function is already extremely close to unity. This simplifies the problem immensely, since the photon distribution is brought into kinetic equilibrium with the matter within a very short time and only evolves rather slowly along a sequence of quasi-stationary stages. Due to the huge entropy of the Universe (there are $\simeq \pot{1.6}{9}$ times more photons than baryons), the DC process is furthermore the crucial source of photons in the $\mu$-era, so that the distortion visibility function is roughly given by $\mathcal{J}_{\rm DC}(z, z')=\expf{-(z/\zmudc)^{5/2}}\expf{(z'/\zmudc)^{5/2}}$, with thermalization redshift $\zmudc\approx\pot{1.98}{6}$ \citep{Sunyaev1970mu, Danese1977, Hu1993}.

For the analysis of future CMB distortion data \citep[see][for discussion of some experimental concepts]{Kogut2011PIXIE, PRISM2013WP, PRISM2013WPII}, it is of interest to improve the approximation for $\Jbb(z, z')$. While numerically, it is straightforward, although time consuming, to compute $\Jbb(z, z')$ precisely \citep[e.g., using {\sc CosmoTherm};][]{Chluba2011therm}, deeper physical insight can be gained analytically. Recently, \citet{Khatri2012b} [KS12, henceforth] took some steps into this direction. They showed that it is easy to analytically include the additional thermalization effect from BR, which becomes significant at $z\lesssim 10^6$. They also added a correction to the shape of the $\mu$-distortion caused by small deviations from quasi-stationarity, showing that at $10^6\lesssim z$ this captures the associated increase in the thermalization efficiency relative to $\mathcal{J}_{\rm DC}(z, z')$ seen numerically.

Here, we revisit this problem with a slightly different approach, basing our analysis on a perturbative expansion of the evolution equation. We show that several terms add at a similar order to the thermalization problem as the non-stationary correction discussed by KS12. We demonstrate that at low frequencies, small frequency-dependent variations of the photon production rate are significant in our approach. The time derivative of the critical frequency, $\xc$, characterizing the transition from effective to ineffective photon production also affects the shape of the distortion at the later stages ($z\lesssim 10^6$). Finally, we extend the validity of the solution for the spectral distortion to high frequencies, showing that it is driven by the time derivative of the electron temperature (see Sect.~\ref{sec:other_terms}), which is in contrast to KS12, who argued that there the time derivative of the chemical potential amplitude is most important.
While the latter does not affect the result for the distortion visibility function by much, as we show here, it is possible to capture this aspect in a simple way, obtaining an $\simeq 0.1\%$ description of the distortion shape at basically all relevant energies.

We fix the integration constants using energetic arguments (Sect.~\ref{sec:eff_T}), finding an $\lesssim 0.1\%-1\%$ agreement of our approximation for the distortion visibility function with the full numerical result. 
We also argue that at late times ($z\lesssim\pot{2}{5}$), photons produced by BR no longer are able to up-scatter efficiently and thus remain trapped at low frequencies (Sect.~\ref{eq:transport_estimate}). This means that the distortion visibility function becomes very close to unity at that point. By also including this effect, we reproduce our numerical result and overcome some of the small differences with respect to the numerical solution seen by KS12 at $z\lesssim 10^6$.

Finally, \citet{Chluba2011therm} included improved approximations for the DC Gaunt factor \citep{Chluba2005, Chluba2007a}. Due to higher order temperature corrections, the DC emissivity reduces relative to the non-relativistic case \citep{Lightman1981, Thorne1981}, making thermalization less efficient. On the other hand, frequency-dependent corrections move the maximal emission towards slightly higher frequencies, increasing the photon production rate. These effects can be captured analytically, giving a net increase in the thermalization efficiency at early times, which is in excellent agreement with our numerical result (Sect.~\ref{sec:DC_correction_Vis}).
This effect is partially counteracted by lowest order temperature corrections to CS, a modification that was not included by previous numerical treatments. Our perturbative approach again allows capturing this effect, providing a simple understanding for the origin of this correction (Sect.~\ref{sec:CS_corrs}). 

The paper is structured as follows: Sect.~\ref{sec:Bose_Einstein} and Sect.~\ref{sec:TimeEvo_phot} provide some of the basic ingredient for computing the evolution of the distortion. Parts of this are very pedagogical and can be skipped by the expert. 
In Sect.~\ref{sec:TimeEvo_Te} and Sect.~\ref{sec:solution_small}, we discuss the solutions for the thermalization problem, assuming small distortions. We also give our discussion for the distortion visibility function, with the main result shown in Fig.~\ref{fig:Vis_plot}. We close our analysis by including lowest order relativistic corrections to DC and CS in Sect.~\ref{sec:rel_corrs}. Our conclusions are given in Sect.~\ref{sec:conclusions}.\footnote{Our reference cosmology is: $\Yp=0.24$, $h=0.71$, $N_{\rm eff}=3.046$, $\Omega_{\rm b}=0.044$, $\Omega_{\rm cdm}=0.216$, $\Omega_\Lambda=0.74$ and $T_0=2.726\,{\rm K}$.}

\section{Non-equilibrium Thermodynamics of Bose-Einstein spectra}
\label{sec:Bose_Einstein}
To formulate the problem, it is useful to first go over the definitions of occupation number, number density, energy density and entropy density of a Bose-Einstein spectrum with small frequency-dependent chemical potential variable\footnote{We call it `variable' because it provides a simple parametrization of the spectrum that appears like a chemical potential, but is not generally identical with the thermodynamic chemical potential, which under equilibrium conditions vanishes for photons. In the text, we will usually drop the term `variable', but it should be kept in mind.}, $\bar\mu(t, x)$. We envision conditions in the early Universe at redshift $z\gtrsim \pot{\rm few}{5}$, when CS is still very efficient in redistributing photons over energy. Any photon distribution can be expressed as\footnote{See Appendix~\ref{app:BE_spect} for general values of $\bar\mu=\rm const$.}
\beal
\label{eq:approx_n}
n(x)&=\frac{1}{\expf{x+\bar\mu(t, x)}-1}\,\approx \nPl(x) - {G}(x) \, \frac{\bar\mu(t, x)}{x} +\mathcal{O}(\bar\mu^2),
\end{align}
where $x=h\nu/k \Tg$ denotes the dimensionless frequency with temperature $\Tg=T_0(1+z)$, which is used to define a reference energy scale, and $T_0=2.726\,{\rm K}$ \citep{Fixsen1996, Fixsen2009}. 
Furthermore, $\nPl(x)=1/[\expf{x}-1]$ is the occupation number of a blackbody at temperature $\Tg$ and ${G}(x)=-x\partial_x\nPl(x)=x\expf{x}/[\expf{x}-1]^2$ describes the spectrum of a simple temperature shift: adding/removing photons with this spectral shape does not create a distortion unless higher order terms become important (i.e., $\Delta T/T$ becomes too large).

Importantly, with $\Tg\propto (1+z)$ no redshifting term appears in the photon Boltzmann equation, which simplifies the problem significantly. Note, however, that $\Tg$ generally is not identical to the effective temperature of the photon field, which for a distorted spectrum can, for instance, be defined in terms of photon number or energy density (cf. Sect.~\ref{sec:eff_T}). It is also  generally not identical to the Rayleigh-Jeans temperature, $\TRJ$, defined at $x\ll 1$, which due to photon emission and absorption processes is very close to the electron temperature, $\Te$, at sufficiently low frequencies.

At any moment, the total CMB spectrum is given by a blackbody part at some temperature, $T_{\rm bb}$, plus a distortion relative to this.
In general, $T_{\rm bb}\neq \Tg$, and if we admit that $T_{\rm bb}$ can change because of energy release, we can write the parametrization
\beal
\label{eq:mu_ansatz}
\bar\mu(t, x)&= x \,\left(\frac{\Tg(t)}{T_{\rm bb}(t)}-1\right) + \mu_\infty(t) \, \hat{\mu}(t, x)
\end{align}
for the chemical potential. Here, the first term just represents the correct temperature shift of $\nPl(x)$ to $\nPl(x\,\Tg/T_{\rm bb})$. The overall amplitude of the effective chemical potential $\mu(t, x)=\mu_\infty(t)\,\hat{\mu}(t, x)$ is defined by $\mu_\infty(t)$, with all frequency-dependent terms captured by $\hat{\mu}(t, x)$. Since generally both $\mu_\infty(t)$ and $\hat{\mu}(t, x)$ depend on time, we still need a convenient normalization condition to uniquely determine the factorization, but we shall return to this point later.

The temperature $T_{\rm bb}$ in principle can be chosen freely, but now its time dependence is generally unknown before the solution of the problem is obtained. 
Since at low frequencies the photon distribution is pushed very close to equilibrium with the electrons by the double Compton and Bremsstrahlung processes, one useful choice is $T_{\rm bb}\equiv \TRJ =\Te$, which we shall use henceforth. This also removes any contribution $\hat\mu(t, x)\propto x$ from the effective chemical potential, as we will see below.

\subsection{Photon number and energy densities}
Setting $T_{\rm bb}= \TRJ=\Te$ in the expressions from above, we can obtain the number, energy and entropy densities of the distorted photon field as \citep[cf.,][]{Sunyaev1970mu}
\bsub
\label{eq:approx_N_rho}
\beal
N_\gamma &\approx N_\gamma^{\rm Pl}(\Tg)
\left[1+3 \frac{\Delta T_{\rm e}}{\Tg} - \mu_\infty \mathcal{M}_2\right],
\\[-1mm]
\label{eq:approx_N_rho_b}
\rho_\gamma &\approx \rho_\gamma^{\rm Pl}(\Tg)
\left[1+4 \frac{\Delta T_{\rm e}}{\Tg} - \mu_\infty \mathcal{M}_3\right],
\\[-1mm]
\label{eq:approx_N_rho_c}
s_\gamma &\approx s_\gamma^{\rm Pl}(\Tg)
\left[1+3 \frac{\Delta T_{\rm e}}{\Tg} - \frac{3}{4}\mu_\infty \mathcal{M}_3\right],
\end{align}
\esub
where $\Delta T_{\rm e}=T_{\rm e}-\Tg$, and $N_\gamma^{\rm Pl}$, $\rho_\gamma^{\rm Pl}$ and $s_\gamma^{\rm Pl}=(4/3)\rho_\gamma^{\rm Pl}/\Tg$ denote the photon number, energy and entropy densities of a blackbody at temperature $\Tg$, respectively (for explicit definition of $s_\gamma$ in terms of photon occupation number, see Appendix~\ref{app:entropy}).
For convenience, we also introduced the integrals $\mathcal{M}_k$ as
\bsub
\label{eq:approx_L_N_rho}
\beal
\mathcal{M}_k
&=\frac{1}{\mathcal{G}^{\rm Pl}_{k}}\,\int x^{k-1} {G}(x)\, \hat{\mu}(t, x)\id x
\stackrel{\stackrel{\hat\mu=1}{\downarrow}}{=}\frac{k\mathcal{G}^{\rm Pl}_{k-1}}{\mathcal{G}^{\rm Pl}_{k}},
\\[-0.8mm]
\label{eq:approx_L_N_rho_c}
\mathcal{G}^{\rm Pl}_{k}&=\int x^k \nPl(x) \id x.
\end{align}
\esub
A few important examples for $\mathcal{G}^{\rm Pl}_{k}$ are $\mathcal{G}^{\rm Pl}_{1}\approx 1.6449$, $\mathcal{G}^{\rm Pl}_{2}\approx 2.4041$, $\mathcal{G}^{\rm Pl}_{3}\approx 6.4939$, $\mathcal{G}^{\rm Pl}_{4}\approx 24.886$ and $\mathcal{G}^{\rm Pl}_{5}\approx 122.08$.
The integrals $\mathcal{M}_k$ directly depend on the shape of the distortion $\hat{\mu}(x, t)$, which also introduces additional time dependence to the problem. Assuming $\hat{\mu}=1$ we have $\mathcal{M}^{\rm c}_2\approx 1.3684$ and $\mathcal{M}^{\rm c}_3 \approx 1.1106$. Henceforth, the superscript `c' will indicate that $\hat{\mu}=1$ was used for the variable.

\subsection{Effective temperatures of the photon field}
\label{sec:eff_T}
One can easily show that for small distortions the effective temperatures\footnote{These are defined by equating the true number and energy densities with the one of a blackbody at the corresponding effective temperature.} of the distorted spectrum with respect to photon number, $T^\ast_{N}$, and energy density, $T^\ast_{\rho}$, are respectively given by
\bsub
\label{eq:effective_temperatures}
\beal
\label{eq:effective_temperatures_a}
T^\ast_N &\approx T_{\rm e}- \frac{1}{3}\Tg\,\mu_\infty \mathcal{M}_2
\approx \Tg\left[1+\frac{\Delta T_{\rm e}}{\Tg}-0.4561\,\mu_\infty\right]
\\
T^\ast_\rho &\approx T_{\rm e}- \frac{1}{4}\Tg\,\mu_\infty \mathcal{M}_3
\approx \Tg\left[1+\frac{\Delta T_{\rm e}}{\Tg}-0.2776\,\mu_\infty\right],
\end{align}
\esub
where for the second approximate sign we assumed constant (independent of frequency) chemical potential. Without distortion one naturally has $T^\ast_N=T^\ast_\rho=\Te$. The effective temperature following from the entropy density, Eq.~\eqref{eq:approx_N_rho_c}, is identical to the one of the energy density.
For positive $\mu_\infty$ and $\hat{\mu}=1$ (constant chemical potential), Eq.~\eqref{eq:effective_temperatures} implies $T^\ast_N<T^\ast_\rho$, or explicitly 
\beal
\label{eq:temperature_difference_rho_N}
T^\ast_\rho-T^\ast_N=\frac{\Tg}{4}\frac{\kappa_\rho(t)\mu_\infty(t)}{3} \approx 0.1785\,\mu^\ast_\infty(t) \Tg,
\end{align}
where we defined $\kappa_\rho(t)=4\mathcal{M}_2-3\mathcal{M}_3$, which for constant chemical potential $\mu(t, x)=\mu_\infty(t)$ gives $\kappa^{\rm c}_\rho=4\mathcal{M}^{\rm c}_2-3\mathcal{M}^{\rm c}_3\approx 2.1419$. Once thermalization creates $\mu_\infty\rightarrow 0$, this also means that $T^\ast_N\rightarrow T^\ast_\rho$, restoring full thermal equilibrium.

In Eq.~\eqref{eq:temperature_difference_rho_N}, we introduced the effective amplitude of the chemical potential, $\mu^\ast_\infty(t)=\hat\kappa_\rho(t)\,\mu_\infty(t)$, with $\hat\kappa_\rho(t)=\kappa_\rho(t)/\kappa^{\rm c}_\rho$. If $\hat\mu=1$, we have $\hat\kappa_\rho(t)=1$ and hence $\mu^\ast_\infty(t)\equiv\mu_\infty(t)$, but generally $\hat\kappa_\rho(t)\neq 1$. Since we still have the freedom to normalize $\hat\mu$, for convenience we choose its normalization such that $\hat\kappa_\rho(t)\equiv 1$ at all times. This simplifies all the expressions discussed below and we can use $\mu^\ast_\infty(t)=\mu_\infty(t)$ and $\kappa_\rho(t)=\kappa^{\rm c}_\rho$ without loss of generality. 

\subsection{Compton equilibrium temperature}
It is well known that in the early Universe after a very short time the free electrons approach a temperature that is dictated by the shape of the (high-frequency) photon distribution and the energy exchange through Compton scattering \citep{Levich1970}:
\beal
\label{eq:Compton_Te}
\Te\approx \Te^{\rm eq}= \Tg \frac{\int n(1+n)x^4\id x}{4\int n x^3\id x}.
\end{align}
For a constant chemical potential $\mu(t, x)$, one has $n(1+n)=-\partial_x n$ and after integration by parts one finds $\Te^{\rm eq}=\Tg\equiv \TRJ$. This is no longer exactly true once photon production starts to revert the distorted spectrum to a blackbody at $x\ll1$. In this case, $\Te$ is pushed away from $\Tg$ and the global energetics have to be considered.

\subsection{Initial state after very short burst of energy release}
After energy release ceases, the comoving energy density of the photon-baryon system remains constant. Assuming that there was no time to produce photons, but that Compton scattering already brought electrons and photons into kinetic equilibrium, we can compute the initial photon temperature and chemical potential. Defining the photon temperature before the energy release as $\Tg^{\rm in}$ and assuming that a total of $\Delta\rho_\gamma/\rho_\gamma\ll 1$ of energy was injected at redshift $z_{\rm in}$, with Eq.~\eqref{eq:effective_temperatures}, assuming $\mu=\mu^{\rm st}_\infty={\rm const}$ (so that $\Te=\Tg=\TRJ$), we find \citep[cf.,][]{Sunyaev1970mu}
\bsub
\label{eq:initial_condistions}
\beal
\label{eq:initial_condistions_a}
\Tg^{\rm st} &\approx \Tg^{\rm in}\left[1+0.4561\,\mu_\infty\right]\equiv \Tg=T_0(1+z_{\rm in})
\\
\label{eq:initial_condistions_b}
\mu^{\rm st}_\infty &
\approx \frac{3}{\kappa^{\rm c}_\rho}\frac{\Delta\rho_\gamma}{\rho_\gamma}
\approx1.401\frac{\Delta\rho_\gamma}{\rho_\gamma}.
\end{align}
\esub
Without additional photon production by the DC and BR processes and without subsequent energy release, this would describe the state of the CMB spectrum even today because the adiabatic expansion of the Universe leaves the photon distribution unaltered once Compton scattering brought things into kinetic equilibrium\footnote{This is not entirely true since the slow Hubble expansion leads to extra Compton cooling of photons caused by the difference in the adiabatic indices of baryons and photons and hence a change of the spectrum by additional Compton scattering \citep{Chluba2011therm}.}.

Expression \eqref{eq:initial_condistions_a} for $\Tg^{\rm st}$ also defines the initial Rayleigh-Jeans temperature of the photon field. Once photon production starts and transport of photons to higher frequencies reduces the value of $\mu_\infty$, one finds $\Tg^{\rm st}> \TRJ=\Te$. On the other hand, the temperature following from the energy density is $T^\ast_\rho \approx \Tg^{\rm in}(1+\frac{1}{4}\Delta \rho_\gamma/\rho_\gamma) \approx \Tg^{\rm st}(1-0.2776\,\mu^{\rm st}_\infty)<\Tg^{\rm st}$. This determines the final temperature of the blackbody (corrected for the effect of redshifting), which is slowly approached as the thermalization process completes. In this process, a total of $\Delta N_\gamma/N_\gamma\approx (3/4)\Delta \rho_\gamma/\rho_\gamma$ photons have to be replenished by BR and DC emission relative to the initial blackbody at temperature $\Tg^{\rm in}$, at least if there was a sufficient amount of time to completely thermalize the distortions. 

\subsection{Evolution equations for $N_\gamma$ and $\rho_\gamma$}
The equations~\eqref{eq:approx_N_rho} are similar to the expressions given by \citet{Sunyaev1970mu}; however, we do not make the assumption that the integrals $\mathcal{M}_k$ are constant and can be computed for $\hat\mu={\rm const}$.
Combining the comoving time derivatives of Eq.~\eqref{eq:approx_N_rho}, we can thus write the more general evolution equations
\bsub
\label{eq:evol_mu_DT}
\beal
\label{eq:evol_mu_DT_a}
\frac{\kappa^{\rm c}_\rho}{\mathcal{M}_2 \mathcal{M}_3} \!\frac{\id}{\id t}\frac{\Te}{\Tg} 
&\approx 
%
\frac{{\rm d} \ln a^4 \rho_\gamma}{\mathcal{M}_3\, {\rm d} t}-\frac{{\rm d} \ln a^3 N_\gamma}{\mathcal{M}_2\, {\rm d} t}
+\mu_\infty \frac{\id}{\id t}\!\ln\left(\frac{\mathcal{M}_3}{\mathcal{M}_2}\right)
\\
\label{eq:evol_mu_DT_b}
\frac{\id\mu_\infty}{\id t} 
&\approx 
\frac{3}{\kappa_\rho^{\rm c}}\frac{{\rm d} \ln a^4 \rho_\gamma}{{\rm d} t}
-\frac{4}{\kappa_\rho^{\rm c}}\frac{{\rm d} \ln a^3 N_\gamma}{{\rm d} t},
\end{align}
\esub
where $a=(1+z)^{-1}$ is the scale factor normalized to unity at $z=0$. As explained in Sect.~\ref{sec:eff_T}, we chose the normalization of $\hat\mu(t, x)$ such that $\kappa_\rho(t)=\kappa_\rho^{\rm c}={\rm const}$. 
The only additional assumption at this point is that the overall amplitude of the distortion is small and only linear order terms need to be considered. 

The terms on the r.h.s. of Eq.~\eqref{eq:evol_mu_DT_b} determine real changes of the photon energy and number density. These need to be obtained from the photon Boltzmann equation, which includes the effect of electron scattering and emission/absorption. After energy release ends, only ${\rm d} \ln a^4 N_\gamma/{\rm d} t$ drives the evolution of $\mu_\infty(t)$ and $\Te(t)$. To compute this term, assuming $\mu(t, x)\approx \mu_\infty(t)$ is {\it insufficient}, since at low frequencies, where most of the photon production happens, emission and absorption processes return the photon distribution to a blackbody very quickly (i.e., $\hat\mu(t, x)\rightarrow 0$ at $x\ll 1$), reducing ${\rm d} \ln a^4 N_\gamma/{\rm d} t$ to a finite (!) value that critically depends on the shape of the low-frequency spectrum. Knowing the exact time dependence of this term is therefore crucial for describing the overall evolution of the distortion (see Sect.~\ref{sec:solution_small}). 

\vspace{-3mm}
\subsection{Consistency relations}
\label{sec:consistency_rels}
The way the problem is set up, we have three unknown functions: $\Te(t)$, $\mu_\infty(t)$ and $\hat\mu(t, x)$. By construction, $\mu_\infty(t)$ and $\hat\mu(t, x)$ define the deviation of the CMB spectrum with respect to the Rayleigh-Jeans temperature, $\TRJ\equiv\Te$. Assuming that only a single injection of $\Delta \rho_\gamma/\rho_\gamma$ occurs at $t_{\rm in}$, we already know the initial state of the photon distribution from Eq.~\eqref{eq:initial_condistions} with $\hat\mu(t_{\rm in}, x)=1$. As thermalization proceeds, $\mu$ and $\TRJ\equiv\Te$ slowly approach $\mu=0$ and $\Te(z)=\Tg^{\rm in}(1+\frac{1}{4}\Delta \rho_\gamma/\rho_\gamma)(1+z)/(1+z_{\rm in})$ after full thermalization. 

In this picture, a few consistency relations ought be fulfilled by the solution at any time. First, if only one episode of energy release occurs, then at any time $t>t_{\rm in}$ we should find 
\beal
\label{eq:energy_cons_relation}
\frac{\Delta \Te(t)}{\Tg}
&\approx\frac{\mathcal{M}_3(t)}{4}\mu_\infty(t)-\frac{3}{4}\frac{\mathcal{M}^{\rm c}_3}{\kappa^{\rm c}_\rho} \frac{\Delta \rho_\gamma}{\rho_\gamma},
\end{align}
reflecting conservation of the comoving energy density\footnote{We neglect the tiny heat capacity of ordinary matter here.} by the system. The coefficient reads $(3/4)(\mathcal{M}^{\rm c}_3/\kappa^{\rm c}_\rho)\approx 0.3889$. 

Similarly, the total number of photons that have been created by DC and BR emission since the energy release should  at any stage of the evolution be given by
\beal
\label{eq:Number_relation}
\frac{\Delta N_\gamma(t)}{N_\gamma}
&\approx
\frac{3}{4}\left(\frac{\Delta \rho_\gamma}{\rho_\gamma}-\frac{\kappa^{\rm c}_\rho\, \mu_\infty(t)}{3}\right),
\end{align}
where $\Delta N_\gamma(t)/\Delta N_\gamma$ is computed with respect to the initial blackbody spectrum before the energy release. This expression already shows that $\kappa^{\rm c}_\rho\, \mu_\infty(t)/3$ can be interpreted as the energy density carried by the non-blackbody part of the spectrum, or explicitly
\beal
\label{eq:mu_infty_def}
\mu_\infty(t)\equiv \frac{3}{\kappa^{\rm c}_\rho}\frac{\Delta \rho_\gamma(t)}{\rho_\gamma}-\frac{4}{\kappa^{\rm c}_\rho}\frac{\Delta N_\gamma(t)}{N_\gamma}.
\end{align}
Here, the energy density and number densities are momentary values computed directly from the distorted photon field. This expression is independent of which temperature is used for the pure blackbody part, as terms $\hat\mu\propto x$ automatically cancel. This is because the distortion is defined with respect to photon deficit (for positive $\mu_\infty$) relative to the photon energy density, also reflected by Eq.~\eqref{eq:evol_mu_DT_b}. This is a very sensible interpretation of the distortion, since scattering processes leave the number of photons unchanged.
In addition, we need two boundary conditions for $\hat\mu(t, x)$ to close the problem. By construction, we have $\hat\mu(t, x)\rightarrow 0$ for $x\ll 1$. This also implies that we need to ensure that $\TRJ = \Te$. The final condition follows from $\kappa_\rho=4\mathcal{M}_2-3\mathcal{M}_3=\kappa_\rho^{\rm c}$, which fixes the overall normalization for $\hat\mu$ as explained in Sect.~\ref{sec:eff_T}.

\vspace{-3mm}
\section{The photon Boltzmann equation} 
\label{sec:TimeEvo_phot}
In this section, we give the Boltzmann equation for photons writing out first-order relativistic corrections to the Compton and DC processes. Although slightly technical, we need these equations to obtain the solutions for the chemical potential. We linearize the expressions assuming small spectral distortions. Discussion of the solutions is presented in Sect.~\ref{sec:solution_small}.
%

\subsection{General form}
\label{sec:PhotonEvo}
In the early Universe, photons undergo many interactions with the
electrons. At redshifts $z\gtrsim \pot{4}{5}$, the most
important processes are CS and DC scattering: DC serves as a source of
soft photons at low frequencies, whereas CS efficiently redistributes photons over frequency. At lower redshifts ($z\lesssim \pot{4}{5}$), BR starts taking over the photon production; however, at that stage, the up-scattering of photons becomes inefficient. 
It is convenient to express all time-scales in units of the Thomson scattering time-scale, $t_{\rm C}=1/\sigT\,\Ne\,c \approx \pot{2.3}{20}(1+z)^{-3} \sec$. Then the Boltzmann equation describing the time
evolution of the photon phase space density $n_{\gamma}(\tau, x)$ in the
expanding Universe is given by
\beal
\label{eq:BoltzEq_Photons}
\pAb{n}{\tau}
=\left.\pAb{n}{\tau}\right|_{\rm CS}
+\left.\pAb{n}{\tau}\right|_{\rm DC}+\left.\pAb{n}{\tau}\right|_{\rm BR}
+\mathcal{S}(\tau, x),
\end{align}
where we introduced the optical depth $\id\tau=\id t/t_{\rm C}$ to Thomson scattering as dimensionless time variable. 
The terms on the r.h.s. of this equation, respectively, describe the effect of CS, DC scattering, BR and additional sources of photons with the source function, $\mathcal{S}(\tau, x)$.
This source function, for example, could include the effect of dissipation of acoustic modes in the early Universe \citep[e.g., see][]{Chluba2012} or photons produced by decaying or annihilating particles.
As explained in the previous section, in $x=h\nu/k\Tg$ we chose $\Tg=T_0(1+z)$ and describe the distortion with respect to a blackbody at the temperature of the electrons $T_{\rm bb}=\Te\equiv \TRJ$.

\subsection{Compton scattering}
\label{sec:CS}
The contribution of CS by thermal electrons to the r.h.s. of Eq.~\eqref{eq:BoltzEq_Photons}, including first-order relativistic corrections, may be compactly written as \citep[compare][]{Challinor1998, Itoh98, Sazonov1998, Chluba2005}:
\beal
\label{eq:Komp_equ_rel_corr}
\left.\Abl{n}{\tau}\right|_{\rm CS}
\!
&=
\frac{\theta_{\rm e}}{x^2}\,
\pAb{}{x}\, x^4 \!
\left[
F
+\theta_{\rm e}\left\{ \frac{5}{2}\,F 
+\frac{21}{5}\,x\,\pAb{F}{x}
\right.\right.\nonumber\\
&\qquad\quad
\left.\left.
+\frac{7}{10}\,x^2\,
\left[
\phi \,F_-\left(\phi-6\,F_-\right)
+\PAb{F_+}{x}{2}
\right]
\right\}
\right]+\mathcal{O}(\The^3)
\Abst{.}
\end{align}
Here, $\theta_{\rm e}=k\Te/\me c^2$, $\phi=\Tg/\Te$, and we defined the photon flux functions $F=\partial_x n_\gamma+\phi \,n(n+1)$ and $F_\pm=F\pm \phi \,n(n+1)$. The first term in the bracket represents the Kompaneets
equation \citep{Kompa56}, whereas the terms proportional to $\theta_{\rm e}^2$ arise from
first-order relativistic corrections, i.e., higher order Doppler effect [$\simeq\mathcal{O}(\The^2)$], recoil effect [$\simeq\mathcal{O}((h\nu/\me c^2)^2)$] and cross terms [$\simeq\mathcal{O}(\The\, h\nu/\me c^2)$]. 
We assumed that the photon distribution is isotropic even if it deviates from the pure blackbody shape. 
One can readily verify that for a blackbody spectrum at temperature $\Tg\equiv \Te$ the flux $F$ vanishes; also the combination of terms related to $F_\pm$ vanishes, so that in equilibrium CS leaves the photon distribution unaffected.

We now linearize Eq.~\eqref{eq:Komp_equ_rel_corr} with respect to $\mu\ll1$. The effect of CS on the blackbody, $\nPl(x\, \Tg/\Te)$, which we chose as a reference (see previous section), vanishes identically.
For terms that involve $\mu(\tau, x)=\mu_\infty(\tau) \,\hat{\mu}(\tau, x)$, as defined in Eq.~\eqref{eq:mu_ansatz}, with $F\rightarrow -{G} \mu/x$, the linearized Compton collision term reads:
\bsub
\label{eq:mu_terms}
\beal
\left.\Abl{n_{\gamma}}{\tau}\right|_{\rm CS}
&\approx
-\Thg\,\frac{{G}}{x}
\left[
x^2 \,\mu'' 
+2  g_1(x) \,x \,\mu'
\right]\left(1+\frac{5}{2}\Thg\right)
\\[-0.25mm]
\nonumber
&\qquad 
-\Thg^2\,\frac{{G}}{x}
\left[
-\frac{42}{5}\,f_1(x)\,x \,\mu'
+\frac{21}{5}\,f_2(x)\,x^2\mu''
\right]
\\[-0.25mm]
\nonumber
&\qquad\quad 
-\Thg^2\,\frac{{G}}{x}
\left[
\frac{28}{5}\,f_3(x)\,x^3\mu'''
+\frac{7}{10} \,x^4\mu''''
\right],
\end{align}
where $\Thg =h\nu/k\Tg$. Here, primes denote derivatives with respect to $x$ and we defined the frequency-dependent functions
\beal
g_1(x)&=-\frac{{Y}_{\rm SZ}}{2{G}}
\approx 1-\frac{x^2}{12}+\frac{x^4}{720}
\\[-0.5mm]
f_1(x)&=x{Y}_{\rm SZ}-x{G}(1+\mathcal{C}_x-5\mathcal{S}_x)
\approx 1+\frac{5x^2}{12}-\frac{29x^4}{720}
\\[-0.5mm]
f_2(x)&=5+\frac{x^2}{6}+2x{G}(1-3\mathcal{S}_x)
\approx 1-\frac{x^2}{2}+\frac{x^4}{60}
\\[-0.5mm]
f_3(x)&=\frac{1}{2}[1+g_1(x)] \approx 1-\frac{x^2}{24}+\frac{x^4}{1440},
\end{align}
\esub
with $\mathcal{C}_x=x\coth(x/2)$, $\mathcal{S}_x=x/\sinh(x/2)$ and the $y$-distortion shape ${Y}_{\rm SZ}(x)={G}(x)[x\coth(x/2)-4]$. For the functions $g_1$ and $f_i$, we also gave the limits $x\ll 1$. These functions are defined such that to leading order in $x\ll1$ they are all equal to unity.

If we look at the terms in Eq.~\eqref{eq:mu_terms} in this limit, we can see that they all are of similar order in $x$. However, the higher derivative terms are suppressed by an extra factor $\Thg \ll 1$.
This allows us to first consider only the lowest order solution, neglecting terms $\propto\Thg^2$.
Relativistic correction can then be added as perturbations to the non-relativistic solution (Sect.~\ref{sec:rel_corrs}).

\subsubsection{Effect on the photon energy density}
By integrating Eq.~\eqref{eq:Komp_equ_rel_corr} over $x^2 \id x$, it is trivial to confirm that the collision terms for CS conserve the photon number. 
To compute the energy exchange between electrons and photons via Compton scattering, we integrate Eq.~\eqref{eq:Komp_equ_rel_corr} over $x^3 \id x$. This yields \citep[see][for similar expressions]{Sazonov2001, Chluba2005}:
\beal
\label{eq:dE_dt_CS}
-\left.
\frac{1}{a^4 \rho_\gamma^{\rm Pl}} \frac{{\rm d} a^4 \rho_\gamma}{{\rm d} \tau}
\right|_{\rm CS}
&\approx
\frac{\mathcal{I}_4}{\mathcal{G}_{3}}\,\Thg-4\,\theta_{\rm e}
-10\,\theta_{\rm e}^2
-\frac{21}{5}\,\frac{\mathcal{G}_{5}}{\mathcal{G}_{3}}\,\Thg^2
\\
\nonumber
&\qquad
+\frac{\mathcal{I}_4}{\mathcal{G}_{3}}\,\left(\frac{47}{2}-\frac{21}{5}\,\frac{\mathcal{H}_{6}}{\mathcal{I}_{4}}
\right)\,\theta_{\rm e}\,\Thg,
\end{align}
where the integrals $\mathcal{G}_i$ are defined by Eq.~\eqref{eq:approx_L_N_rho_c} with $\nPl(x) \rightarrow n(x)$. For convenience, we also introduced $\mathcal{I}_i=\int x^i n(1+n)\id x$ and $\mathcal{H}_{i}=\int x^i (\partial_x n)^2\id x$.
The first two terms in Eq.~\eqref{eq:dE_dt_CS} determine the usual contributions in the non-relativistic limit, while the other terms capture relativistic corrections.
From this expression, the Compton equilibrium temperature can be given by
\bsub
\beal
\label{eq:Te_CS}
\theta_{\rm eq}&\approx \theta^{\rm nr}_{\rm eq}
\left[1-\frac{21}{5}\theta^{\rm nr}_{\rm eq} 
\left\{\frac{4\mathcal{G}_{3}\mathcal{G}_{5}}{\mathcal{I}_{4}^2}+\frac{\mathcal{H}_{6}}{\mathcal{I}_{4}}-5
\right\}
\right]
\nonumber
\\
&\approx \Thg\left[
1+\delta \gamma_1
-\Thg \,\delta \gamma_2 
\right],
\end{align}
using the fact that the non-relativistic Compton equilibrium temperature is small, i.e., $\theta^{\rm nr}_{\rm eq}=\Thg\mathcal{I}_{4}/4\mathcal{G}_{3}\ll 1$, and keeping terms up to second order of $\theta^{\rm nr}_{\rm eq}$ only.
In the second line, we linearized the problem for small distortion.
The corrections to the integrals thus take the form
\beal
\label{eq:Te_CS_gamma_i}
\delta \gamma_1&=
\frac{\delta \mathcal{I}_{4}}{4\mathcal{G}^{\rm Pl}_{3}}-\frac{\delta \mathcal{G}_{3}}{\mathcal{G}^{\rm Pl}_{3}} 
\\ \nonumber
\delta \gamma_2&=\frac{21}{20}
\left[
\left(\frac{\delta \mathcal{G}_{5}}{\mathcal{G}^{\rm Pl}_{3}}-\frac{\mathcal{G}^{\rm Pl}_{5}}{\mathcal{G}^{\rm Pl}_{3}}\frac{\delta \mathcal{G}_{3}}{\mathcal{G}^{\rm Pl}_{3}}\right) 
+4\left(\frac{\delta \mathcal{H}_{6}}{\mathcal{I}^{\rm Pl}_{4}}-\frac{\mathcal{H}^{\rm Pl}_{6}}{\mathcal{I}^{\rm Pl}_{4}}\frac{\delta \mathcal{I}_{4}}{\mathcal{I}^{\rm Pl}_{4}}\right)
\right]
-4\pi^2\gamma_1,
\end{align}
\esub
where we only have to compute the terms for the spectral distortions, as indicated by `$\delta$'. These do, however, generally include the difference in the spectrum due to $\TRJ=\Te\neq \Tg$, so that for example $\delta \gamma_1\approx \Delta\Te/\Tg + \delta\gamma^\mu_1$, where now the second term only includes contributions $\propto \hat\mu$.
%
%
Note also that $\mathcal{I}^{\rm Pl}_{4}\equiv 4\mathcal{G}^{\rm Pl}_{3}$.
%
%

\noindent
With Eq.~\eqref{eq:Te_CS} one can recast Eq.~\eqref{eq:dE_dt_CS} into the form
\beal
\label{eq:dE_dt_CS_new}
-\left.
\frac{{\rm d} \ln a^4 \rho_\gamma}{{\rm d} \tau}
\right|_{\rm CS}
&\approx
4\left(\theta_{\rm eq}-\The \right)
\left[1+\theta^{(0)}_{\rm eq}\left(\frac{21}{5}\,\frac{\mathcal{H}_{6}}{\mathcal{I}_{4}}
-\frac{37}{2}\right)\right]
\nonumber\\
&\approx
4\left(\theta_{\rm eq}-\The \right)\left[1-17.239\,\Thg\right],
\end{align}
where in the second line we again linearized the problem for small distortion, with $2\pi^2-5/2\approx 17.239$.
This shows that at the lowest order, the distortions only affect the Compton equilibrium temperature.
Furthermore, the time-scale on which Compton equilibrium is achieved is increased by $\simeq[1-17.239\, \Thg]^{-1}$, which at $z\simeq \pot{2}{6}$ implies an $\simeq 2\%$ effect. The change in the equilibration time-scale and even the exact value for the equilibrium temperature are, however, not relevant to the final solution for the chemical potential, and only enter the problem at higher perturbation order.

\subsection{Double Compton and Bremsstrahlung emission}
\label{sec:DC_em}
The contribution of DC scattering and BR to the r.h.s. of the photon Boltzmann equation \eqref{eq:BoltzEq_Photons} can be
written in the form \citep[cf.,][]{Rybicki1979, Lightman1981, Thorne1981}:
\beal
\label{eq:DC_term}
\left.\Abl{n_{\gamma}}{\tau}\right|_{\rm DC+BR} 
&= \frac{
  \expf{-x}}{x^3}\Big[1-n_{\gamma}\,(\expf{\phi\,x}-1)\Big] \Lambda(x, \Thg, \The) 
\end{align}
where the emission coefficient $\Lambda$ is given by the sum of the contribution due
to double Compton scattering and Bremsstrahlung,
$\Lambda = \Lambda_{\rm DC}+ \Lambda_{\rm BR}$.
As this expression shows, the leading order emission term scales $\propto x^{-3}$. This is the reason why at low frequencies the spectrum returns to a blackbody in equilibrium with the electrons after a very short time.

At lowest order in the distortion, we have
\beal
\label{eq:DC_term}
\left.\Abl{n_{\gamma}}{\tau}\right|_{\rm DC+BR} 
&\approx \frac{\Lambda}{x^4}(1-\expf{-x})\,\mathcal{G} \,\mu
\end{align}
where we inserted Eq.~\eqref{eq:approx_n} and \eqref{eq:mu_ansatz}. Note that the emission coefficient, $\Lambda$, is evaluated at the background level, i.e., it just depends on $x$ and $\Thg$. At low frequencies, its dependence on $x$ is rather weak, a fact that can be used to simplify the problem.
Also, since we chose the temperature of the electrons as reference ($T_{\rm bb}=\Te$), no additional emission/absorption term related to the blackbody part arises, one of the reasons for this definition.

\subsubsection{Double Compton scattering}
Due to the large entropy of the Universe, DC emission dominates over BR at high redshifts ($z\gtrsim \pot{4}{5}$). The DC scattering emission coefficient can be expressed as
\beal
\label{eq:K_DC}
\Lambda_{\rm DC}(x, \Thg)
&=\frac{4\alpha}{3\pi}\, \Thg^2 \, \expf{x} g_{\rm dc}(x, \Thg)
\Abst{,}
\end{align}
where $\alpha$ is the fine structure constant and $g_{\rm dc}(x, \Thg)$
is the effective DC Gaunt factor.  
It has the form \citep{Chluba2005, Chluba2007a, Chluba2011therm}:
\beal
\label{eq:g_DC}
g_{\rm dc}(x, \Thg)
&\approx \frac{\mathcal{I}^{\rm Pl}_4\,H^{\rm Pl}_{\rm dc}(x)}{1+14.16\, \Thg} 
\Abst{,}
\end{align}
where $\mathcal{I}^{\rm Pl}_4=\int x^4 \nPl(\nPl+1)\id x= 4\pi^4/15\approx 25.976$, 
and 
\beal
\label{eq:H_DC_appr}
H^{\rm Pl}_{\rm dc}(x)
&\approx \expf{-2x } \left[ 1+\frac{3}{2}x+\frac{29}{24} x^2+\frac{11}{16} x^3+\frac{5}{12} x^4\right].
\end{align}
Here, we included the first-order relativistic correction in the photon temperature.
At low frequencies, $\Lambda_{\rm DC}(x, \Thg)\propto (1+\frac{1}{2} x)$. In the non-relativistic limit, we have $H^{\rm Pl}_{\rm dc}(x)\approx\expf{-x}$ and $g_{\rm dc}(x, \Thg)\approx\mathcal{I}^{\rm Pl}_4\,\expf{-x}$.

\subsubsection{Bremsstrahlung}
At low redshifts ($z\lesssim  \pot{4}{5}$), Bremsstrahlung becomes the main
source of soft photons. One can define the Bremsstrahlung emission coefficient
by \citep[cf.][]{Burigana1991, Hu1993}
\beal
\label{eq:K_BR}
\Lambda_{\rm BR}(x, \theta_{\rm e})
&=\frac{\alpha\,\lambda_{\rm e}^3}{2\pi\sqrt{6\pi}}\;\frac{\theta_{\rm e}^{-7/2}\,\expf{x(1-\phi)}}{\phi^3}
\sum_{\rm i} Z^2_{\rm i}\,N_{\rm i}\, g_{\rm ff}(Z_{\rm i},x,\theta_{\rm e})
\Abst{.}
\end{align}
Here, $\lambda_{\rm e}=h/\me\,c$ is the Compton wavelength of the electron, and
$Z_{\rm i},\,N_{\rm i}$ and $g_{\rm ff}(Z_{\rm i},x,\theta_{\rm e})$ are the
charge, the number density and the BR Gaunt factor for a nucleus of the atomic species i,
respectively. 

Various simple analytical approximations exist \citep{Rybicki1979}, but nowadays more accurate fitting formulae, valid over a wide range of temperatures and frequencies, may, for example, be found in \citet{Nozawa1998} and \citet{Itoh2000}.
We find, however, that the differences introduced by the various approximations for $\Lambda_{\rm BR}$ are not very important, both for the shape of the distortion and the distortion visibility function. It is pretty straightforward to include them consistently, so that for any of the computations we just use the expressions of \citet{Itoh2000}.


\begin{figure}
\centering
\includegraphics[width=\plotwd]{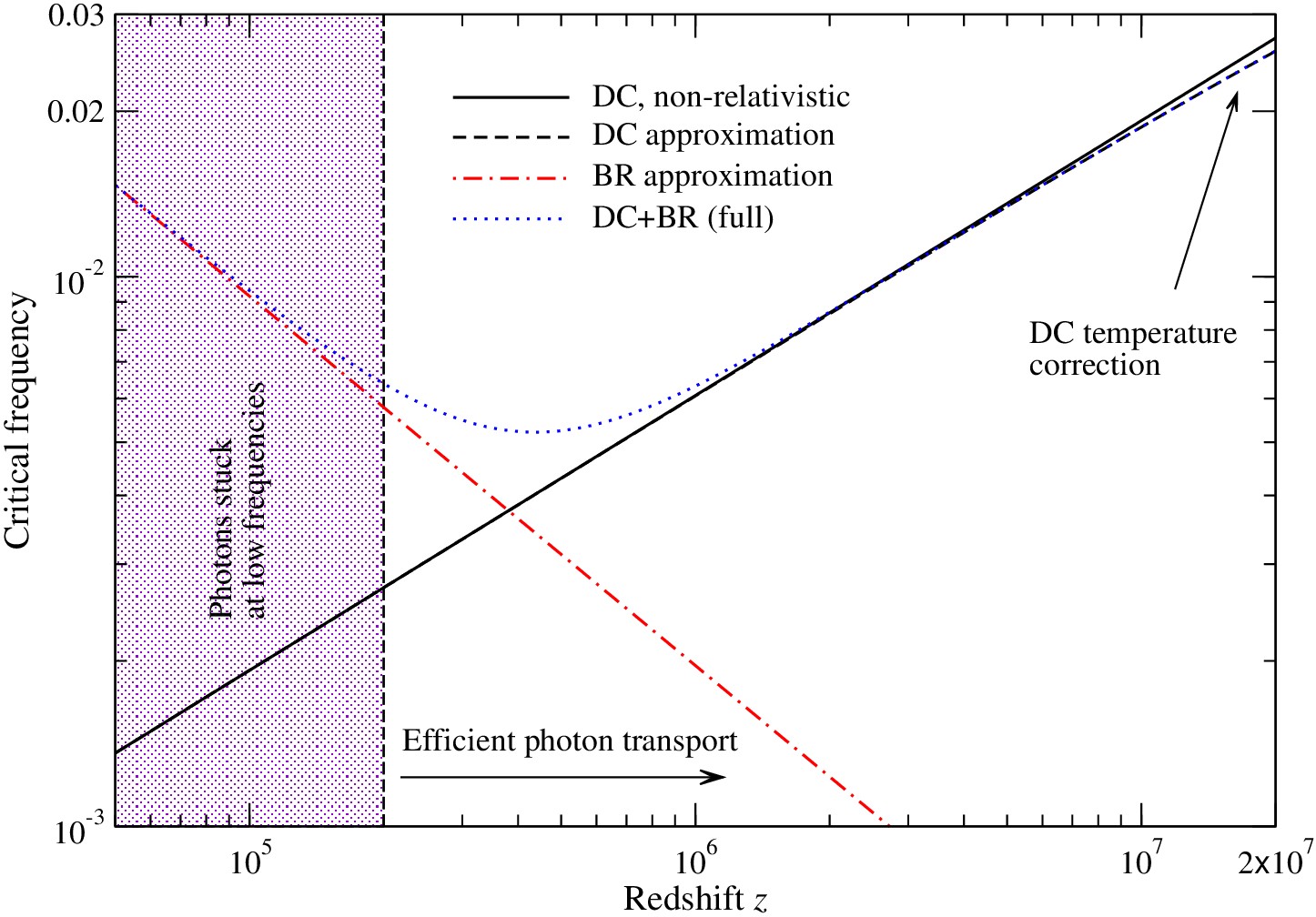}
\caption{Critical frequency, $\xc$, defined by Eq.~\eqref{eq:xc_def} as a function of $z$. Photon transport is inefficient below $z\simeq \pot{2}{5}$ so that the distortion visibility function quickly approaches unity. DC temperature corrections become noticeable at $z\gtrsim 10^6$. The approximations are from Eq.~\eqref{eq:xc_DC_appr} and \eqref{eq:xc_BR_appr}.}
\label{fig:xc_fig}
\end{figure}
\subsubsection{Critical frequency}
\label{eq:xc_defs}
For the computations below, we need the critical frequency determined by \citep[compare][]{Sunyaev1970mu, Danese1982, Burigana1991, Hu1993}
\beal
\label{eq:xc_def}
\xc(\tau)\equiv \sqrt{\Lambda(\xc)/\Thg}.
\end{align}
Close to $\xc$, the rate of Compton scattering equals that of photon emission/absorption, roughly defining the maximum of the photon production, with some corrections which are important for the late-time evolution (see Sect.~\ref{sec:classical_sol}). 
For DC alone, one has
\bsub
\label{eq:xc_DC_appr}
\beal
\label{eq:xc_DC_appr_a}
\xc^{\rm DC, 0}&\approx \sqrt{\frac{4\alpha}{3\pi}\, \Thg \mathcal{I}^{\rm Pl}_4} \approx \pot{8.60}{-3} \left[\frac{1+z}{\pot{2}{6}}\right]^{1/2}
\\[1mm]
\label{eq:xc_DC_appr_b}
\xc^{\rm DC}&\approx\xc^{\rm DC, 0}\left(1+\frac{1}{2}\xc^{\rm DC, 0}\right)^{1/2}/\left(1+14.16\, \Thg\right)^{1/2},
\end{align}
\esub
where for $\xc^{\rm DC, 0}$ we neglected relativistic corrections to the DC Gaunt factor, while we included them for $\xc^{\rm DC}$.

The critical frequency for DC is shown in Fig.~\ref{fig:xc_fig}. One can see that at high redshifts, the temperature correction reduces the critical frequency notably. 
As we will see below, this implies that thermalization should be less efficient, since the frequency at which most photons are produced decreases. 
At the thermalization redshift $z\simeq \pot{2}{6}$, the temperature correction to the critical frequency $\xc$ is roughly $0.5\%$ and it reaches $\simeq 1\%$ at $z\simeq \pot{4}{6}$. Although this appears to be small, since the critical frequency enters the problem through an integral, the cumulative effect matters so that the correction is amplified and hence significant (see below). 
At $z\gtrsim \pot{\rm few}{7}$, the number of electrons and positrons becomes comparable to the number of photons, so that there the thermalization efficiency increases vastly. A treatment of the thermalization problem in that era is, however, beyond the scope of this paper.

For BR alone, we determined the critical frequency numerically using the expressions from 
 \citet{Itoh2000} and assuming a helium mass fraction of $\Yp=0.24$. We find\footnote{Note that again we can evaluate $\Lambda$ assuming $\Te = \Tg$.}
\beal
\label{eq:xc_BR_appr}
\xc^{\rm BR}&\approx \pot{1.23}{-3}\,\left[\frac{1+z}{\pot{2}{6}}\right]^{-0.672}
\end{align}
to work very well. Comparing with Eq.~\eqref{eq:xc_DC_appr_a}, we can see that at the thermalization redshift $z\simeq \pot{2}{6}$ BR contributes about $10\%$ to the value of the critical frequency. However, the contribution drops rapidly towards higher redshifts (Fig.~\ref{fig:xc_fig}).
To percent precision, the total critical frequency is $\xc^2\approx (\xc^{\rm DC})^2+(\xc^{\rm BR})^2$ \citep{Hu1993}.

\subsubsection{Transport of photons towards higher frequencies}
\label{eq:transport_estimate}
Photons, produced by DC and BR at low frequencies, can only help thermalizing the distortion if they actually reach the high-frequency part of the spectrum before being reabsorbed or before no time to up-scatter is left. We thus need to estimate until when the photon redistribution by Compton scattering is efficient. The evolution of a narrow line within an ambient blackbody radiation field can be computed for any value of the Compton $y$-parameter $y_\gamma=\int \Thg \id \tau$ using the analytic solutions from \citet{Chluba2008d}. These are valid as long as electron recoil remains negligible. They include the extra drift of photons towards lower frequencies caused by stimulated effects. The average photon energy thus increases like $\nu\simeq \nu_0 \,\expf{2y_\gamma}$ \citep[the classical result of][does not include stimulated effects and thus gives $\nu\simeq \nu_0 \expf{4y_\gamma}$]{Zeldovich1969}.

Assuming that photons start their journey around $x\simeq \xc\ll 1$, we can estimate the time it takes for the average photon distribution to reach $x\simeq 1$. This then implies that we need $y_\gamma\gtrsim (1/2)\ln\xc^{-1}$. Estimating the total $y$-parameter from some initial redshift $z$ until today using $y_\gamma\approx 4.3 [(1+z)/\pot{3}{5}]^2$ (radiation domination) then implies that at $z\lesssim 10^5\sqrt{\ln\xc^{-1}}$ photons can no longer be sufficiently up-scattered. This implies that around $z\simeq \pot{2}{5}$, photon transport no longer is efficient enough to replenish the high-frequency photon deficit. At that moment, the low- and high-frequency parts of the photon distribution practically decouple, and the effective chemical potential (or more precisely the high-frequency photon number) freezes in. Below $z\simeq \pot{2}{5}$, basically all the released energy remains in the distortion, and the distortion visibility function thus becomes unity. This statement is in fact irrespective of the shape of the distortion at late times ($\mu$, $y$ and residual distortion). Our numerical computations confirm this statement (see Fig.~\ref{fig:DVis_plot}).

\subsubsection{Effect on number and energy density}
\label{sec:DNN_DEE}
In contrast to Compton scattering, DC and BR change the photon number density. To compute the effective photon production and associated change in the photon energy density caused by this, we can simply integrate Eq.~\eqref{eq:DC_term} over $x^2 \id x$ and $x^3 \id x$, finding:
\beal
\label{eq:DC_BR_dN_dt_final}
\left.
 \frac{{\rm d} \ln a^3 N_\gamma}{{\rm d} \tau}
\right|_{\rm DC,BR}
&\approx \frac{1}{\mathcal{G}^{\rm Pl}_2}\,\int \frac{\Lambda(x)}{x[\expf{x}-1]}
\,\mu\id x
\nonumber\\
\left.
\frac{{\rm d} \ln a^4 \rho_\gamma}{{\rm d} \tau}
\right|_{\rm DC,BR}
&\approx \frac{1}{\mathcal{G}^{\rm Pl}_3}\,\int \frac{\Lambda(x)}{\expf{x}-1}
\,\mu\id x
=\mathcal{H}_{\rm em}.
%
\end{align}
In general these integrals have to be performed numerically once the solution for the low-frequency distortion is known. 
In particular, the non-trivial frequency dependence of the BR Gaunt factor renders it difficult to find accurate analytic approximations. This is because at the percent level the integrals picks up small contributions even at higher frequencies and the assumption of constant $\Lambda$ is not well justified.
The integrand scales like $\Lambda(x) \mu(x)/x^2$ at low frequencies, so that for $\Lambda(x)\approx \rm const$ we need $\mu$ to vanish faster than $\simeq x^2$ to obtain a finite result for ${\rm d} \ln a^3 N_\gamma/{\rm d} \tau$. For ${\rm d} \ln a^3 \rho_\gamma/{\rm d} \tau$, we find $\mu\propto x$ is sufficient, at least to first order in perturbations. This shows that the integrals in Eq.~\eqref{eq:DC_BR_dN_dt_final} both diverge for constant $\mu$, but the lowest order solution (see Sect.~\ref{sec:classical_sol}) $\mu=\expf{-\xc/x}$ is sufficient to regularize these expressions.

We also mention a slight inconsistency of the formulation that is present in all treatments of the problem so for. The energy needed for the production of DC photons is taken partially (in the limit of resting electron in which the DC Gaunt factor is derived in fact fully!) from the photon field itself: the incoming high-energy photons scatter and redistribute in energy, giving rise to a small correction to the Compton process and heat exchange with the electrons in addition to the photon emission. This implies that the emission integral $\mathcal{H}_{\rm em}$ should be slightly smaller; we are, however, going to neglect this effect, given that the total energy used up by photon production is small. For additional discussion, see \citet{Chluba2005}.

\section{Evolution of the matter temperature} 
\label{sec:TimeEvo_Te}
At high redshifts, electrons and photons rapidly exchange energy via Compton scattering, while Coulomb interactions keep electrons and baryons in equilibrium at one temperature, $T_{\rm m}=\Te$.
The electrons also cool by BR and DC emission; in addition, the matter (electrons plus baryons) in the Universe cools because of the adiabatic expansion of the medium. And finally, energy release can heat the matter and thereby increase its temperature.

Including all these processes, the evolution equation for the matter temperature can be written as \citep[e.g.,][]{Chluba2011therm}
\beal
\label{eq:Te_equation}
\Abl{\The/\Thg}{\tau}&=
\frac{\dot{\mathcal{Q}}_{\rm e}}{\alpha_{\rm h}\Thg}
+\frac{4\tilde{\rho}^\ast_\gamma}{\alpha_{\rm h}\Thg}[\theta_{\rm eq}-\The] 
- \frac{\tilde{\rho}_\gamma}{\alpha_{\rm h}\Thg}\mathcal{H}_{\rm em} 
-H\,t_{\rm C} \frac{\alpha^{\rm nr}_{\rm h}-\xi_{\rm h}}{\alpha_{\rm h}\Thg}\The,
\nonumber
\\
\alpha_{\rm h}&=\alpha^{\rm nr}_{\rm h}+\xi_{\rm h},
\quad
\xi_{\rm h}=\frac{15}{4}\,\The\frac{1-\frac{1}{2}\Yp}{1-\Yp} N_{\rm e}
\approx 4.342 \, N_{\rm e} \, \The.
\end{align}
Here, $k\alpha^{\rm nr}_{\rm h}=\frac{3}{2}k[N_{\rm e}+N_{\rm H}+N_{\rm He}]$ denotes the matter heat capacity, and $\xi_{\rm h}$ defines its lowest order temperature correction\footnote{Only the contribution from electrons matters here, since the terms for the baryons are suppressed by ratios of the masses \citep{Chluba2005}.}. 
The number densities, $N_{\rm i}$, are for free electrons (i=`e'), total number of hydrogen (i=`H') and helium (i=`He') nuclei, and $\Yp\simeq 0.24$ is the helium mass fraction.
Furthermore, we introduced the energy injection term, $\dot{\mathcal{Q}}_{\rm e}\equiv(\me c^2)^{-1}\id{Q}_{\rm e}/\id \tau$, which for example could be caused by some decaying or annihilating particles; 
the energy density of the photon field in units of electron rest mass is defined as $\tilde{\rho}_\gamma=\kappa_\gamma\Thg^4\,\mathcal{G}_{3}$ with $\kappa_\gamma=8\pi \lambda_{\rm e}^{-3}\approx \pot{1.760 }{30}\,{\rm cm}^{-3}$ and $\lambda_{\rm e}$ denoting the Compton wavelength of the electron.
Finally, the Compton equilibrium temperature, $\theta_{\rm eq}$, is given by Eq.~\eqref{eq:Te_CS}, which includes first-order relativistic corrections to the problem and the effect of spectral distortions. 
We also used expression~\eqref{eq:dE_dt_CS_new} for the Compton energy exchange between electrons and photons. This is indicated by the asterisk on $\tilde{\rho}_\gamma$, which means $\tilde{\rho}^{\ast}_\gamma=\tilde{\rho}_\gamma[1-17.239\,\Thg]$.
The cooling by DC and BR is defined as $\mathcal{H}_{\rm em}\equiv \left.\id \ln a^4 \rho_\gamma/\id \tau\right|_{\rm DC, BR}$.

\subsection{Perturbative solution for $\Te$}
\label{sec:solution_Te_pert}
Without knowing the exact solution of the CMB spectrum, it is still possible to obtain a solution for $\Te$, because for conditions in the early Universe it evolves along a sequence of quasi-stationary stages. Assuming that only the matter is heated directly by the energy release, this then allows us to eliminate $\id \ln a^4 \rho_\gamma/\id \tau$ in Eq.~\eqref{eq:evol_mu_DT_b}.
Setting $\id (\The/\Thg)/\id \tau\approx 0$, from Eq.~\eqref{eq:Te_equation} we thus find
\beal
\label{eq:Te_sol}
\The^{(0)}&\approx
\theta^{(0)}_{\rm eq}
+\frac{\dot{Q}_{\rm e}}{4 \rho^{\ast}_\gamma}
-\frac{\tilde{\rho}_\gamma}{4\tilde{\rho}^{\ast}_\gamma}\mathcal{H}^{(0)}_{\rm em} 
-\frac{H\,t_{\rm C}\alpha^{\rm nr}_{\rm h}\Thg}{4\tilde{\rho}^{\ast}_\gamma}
\left[1-\lambda_{\rm h}\Thg\right].
\end{align}
with $\lambda_{\rm h} \Thg =\xi_{\rm h}(\theta)/\alpha^{\rm nr}_{\rm h}\approx 1.39\,\Thg$.
This means that at the lowest order in perturbations, the Compton energy-exchange term reads
\beal
\label{eq:dE_dt_CS_new_II}
\left.
\frac{{\rm d} \ln a^4 \rho^{(0)}_\gamma}{{\rm d} \tau}
\right|_{\rm CS}
&\approx
\frac{\dot{Q}_{\rm e}}{\rho_\gamma}
- \mathcal{H}^{(0)}_{\rm em}
-\frac{H\,t_{\rm C}\alpha^{\rm nr}_{\rm h}\Thg}{\tilde{\rho}_\gamma}
\left[1-\lambda_{\rm h}\Thg\right].
\end{align}
Therefore the net change in the photon energy density is
\beal
\nonumber
\frac{{\rm d} \ln a^4 \rho^{(0)}_\gamma}{{\rm d} \tau}
&\approx
\left.
\frac{{\rm d} \ln a^4 \rho_\gamma}{{\rm d} \tau}
\right|_{\rm CS}+\mathcal{H}^{(0)}_{\rm em}
\approx
\frac{\dot{Q}_{\rm e}}{\rho_\gamma}
-\frac{H\,t_{\rm C}\alpha^{\rm nr}_{\rm h}\Thg}{\tilde{\rho}_\gamma}
\left[1-\lambda_{\rm h}\Thg\right].
\end{align}
The emission integral and also the precise value of the Compton equilibrium temperature dropped out of the problem.
At lowest order perturbation theory, the spectral distortion therefore does not directly affect the heat exchange between electrons and photons.
Higher order temperature corrections only enter the thermalization problem via the heat capacity of the electrons. This reduces the distortion created by the adiabatic cooling of matter \citep{Chluba2011therm} by a small amount. However, the effect only reaches 1 percent at $z\simeq \pot{1.6}{7}$, where the distortion visibility is already extremely small (see next Section). This effect can therefore be safely neglected.

Neglecting higher order corrections to the adiabatic cooling term, and realizing that the corrections to $\mathcal{H}_{\rm em}$ are always canceled by the corresponding photon emission term (a consequence of energy conservation), we find the correction to the electron-photon energy-exchange term
${\rm d} \ln a^4 \rho^{(1)}_\gamma/{\rm d} \tau\approx 
-(\alpha_{\rm h}\Thg/\tilde{\rho}_\gamma) \partial_\tau(\Te^{(0)}/\Tg)$.
Knowing the solution for the spectral distortion at lowest order in perturbation theory, one can compute $\id(\Te^{(0)}/\Tg)/\id\tau$ using Eq.~\eqref{eq:evol_mu_DT_a} to close the system of equations. This in principle allows us to obtain the next-order correction to the photon distribution. 
As we will see below, $\id (\Te^{(0)}/\Tg)/\id \tau \approx \mathcal{O}(\xc)$. In comparison with any of the external heating terms, $\dot{Q}_{\rm e}$, the correction is suppressed by an additional factor of the photon-to-baryon ratio, so that from the practical point of view this correction again can be neglected.
Henceforth, we will thus simply write the photon heating term as $\id \ln a^4 \rho_\gamma/\id \tau\approx \dot{Q}_{\rm e}/\rho_\gamma-\left[H\,t_{\rm C}+\id (\Te^{(0)}/\Tg)/\id \tau\right]\alpha_{\rm h}\Thg/\tilde{\rho}_\gamma\equiv  \dot{Q}^\ast_{\rm e}/\rho_\gamma$. This defines the effective photon heating rate caused by energy release that initially only affects the temperature of ordinary matter in the Universe. The second term will be neglected in our discussion.

\section{Solution for the spectral distortions in the limit of small chemical potential}
\label{sec:solution_small}
In this section, we develop a perturbative treatment for the approximation to the chemical potential, $\mu(\tau, x)$. Before adding higher order corrections, we briefly recap the classical solution obtained by \citet{Sunyaev1970mu}. The basic ansatz is that time- and frequency-dependent parts of the solution can be approximately separated: $\mu(\tau, x)\approx \mu_\infty(\tau) \, \hat{\mu}(x)$. This means that $\mu(\tau, x)$ only evolves very {\it slowly} with time, moving along a sequence of quasi-stationary stages, with the main time dependence being captured by an overall amplitude factor, while the {\it shape} of the distortion is fixed to $\simeq \hat{\mu}(x)$. Here, we go beyond this approximation. 
It is furthermore clear, that DC and BR emission are effective only at rather small $x$, so that one can expect to find the main frequency dependence of the solution there, while at much higher frequencies the solution varies only slowly with $x$. This introduces an energy scale, suggesting the scaling $x\rightarrow \xc \xi$, where $\xc$ is a critical frequency at which the spectrum changes rapidly. This provides a natural perturbation parameter, $\xc$, with corrections being ranked by their order in $\xc\ll 1$.

\subsection{Integral solution}
With the general ansatz $\mu(\tau, x)= \mu_\infty(\tau) \, \hat{\mu}(\tau, x)$, we can already write an integral solution to Eq.~\eqref{eq:evol_mu_DT_b}. In contrast to the original works, we do not assume that $\hat{\mu}(\tau, x)\approx \hat{\mu}(x)$, but explicitly include slow time dependence in the shape of the distortion. Scaling out the main terms, with Eq.~\eqref{eq:DC_BR_dN_dt_final} the photon production rate is given by 
\beal
\label{eq:DC_BR_I_mu}
&\left.
 \frac{{\rm d} \ln a^3 N_\gamma}{{\rm d} \tau}
\right|_{\rm DC,BR} \approx \frac{\Thg \xc}{\mathcal{G}^{\rm Pl}_2}\,\mu_{\infty} \, \mathcal{I}_{\hat{\mu}}
\nonumber\\
&\mathcal{I}_{\hat{\mu}}(\tau)
=\int \frac{\Lambda(x)}{\Lambda(\xc)}\frac{\xc \,\hat{\mu}(\tau, x)}{x[\expf{x}-1]}\id x
\end{align}
Generally, $\mathcal{I}_{\hat{\mu}}$ is very close to unity with corrections of $\mathcal{O}(\xc)$ [i.e., higher order terms].
Inserting this into Eq.~\eqref{eq:evol_mu_DT_b}, with the definition of the effective heating rate $\dot{Q}^\ast_{\rm e}$ given in Sect.~\ref{sec:solution_Te_pert} we find
\beal
\label{eq:mu_infty_final_eq}
&\Abl{\mu_\infty}{\tau}
\approx \gamma_\rho
\frac{\dot{Q}^\ast_{\rm e}}{\rho_\gamma} -\gamma_N\, \Thg \xc \, \mathcal{I}_{\hat{\mu}}
\,\mu_\infty
\nonumber\\
&\gamma_\rho=3/\kappa^{\rm  c}_\rho\approx 1.401,
\quad
\gamma_N =4/(\mathcal{G}^{\rm Pl}_2 \kappa^{\rm c}_\rho)\approx 0.7769,
\end{align}
where $\kappa^{\rm c}_\rho=2.1419$.
Then, by introducing the optical depth
\bsub
\label{eq:tau_mu}
\beal
\tau_\mu(z)&=\tau_{\mu, 0}(z)+ \Delta\tau_\mu(z),
\nonumber\\
\label{eq:tau_mu_a}
\tau_{\mu, 0}(z)
&=\gamma_N \int^z_0 \frac{\Thg \xc}{t_{\rm C}} \frac{\id z'}{H(1+z')},
\\
\label{eq:tau_mu_b}
\Delta\tau_\mu(z)
&=\gamma_N \int^z_0 \frac{\Thg \xc}{t_{\rm C}} 
\left(\mathcal{I}_{\hat{\mu}}-1
\right)
\frac{\id z'}{H(1+z')},
\end{align}
\esub
and assuming that there is no initial distortion at very early times, we can finally write
\beal
\label{eq:mu_sol}
\mu_\infty(z)
&\approx
1.401\int_z^\infty 
\frac{\dot{Q}^\ast_{\rm e}}{\rho_\gamma}\frac{\expf{-\tau_\mu(z', z)}\id z'}{H(1+z')}
\end{align}
with $\tau_\mu(z, z')=\tau_\mu(z)-\tau_\mu(z')$. 
Equation~\eqref{eq:mu_sol} gives the formal solution for  $\mu_\infty(z)$ at high redshifts, including all important effects that affect the integrated energy and number density of the photon field when heating the ordinary matter by some process occurs.
The optical depth, $\tau_\mu$, is affected by (i) the shape of the spectral distortion at low frequencies (which enters $\mathcal{I}_{\hat\mu}$; see Fig.~\ref{fig:DImu_plot}) and (ii) the precise redshift dependence of the critical frequency, $\xc$ (see Fig.~\ref{fig:xc_fig}). At lowest order in $\xc$, we have $\mathcal{I}_{\hat{\mu}}\approx 1$, so that $\Delta \tau_\mu(z)\approx 0$.

\subsubsection{Total photon emission}
Inserting the solution for $\mu$ into Eq.~\eqref{eq:DC_BR_I_mu}, we can directly compute the total change in the number of photons over the energy release history. This yields the simple expression
\beal
\label{eq:total_Ndot}
\frac{\Delta N_\gamma}{N_\gamma} 
&\approx \frac{3}{4} \int_z^\infty \frac{\partial \tau_\mu(z')}{\partial z'}
\left(\int_{z'}^\infty 
\frac{\dot{Q}^\ast_{\rm e}}{\rho_\gamma}\frac{\expf{-\tau_\mu(z'', z')}\id z''}{H(1+z'')}\right) \id z'
\nonumber\\
&=\frac{3}{4} \int_z^\infty \frac{\partial \expf{-\tau_\mu(z')}}{\partial z'}
\left(\int_{z'}^\infty 
\frac{\dot{Q}^\ast_{\rm e}}{\rho_\gamma}\frac{\expf{-\tau_\mu(z'')}\id z''}{H(1+z'')}\right) \id z'
\nonumber\\
&=\frac{3}{4} \int_{z}^\infty \frac{\dot{Q}^\ast_{\rm e}}{\rho_\gamma}
\left(1-\expf{-\tau_\mu(z', z)}\right)
\frac{\id z'}{H(1+z')}
\nonumber\\
&=\frac{3}{4}
\left(
\frac{\Delta \rho_\gamma}{\rho_\gamma}-\left.\frac{\Delta \rho_\gamma}{\rho_\gamma}\right|_{\rm dist}
\right)
\equiv \frac{3}{4}\left.\frac{\Delta \rho_\gamma}{\rho_\gamma}\right|_{T},
\end{align}
where we used the definition of $\tau_\mu$ given in Eq.~\eqref{eq:tau_mu} and identified the energy density changes 
\bsub
\label{eq:rho_defs}
\beal
\label{eq:rho_defs_a}
&\frac{\Delta \rho_\gamma}{\rho_\gamma}
=\int_{z}^\infty \frac{\dot{Q}^\ast_{\rm e}}{\rho_\gamma}\frac{\id z'}{H(1+z')}
\\
\label{eq:rho_defs_b}
&\left.\frac{\Delta \rho_\gamma}{\rho_\gamma}\right|_{\rm dist}
=
\int_{z}^\infty \frac{\dot{Q}^\ast_{\rm e}}{\rho_\gamma}\expf{-\tau_\mu(z', z)}
\frac{\id z'}{H(1+z')}
\\
\label{eq:rho_defs_c}
&\left.\frac{\Delta \rho_\gamma}{\rho_\gamma}\right|_{T}
=\frac{\Delta \rho_\gamma}{\rho_\gamma}-\left.\frac{\Delta \rho_\gamma}{\rho_\gamma}\right|_{\rm dist}.
\end{align}
\esub
These expressions give the following picture: the total energy release branches into temperature shift and distortion. At any moment, part of the total energy release, $\Delta \rho_\gamma/\rho_\gamma$, is stored in the distortion (non-blackbody) and carries an energy density change $\Delta \rho_\gamma/\rho_\gamma|_{\rm dist}$. The remainder $\Delta \rho_\gamma/\rho_\gamma|_{T}$ is carried by the blackbody and is associated with a shift of the initial blackbody temperature $\Tgin$ by 
$(T_N^\ast-\Tg^{\rm in})/\Tg^{\rm in}=\frac{1}{3}\Delta N_\gamma/N_\gamma=\frac{1}{4}\Delta \rho_\gamma/\rho_\gamma|_{T}$. 
The branching ratio between the distortion and temperature parts depends on time and the efficiency of the thermalization process. This defines the distortion visibility function, $\Jbb(z', z)=\expf{-\tau_\mu(z', z)}$, which determines the energy branching ratio at $z$ given that the heating occurred at $z'$.

\subsubsection{Single energy release}
Assuming that the distortion caused by the adiabatic cooling of matter is negligible, from Eq.~\eqref{eq:mu_sol} we find 
\citep[cf.][]{Sunyaev1970mu}
\beal
\label{eq:mu_sol_sing}
\mu_\infty(z)
&\approx 
1.401\frac{\Delta\rho_\gamma}{\rho_\gamma} 
\expf{-\tau_\mu(z_{\rm h}, z)}=\mu_{\infty}^{\rm st}\,\Jbb(\zh, z)
\end{align}
for a single energy release of $\Delta\rho_\gamma/\rho_\gamma$ at heating redshift $z_{\rm h}$.
Here, we used $\mu_{\infty}^{\rm st}=1.401\Delta \rho_\gamma/\rho_\gamma$. The factor $\expf{-\tau_\mu(z_{\rm h}, z)}$ is the spectral distortion visibility between the heating redshift $\zh$ and $z$.
Explicitly, the amount of energy stored in distortions (independent of its specific shape actually) at any moment is
\beal
\label{eq:rho_dist}
\left.\frac{\Delta\rho_\gamma}{\rho_\gamma} \right|_{\rm dist}
&\equiv \frac{\Delta\rho_\gamma}{\rho_\gamma}  \expf{-\tau_\mu(z_{\rm h}, z)}.
\end{align}
In the above formulation, only $\tau_\mu(z_{\rm h}, z)$ has to be computed precisely to obtain approximations for $\mu_\infty(z)$. Both quantities require a solution for $\hat{\mu}$, which we shall discuss below. 
According to Eq.~\eqref{eq:Number_relation}, the added number of photons therefore is 
\beal
\label{eq:Number_sol}
\frac{\Delta N_\gamma(t)}{N_\gamma}
&\approx
\frac{3}{4}\frac{\Delta \rho_\gamma}{\rho_\gamma}\left(1-\expf{-\tau_\mu(z_{\rm h}, z)}\right)\equiv \frac{3}{4}\left.\frac{\Delta \rho_\gamma(t)}{\rho_\gamma}\right|_{T},
\end{align}
again with respect to the initial blackbody at temperature $\Tg^{\rm in}<\Tg$. From Eq.~\eqref{eq:energy_cons_relation}, it also immediately follows that 
\beal
\label{eq:Te_sol}
\frac{\Delta \Te(t)}{\Tg}
&\approx \frac{3}{4}\frac{\mathcal{M}^{\rm c}_3}{\kappa^{\rm c}_\rho}\left(\frac{\mathcal{M}_3(t)}{\mathcal{M}^{\rm c}_3}\expf{-\tau_\mu(z_{\rm h}, z)}-1\right)\frac{\Delta \rho_\gamma}{\rho_\gamma}.
\end{align}
Initially $\Delta \Te(t)/\Tg=0$ and as thermalization proceeds, $\Te=\TRJ$ decreases until $\Delta \Te(t)/\Tg\approx -0.3889\Delta \rho_\gamma/\rho_\gamma$, so that in the final stage $T^\ast_N=T^\ast_\rho=\Tg^{\rm in}(1+\frac{1}{4} \Delta \rho_\gamma/\rho_\gamma)$. These relations show that $\Jbb(\zh, z)$ and $\hat\mu(t, x)$ fully characterize the solution of the thermalization problem in terms of photon energy and number density.

\subsubsection{Numerical computation of $\Jbb(z', z)$}
\label{sec:Jbb_def}
Equations~\eqref{eq:mu_sol_sing}--\eqref{eq:Number_sol} suggest a procedure to numerically compute the distortion visibility function {\it without} directly relying on the amplitude of the chemical potential at different frequencies: in the picture given above, $\Jbb(z_{\rm h}, z)$ determines how much of the injected energy is available for distortions at redshift $z$. 
Numerically, this means: (i) compute the effective temperature of the photon distribution with respect to the photon number density, $T^\ast_N$ [Eq.~\eqref{eq:effective_temperatures_a}] and (ii) subtract the corresponding energy density based on this temperature from the total injected energy density, $\Delta\rho_\gamma/\rho_\gamma$. The remainder determines the total amount of energy that went into distortions, $\Delta\rho_\gamma/\rho_\gamma |_{\rm dist}$, and from the final ratio of the injected energy densities one can obtain 
\beal
\label{eq:Jbb_num}
\Jbb(z_{\rm h}, z)\approx \frac{\Delta\rho_\gamma/\rho_\gamma|_{\rm dist}}{\Delta\rho_\gamma/\rho_\gamma}.
\end{align}
Since the high-frequency spectrum varies logarithmically with $x$ [see Eq.~\eqref{eq:mu_high_1_II}], this approach provides a more robust (valid even after the $\mu$-era) definition for the visibility function, removing possible ambiguities introduced by comparing the chemical potential at some fixed frequency and different times to construct the visibility function. 
We will now derive analytic expressions for $\Jbb(z_{\rm h}, z)$ in different limits and then compare them with numerical results.

\subsection{Classical solution of Sunyaev \& Zeldovich (1970)}
\label{sec:classical_sol}
We already mentioned that for the evolution of $\mu_\infty(\tau)$, we need to obtain an approximation for ${\rm d} \ln a^3 N_\gamma/{\rm d} \tau$ in terms of $\hat{\mu}(x)$. From Eq.~\eqref{eq:mu_terms} and \eqref{eq:DC_term} with $\id n/\id \tau=-({G}/x)\, \partial_\tau \mu$, we have the general evolution equation for the chemical potential
\beal
\label{eq:mu_equation_full}
\pAb{\mu}{\tau}-x\, \partial_\tau \frac{\Te}{\Tg}
&\approx
\Thg
\left[
x^2 \,\mu'' 
+2  g_1(x) \,x \,\mu' 
\right] 
-\frac{\Lambda}{x^3}(1-\expf{-x}) \mu.
\end{align}
We discuss terms of $\mathcal{O}(\Thg^2)$ in Sect.~\ref{sec:rel_corrs}. Following \citet{Sunyaev1970mu}, we set the l.h.s. of this equation to zero and go to the limit $x\ll 1$, finding
\beal
\label{eq:mu_equation_full_lowest_order}
0&\approx
x^2 \,\mu'' +2 \,x \,\mu' -\frac{\Lambda/\Thg}{x^2} \mu = \partial_x x^2 \partial_x \mu - \frac{\Lambda/\Thg}{x^2} \mu.
\end{align}
Since at low frequencies, $\Lambda$ varies only slowly with $x$ (cf. Sect.~\ref{sec:DC_em}), we can replace it by a constant, $\Lambda(x)\approx \Lambda(\xc)=\Thg \xc^2$. The idea is that $\Lambda(x)$ evaluated at $\xc$ roughly determines the maximum of the emission. The lowest order solution found by \citet{Sunyaev1970mu} therefore reads $\mu^{(0)}(\tau, x)=\mu_\infty(\tau)\,\expf{-\xc(\tau)/x}$. This solution becomes constant at high frequencies and vanishes at low frequencies. It does, however, not follow our normalization condition $\kappa_\rho=\kappa^{\rm c}_\rho$, but the deviation is of higher order in $\xc$ and thus is neglected now.\footnote{A small improvement is in principle possible here. Using $\mu^{(0)}(x, \tau)$ one can show that for nearly frequency independent $\Lambda(x)$ the maximum emission arises around $x_{\rm m}\simeq \xc/2$ instead of $\xc$. Therefore, instead of Eq.~\eqref{eq:xc_def} one should use $\Lambda(\xc/2)(1-\expf{-\xc/2})/(\xc/2)=\Thg \xc^2$ to determine $\xc$. This aspect becomes noticeable during the BR-era, but for simplicity we shall treat the associated difference as a correction.}

\subsubsection{Lowest order solution for $\mu_\infty(\tau)$}
To lowest order in $\xc$, we can set $\mathcal{I}_{\hat{\mu}}\simeq 1$. Assuming that energy release occurs only at one single heating redshift, $\zh$, we then have $\tau_{\mu, 0}(\zh, 0)\approx\gamma_\mu \int^{\zh}_0 (1+z)\,\xc(z)  \id z$. Using $H(1+z)/\sigT \Ne c \approx 4.79$ for $H\approx \pot{2.09}{-20}(1+z)^2\sec^{-1}$ [radiation-dominated era], the coefficient $\gamma_\mu$ is given by
\beal
\gamma_\mu
&\approx \gamma_N
\,\frac{c N_{\rm e, 0}\sigT}{H_{\rm rad, 0}} \frac{kT_0}{\me c^2} \approx \pot{7.45}{-11},
\end{align}
where $N_{\rm e, 0}\approx \pot{1.12}{-5}\,\Omega_{\rm b} h^2(1-\Yp/2)\,{\rm cm}^{-3}$ is the electron number density at $z=0$ and $H_{\rm rad, 0}=\Omega_{\rm rel}^{1/2} H_0\approx \pot{2.09}{-20} \sec^{-1}$. Here, $\Omega_{\rm rel}$ is the density parameter of relativistic species (radiation + neutrinos) and $H_0$ denotes the Hubble parameter today.
Neglecting BR [i.e. $\xc \propto (1+z)^{1/2}$] then yields
\beal
\mu_\infty(z=0) 
&\approx 1.401 \frac{\Delta \rho_\gamma}{\rho_\gamma}\,\expf{-(\zh/\zmudc)^{5/2}},
\end{align}
or $\tau_{\mu, 0}^{\rm DC}(z)=(z/\zmudc)^{5/2}$, where the DC thermalization redshift is given by $\zmudc = \left[(2/5)\gamma_\mu \xc^{\rm DC, 0}(z=0)\right]^{-2/5}\approx \pot{1.98}{6}$. The exponential factor is the distortion visibility function, $\mathcal{J}_{\rm DC}=\expf{-(\zh/\zmudc)^{5/2}}$, discussed above. 
Using $\xc=\xc^{\rm BR}$ from Eq.~\eqref{eq:xc_BR_appr} instead, we find
\beal
\label{eq:tau_mu_BR}
\tau^{\rm BR}_{\mu,0}(z)&\approx 
\gamma_\mu
\int^z_0 (1+z)\,\xc^{\rm BR}  \id z \approx \left(\frac{1+z}{\pot{5.27}{6}}\right)^{1.328}.
\end{align}
In the classical result, given first by \citet{Sunyaev1970mu}, the power-law coefficient is $5/4=1.25$ because a different approximation for the BR Gaunt factor was utilized.
Similar expressions were also given by \citet{Danese1982} and \citet{Hu1993}. This shows that the thermalization redshift is significantly higher when only BR is included. In addition, the distortion visibility function is less steep at $z\gtrsim \pot{5.27}{6}$.

Since for energy release at high redshifts the photon distribution evolves through both the DC- and BR-era, we need to take the full expression for $\xc$ into account when computing $\tau_\mu(z, z')$. With $\xc\approx [(\xc^{\rm DC})^2+(\xc^{\rm BR})^2]^{1/2}$, the integral is rather simple and given by 
\beal
\label{eq:tau_mu_DCBR_high}
\tau_{\mu,0}(z)
&\approx \frac{5}{5-\gamma}\,\left(\frac{z}{\zmudc}\right)^{5/2}\sqrt{1+(z/\zmubr)^{-\gamma} } 
\\ \nonumber
&\qquad \times\left[1-\frac{\gamma \,(z/\zmubr)^\gamma}{5+\gamma} \!\left._2F_1\left(1, 1+\frac{5}{2\gamma}, \frac{3}{2}+\frac{5}{2\gamma}, -(z/\zmubr)^\gamma\right)\right.\!\right],
\end{align}
where we introduced $\gamma=2\times 0.672+1=2.344$, $\zmudc=\pot{1.98}{6}$, $\zmubr=\left[\xc^{\rm BR, 0}(z=0)/\xc^{\rm DC, 0}(z=0)\right]^{2/\gamma}\approx \pot{3.81}{5}$ and $\left._2F_1(a, b, c, x)\right.$ is the hypergeometric function. Here, we neglected DC temperature and frequency corrections. All the coefficients just follow from the expressions for the critical frequency. 

\begin{figure}
\centering
\includegraphics[width=\plotwd]{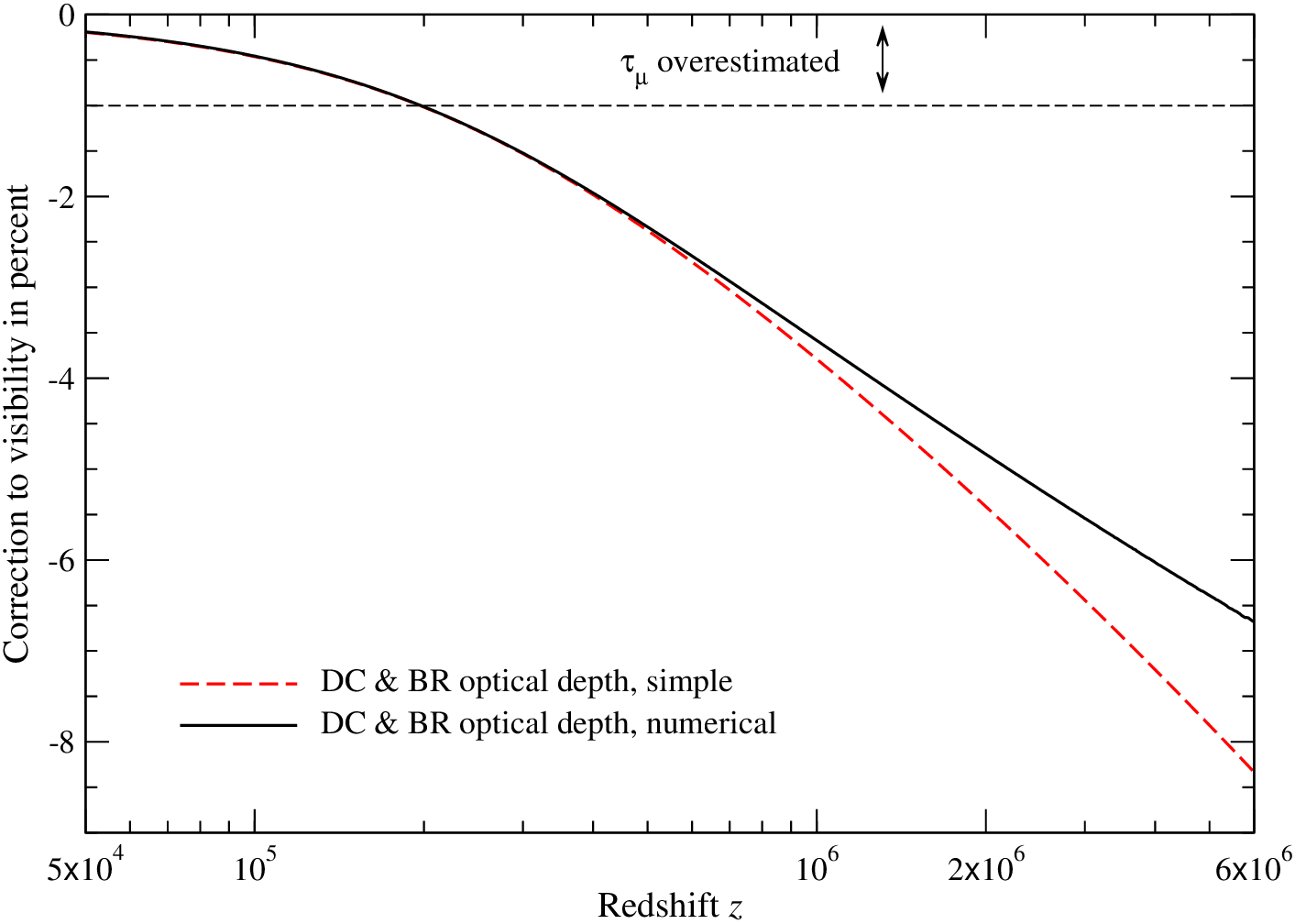}
\caption{Change in the distortion visibility, $\Jbb=\expf{-\tau_{\mu, 0}(z, 0)}$, when using the total optical depth including DC and BR. The dotted line is for the simple approximation Eq.~\eqref{eq:tau_mu_DCBR_high}, while the solid line is obtained by evaluating Eq.~\eqref{eq:tau_mu_a} numerically. We compared to $\mathcal{J}_{\rm DC}=\exp(-[z/\zmudc]^{5/2})$ with thermalization redshift $\zmudc\approx \pot{1.98}{6}$. The thermalization optical depth correction was computed between $z$ and $z=0$, so that the correction is slightly overestimated (see text).}
\label{fig:Vis_non_rel}
\end{figure}

In Fig.~\ref{fig:Vis_non_rel}, we illustrate the effect on the visibility function, comparing with $\mathcal{J}_{\rm DC}$.
Close to the thermalization redshift $\zmudc\simeq \pot{2}{6}$, the visibility of spectral distortions is reduced by $\simeq 5\%$ in comparison to the DC only approximation.
This is in good agreement with the recent findings of KS12.
We only show the correction to the visibility function up to $\sim 3$ times the thermalization redshift, since there $\Jbb\simeq \pot{1.7}{-7}$, which for $\Delta \rho/\rho \simeq 1\%$ could still lead to a detectable $\mu$-distortion for PRISM. We see that at high redshifts the approximation, Eq.~\eqref{eq:tau_mu_DCBR_high}, starts to break down. This is because especially around $z\simeq \pot{4}{5}$ the true critical frequency deviates slightly from the approximation $\xc\approx [(\xc^{\rm DC})^2+(\xc^{\rm BR})^2]^{1/2}$, which leads to degradation of the total integral for very large $\zh$. 
Full numerical determination of $\xc$ is trivial and Eq.~\eqref{eq:tau_mu_DCBR_high} will thus only be used for estimates.

We also already explained in Sect.~\ref{eq:transport_estimate} that photon transport ceases at $z\lesssim \pot{2}{5}$. Photons produced by BR below this redshift are stuck at low frequencies and no longer help thermalizing the full spectrum. The total DC thermalization optical depth between $z\simeq \pot{2}{5}$ and $z=0$ is only $\tau_\mu\approx 0.003$, but when including  BR, from Eq.~\eqref{eq:tau_mu_BR} we find $\tau_\mu\simeq 0.01$. This leads to an $\simeq 1\%$ overestimation of the thermalization efficiency and hence a similar error in the distortion visibility function. Corrections to the thermalization optical depth should thus only be computed at $\pot{2}{5}\lesssim z$, a modification that is straightforward to include but was omitted before.

\subsubsection{Compton equilibrium temperature}
\label{sec:TRJ_Te_condition}
To check the consistency of the solution, we briefly turn to the condition $\Te=\TRJ$. Using $\mu^{(0)}=\mu_\infty(\tau)\,\expf{-\xc(\tau)/x}$, we can readily compute the Compton equilibrium temperature in the distorted radiation field. From Eq.~\eqref{eq:Te_CS}, neglecting terms $\mathcal{O}(\Thg^2)$ it is given by 
\beal
\label{eq:Te_eq_sol}
\Te^{\rm eq, \rm (0)}\approx \TRJ^{(0)}-\frac{\Tg}{4\mathcal{G}^{\rm Pl}_3}\int_0^\infty x^2 {Y}_{\rm SZ}(x) \, \mu^{(0)}(\tau, x)\id x.
\end{align}
For frequency independent $\mu^{(0)}(\tau, x)$, this immediately implies $\Te^{\rm eq, \rm (0)}= \TRJ^{(0)}$ because $\int x^2 {Y}_{\rm SZ}(x) \id x$ vanishes; however, when inserting the lowest order solution, the integral no longer vanishes and $\Te^{\rm eq, \rm (0)}$ deviates from $\TRJ^{(0)}$ by $\mathcal{O}(\xc/10)$. Since after the energy release stopped, at lowest perturbation order one should find $\Te^{(0)}=\Te^{\rm eq, (0)}\equiv \TRJ^{(0)}$, this means that the solution is slightly inconsistent, but the discrepancy is indeed of higher order in $\xc$. 

\subsection{First-order corrections}
\label{sec:corrs_O1}
At this point, we have not included any additional physics but simply kept all lowest order terms, consistent with $\mathcal{O}(\xc)$ in the evolution equation for $\mu$ and definition of $\tau_\mu$. The derived corrections were already present in previous numerical calculations and even without improvements of the BR and DC Gaunt factors introduced in {\sc CosmoTherm} could have been obtained with little effort.
The next step is to obtain the corrections to $\mathcal{I}_{\hat{\mu}}$ and then evaluate the changes to the optical depth term, $\Delta\tau_\mu(z)$, given by Eq.~\eqref{eq:tau_mu_b}, adding terms of $\mathcal{O}(\xc^2)$ to the integrant.

\subsubsection{Correction to $\hat{\mu}(\tau, x)$ at order $\mathcal{O}(\xc)$}
Since the photon production integral, Eq.~\eqref{eq:DC_BR_I_mu}, is of $\mathcal{O}(\xc)$, we have to improve the solution of $\hat\mu(\tau, x)$ to order $\mathcal{O}(\xc)$. From the lowest order solution, we know already that the time derivatives $\partial_\tau \mu$ and $\partial_\tau (\Te/\Tg)$ are both of order $\mathcal{O}(\xc)$. Similarly, $\Lambda(x)/\Thg\simeq \xc^2$. We furthermore already understand that the main frequency dependence of the solution is found around $x\simeq \xc\ll 1$. Defining $\sigma=\xc y = \xc \int \Thg \id \tau$, and re-scaling the frequency as $x=\xc\,\xi$, we have the evolution equation:
\beal
\nonumber
\xc \pAb{{\mu}}{\sigma}-\xc^2 \xi\, \partial_\sigma \frac{\Te}{\Tg}
&\approx
\xi^2 {\mu}'' 
+2  g_1(\xc \xi) \,\xi  {\mu}' 
-\frac{\lambda(\xc\xi)}{\xi^2}\frac{(1-{\rm e}^{-\xc \xi})}{\xc\xi} {\mu},
\end{align}
where primes now denote derivatives with respect to $\xi$ and we defined $\lambda=\Lambda(x)/[\Thg  \xc^2]=1+\Delta \lambda$ [for double Compton $\Delta \lambda = 0$ when neglecting frequency corrections to the Gaunt factor]. Scaling the equations in this way shows that both the temperature term on the l.h.s. of this equation and the higher order Compton correction [$\propto 2(g_1-1)\simeq -x^2/6[1-x^2/60]$ for $x\ll 1$] enter the problem at higher order in perturbation theory, so that we neglect them for now\footnote{As we will see in Sect.~\ref{sec:other_terms}, this is too naive and the Compton terms give $\mathcal{O}(\xc)$ contributions at intermediate frequencies.}.
For the correction to the emission term, we have $\lambda(\xc\xi)(1-{\rm e}^{-\xc \xi})/(\xc\xi)\approx 1-\xc \xi/2+\mathcal{O}(\xc^2)$ in the DC-era. During the BR-era, matters are complicated by the logarithmic dependence of the Gaunt factor on $x$. This means that in this case deviations from $\Lambda(x)=\rm const$ enter at order $\simeq \xc\ln(\xc)$. For simplicity we define
\beal
\label{eq:alpha_emission}
\alpha_{\rm em}(x)=\Lambda(x)(1-{\rm e}^{-x})/(\Thg  \xc^2 x)-1,
\end{align}
and include all frequency correction terms simultaneously at the first perturbation order. For the time derivative of $\mu^{(0)}(x, \tau)$, we have 
\beal
\label{eq:mu_O0_dy}
\pAb{{\mu}^{(0)}(\xc(\tau) \xi, \tau)}{y}
&\approx {\mu}^{(0)}
\left[\pAb{\ln \mu_\infty^{(0)}(\tau)}{y}-\frac{\xc}{x}\,\pAb{\ln \xc(\tau)}{y}\right].
\end{align}
The first term was recently considered by KS12, although there it was treated as ${\mu}^{(0)}\partial_y \ln \mu_\infty^{(0)}(\tau)\rightarrow {\mu}\,\partial_y \ln \mu_\infty^{(0)}(\tau)$, giving a modified Bessel function solution for $\mu$. The second term leads to a small correction at high redshifts ($z\gtrsim 10^6$), but in our approach it does become significant later. 

Put together, this then determines the evolution equation for the first correction $\mu^{(1)}(\tau, x)=\mu^{(0)}_\infty(\tau) \hat\mu^{(1)}(\tau, x)+\mu^{(1)}_\infty(\tau) \hat\mu^{(0)}(\tau, x)$:
\beal
\label{eq:mu_O1_correction}
S^{(1)}(x)=\left[\frac{\partial \ln \mu^{(0)}}{\partial y}+\alpha_{\rm em}(x)\right]\hat\mu^{(0)}
&\approx\partial_x x^2 \partial_x \hat\mu^{(1)}-\frac{\xc^2}{x^2} \hat\mu^{(1)}.
\end{align}
The general solution of this equation reads
\beal
\label{eq:mu1_sol}
\hat\mu^{(1)}&={\rm C}_1\,{\rm e}^{-\xc/x}+{\rm C}_2\,{\rm e}^{\xc/x}
+\int_0^x \sinh\left(\frac{\xc}{x'}-\frac{\xc}{x}\right) S^{(1)}(x')\frac{\id x'}{\xc},
\end{align}
where ${\rm C}_1$ and ${\rm C}_2$ are fixed by the boundary conditions. Inserting the expression for $S^{(1)}(x)$, then gives the first-order correction to the chemical potential as
\beal
\label{eq:mu_sol_upto_O1}
\hat\mu^{(1)}&\approx {\rm C}_1\,{\rm e}^{-\xc/x} 
+\ln (x/\xc) \,{\rm e}^{-\xc/x} \partial_y\ln \mu^{(0)}_\infty
\\ \nonumber
&\quad
+D_\mu(\xc/x)\left[\partial_y\ln \mu^{(0)}_\infty+\frac{1}{2}\partial_y\ln\xc\right] + D_{\rm em}(\xc, \xc/x) \,\xc.
\end{align}
The new integration constant can be fixed by requiring $\kappa_\rho=\kappa^{\rm c}_\rho$ (see Sect.~\ref{sec:const_C1}). We directly absorbed any contribution from the source term leading to an asymptotic behavior $\propto {\rm e}^{-\xc/x}$ at $x\ll 1$ into this integration constant. The functions $D_\mu$ and $D_{\rm em}$ are defined as
\beal
\label{eq:D_i_corrs}
&D_\mu(\zeta) ={\rm e}^{-\zeta} \left[\ln (2\zeta)+\gamma_E-{\rm e}^{2\zeta}\,{\rm Ei}(-2\zeta)\right]
\nonumber\\ 
&D_{\rm em}(\xc, \zeta) = F_{\rm em}(\xc, 0)\,{\rm e}^{-\zeta}-F_{\rm em}(\xc, \zeta)
\\ \nonumber
&F_{\rm em}(\xc, \zeta) =\frac{{\rm e}^{-\zeta}}{2\xc} \!\int_0^{x} \zeta' \alpha_{\rm em}(x')\,\left[{\rm e}^{2(\zeta-\zeta')}-1\right]\frac{\id  x'}{x'},
\end{align}
where in the expression for $F_{\rm em}$ we use $x=\xc/\zeta$ and $x'=\xc/\zeta'$. The integral for $F_{\rm em}$ can be carried out numerically very efficiently.

\begin{figure}
\centering
\includegraphics[width=\plotwd]{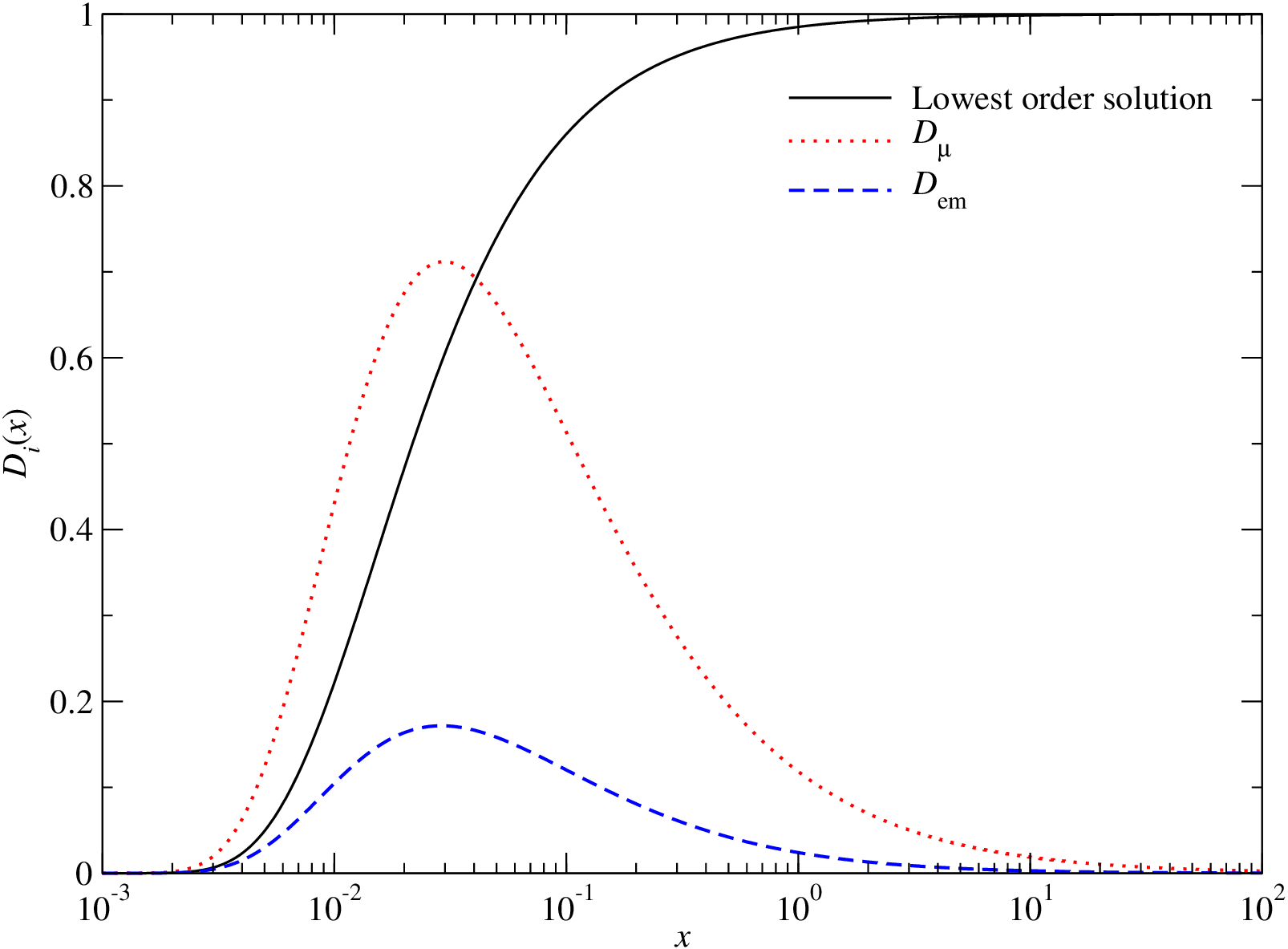}
\\[3mm]
\includegraphics[width=\plotwd]{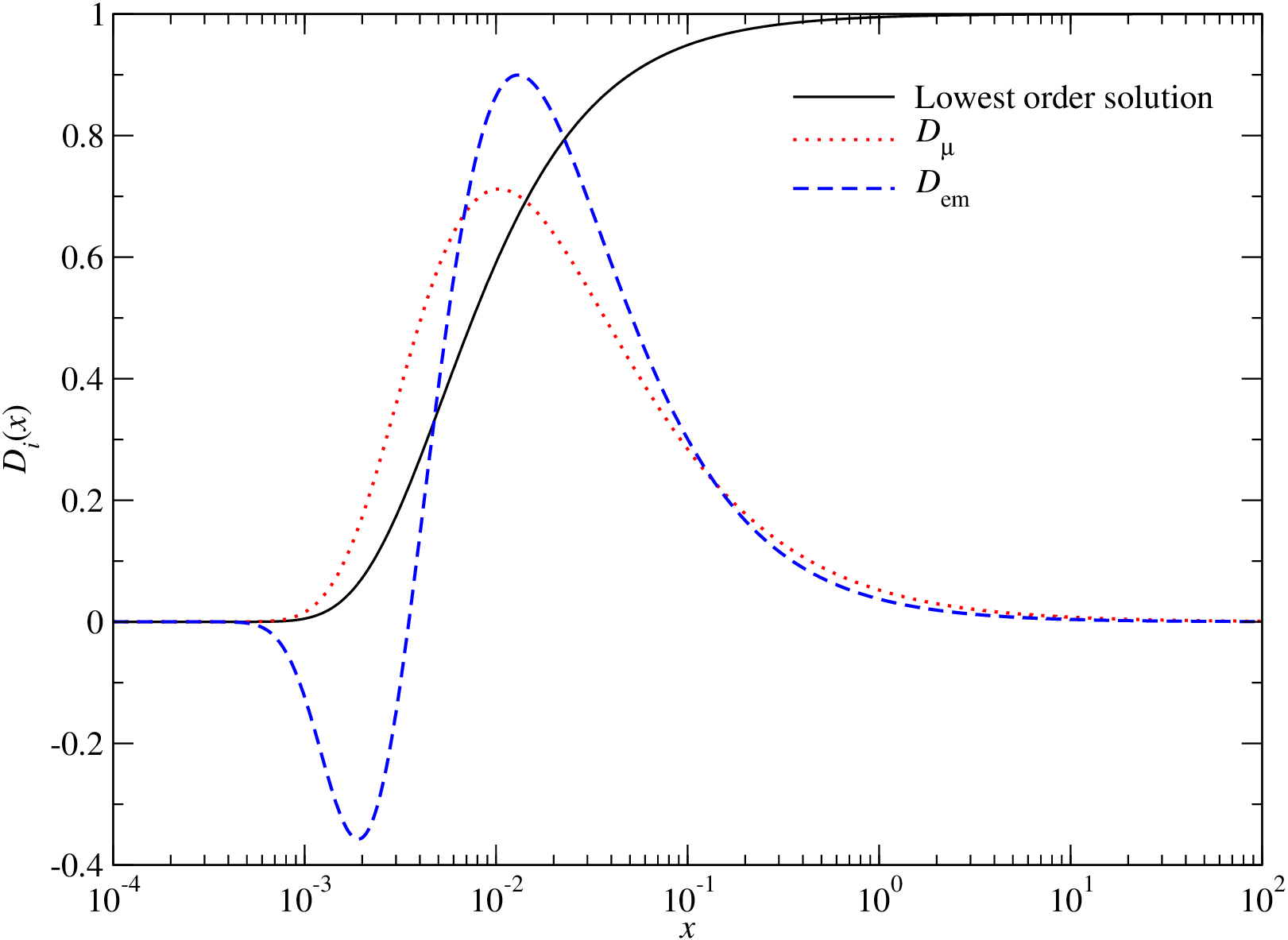}
\caption{Frequency dependence of the correction functions $D_\mu$ and $D_{\rm em}$ in comparison to $\hat{\mu}^{(0)}=\expf{-\xc/x}$. The upper panel illustrates the case for $\xc \simeq 0.015$ ($z\simeq \pot{6}{6}$). $D_\mu$ and $D_{\rm em}$ have a very similar shape at low frequencies, although rescaling them to coincide around the maximum reveals small differences at $x\gtrsim 1$. In the lower panel, we show the functions for $\xc\simeq \pot{5.3}{-3}$ ($z\simeq \pot{5}{5}$), for which, due to the logarithmic dependence of the BR Gaunt factor on frequency, $D_\mu$ and $D_{\rm em}$ differ significantly.}
\label{fig:D_mu_D_em}
\end{figure}

We scaled the two correction function so that they are comparable in amplitude. Their shapes are illustrated in Fig.~\ref{fig:D_mu_D_em}.
\begin{figure}
\centering
\includegraphics[width=\plotwd]{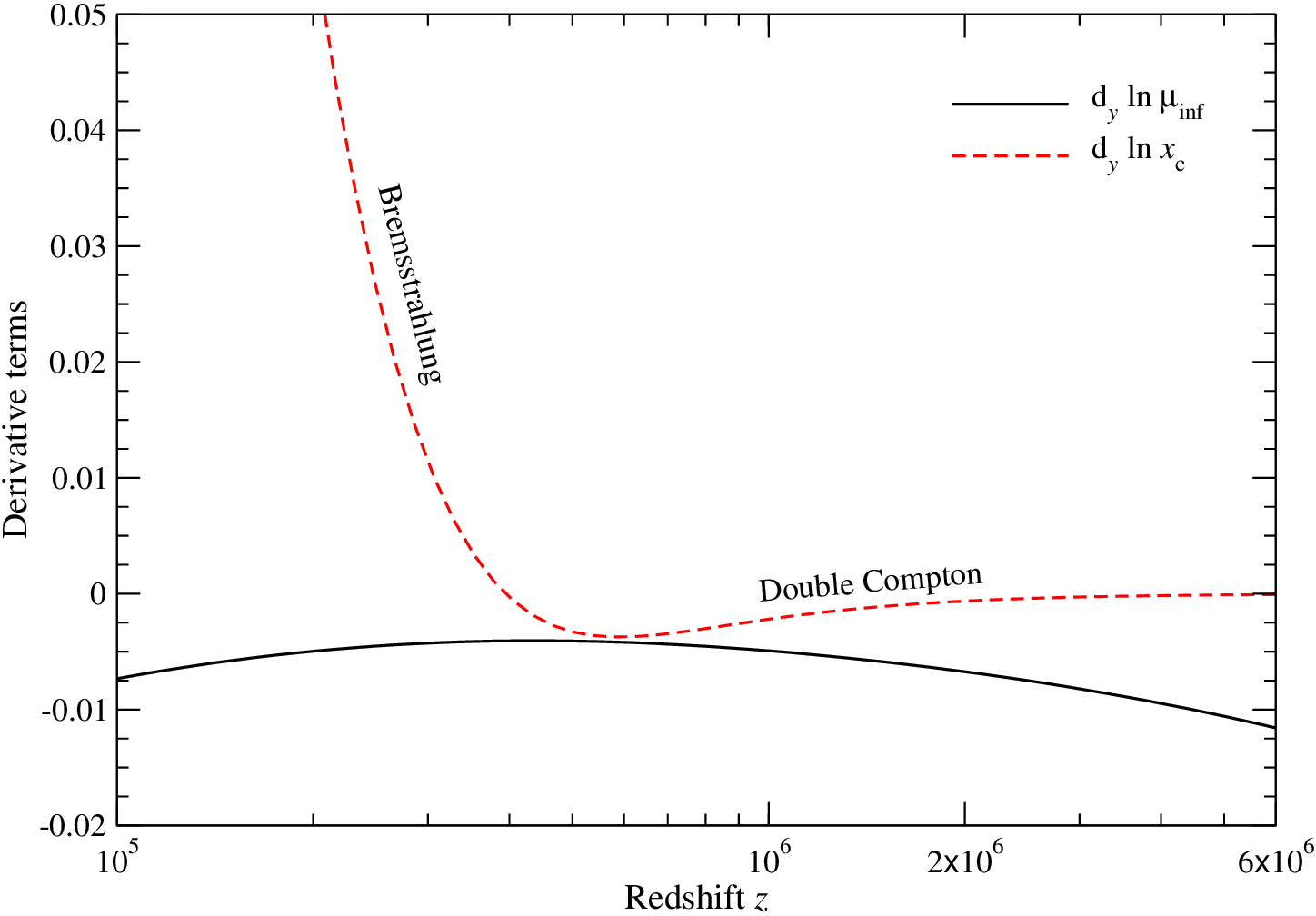}
\caption{Contributions to the time derivative of the lowest order solution to $\mu_\infty$. At late times, the low-frequency spectrum changes mostly because of emission and absorption processes ($\partial_y \ln \xc$ becomes significant), so that scattering-driven quasi-stationarity no longer is a good assumption.}
\label{fig:derivatives_mu_xc}
\end{figure}
The main correction appears around $x\simeq 2\xc$, where also most of the photon emission comes from. This also implies that, while these two correction functions change little for the total energetics, they directly affect the thermalization efficiency.  
In particular at $z\lesssim 10^6$, when BR starts dominating, the correction related to $D_{\rm em}$ becomes significant, giving rise to a nontrivial dependence on frequency. This is partially due to the small mismatch of $\xc$ with the real position of the emission maximum caused by the frequency dependence of the DC and BR Gaunt factors, but also the $\simeq (1-\expf{-x})/x$ modulation of the emission term in Eq.~\eqref{eq:DC_term}.

To obtain the final solution, we can use $\partial_y\ln \mu^{(0)}_\infty \approx - 0.7769 \xc$, which follows from Eq.~\eqref{eq:mu_infty_final_eq}. 
The derivative of the critical frequency with respect to $y$ is approximately given by
\beal
\label{eq:deriv_xc}
\frac{\partial_\tau \xc}{2\xc \Thg}
&\approx \frac{\pot{2.01}{-2}}{1+z}\frac{\partial_\tau z}{\xc^2}
\left(1-\pot{2.75}{-2}\left[\frac{1+z}{\pot{2}{6}}\right]^{-2.344}\right),
\end{align}
where we used the expression given in Sect.~\ref{eq:xc_defs}, but neglected relativistic corrections. During the radiation-dominated era ($z\gtrsim 3300$), we have $\partial_\tau z=-H(1+z)/\sigT \Ne c \approx - 4.80$.
The second term in parentheses arises because of BR, which can be neglected at high redshifts. 
At $z=\pot{2}{6}$, we find $\frac{1}{2}\partial_y \ln \xc\approx - \pot{6}{-4}\simeq - 0.07\xc$; however, at $z\simeq\pot{2}{5}$ we have $\frac{1}{2}\partial_y \ln \xc\simeq \pot{6}{-2}\simeq 9.4\xc$ (see Fig.~\ref{fig:derivatives_mu_xc} for more details), which shows that at low frequencies the quasi-stationary approximation starts to break down in this regime. This behavior is expected, since the efficiency of Compton scattering decreases with redshift, so that full kinetic equilibrium between photons and electrons can no longer be established. 
We thus do not expect to obtain very accurate analytic approximations at $z\lesssim \pot{2}{5}$ using our perturbative approach. 

At $z\lesssim \pot{2}{5}$, photons produced by BR are furthermore stuck at low frequencies and no longer up-scatter strongly (see Sect.~\ref{eq:transport_estimate}). In this regime, the shape of the low-frequency spectrum is fully determined by photon emission and absorption, and the effect of scattering can be added as a perturbation. Since here we are mainly interested in the $\mu$-era, we leave a more detailed discussion of this problem for another paper.

\subsubsection{Fixing the integration constant ${\rm C}_1$}
\label{sec:const_C1}
To give the full first-order solution, we still need to determine the integration constant ${\rm C}_1$ in Eq.~\eqref{eq:mu_sol_upto_O1}. It simply follows from our normalization condition $\kappa_\rho\equiv \kappa_\rho^{\rm c}$.
Since the lowest order solution is $\hat\mu=\expf{-\xc/x}$, we can just set $1+{\rm C}_1=A(\xc)$ in Eq.~\eqref{eq:mu_sol_upto_O1} and directly discuss the full solution up to first order in $\xc$.
Just using $\hat\mu=A_0(\xc)\,\expf{-\xc/x}$, we find that
\beal
\label{eq:kappa_zero}
A_0(\xc)&\approx [1 - 1.85 \, \xc^{0.83}/(1+2.44\xc)]^{-1}
\end{align}
fulfills the normalization condition for $10^{-3}\leq \xc\leq 0.05$ very well. 

For $\hat\mu=A(\xc)\,\expf{-\xc/x}+\ln(x/\xc)\,\expf{-\xc/x} \partial_y\ln \mu^{(0)}_\infty$, we obtain the correction
\beal
\label{eq:kappa_zero_log}
\Delta A_{\ln}(\xc)&\approx 1.06 \,A_0(\xc)(1 + 0.98\ln\xc) \, \partial_y\ln \mu^{(0)}_\infty
\end{align}
to $A_0(\xc)$. Since the term associated with $D_\mu$ can become significant at late times, we have the additional contribution
\beal
\label{eq:kappa_zero_log_Dmu}
\Delta A_\mu(\xc)&\approx 1.15\xc\,A_0(\xc)[1+22.7(1-0.97 \xc^{0.029})\ln\xc]\, \alpha(\tau),
\end{align}
with $\alpha(\tau)= \partial_y[\ln \mu^{(0)}_\infty+(1/2)\ln \xc]$, which works well when using $\hat\mu=A(\xc)\,\expf{-\xc/x}+\ln(x/\xc)\,\expf{-\xc/x} \partial_y\ln \mu^{(0)}_\infty+D_\mu(\xc/x) \alpha(\tau)$. Note that in terms of perturbations, $\Delta A_\mu(\xc)$ formally is of higher order in $\xc$ at $z\gtrsim \pot{4}{5}$; however, at late times it becomes pretty significant, so that we generally include it.

\begin{figure}
\centering
\includegraphics[width=\plotwd]{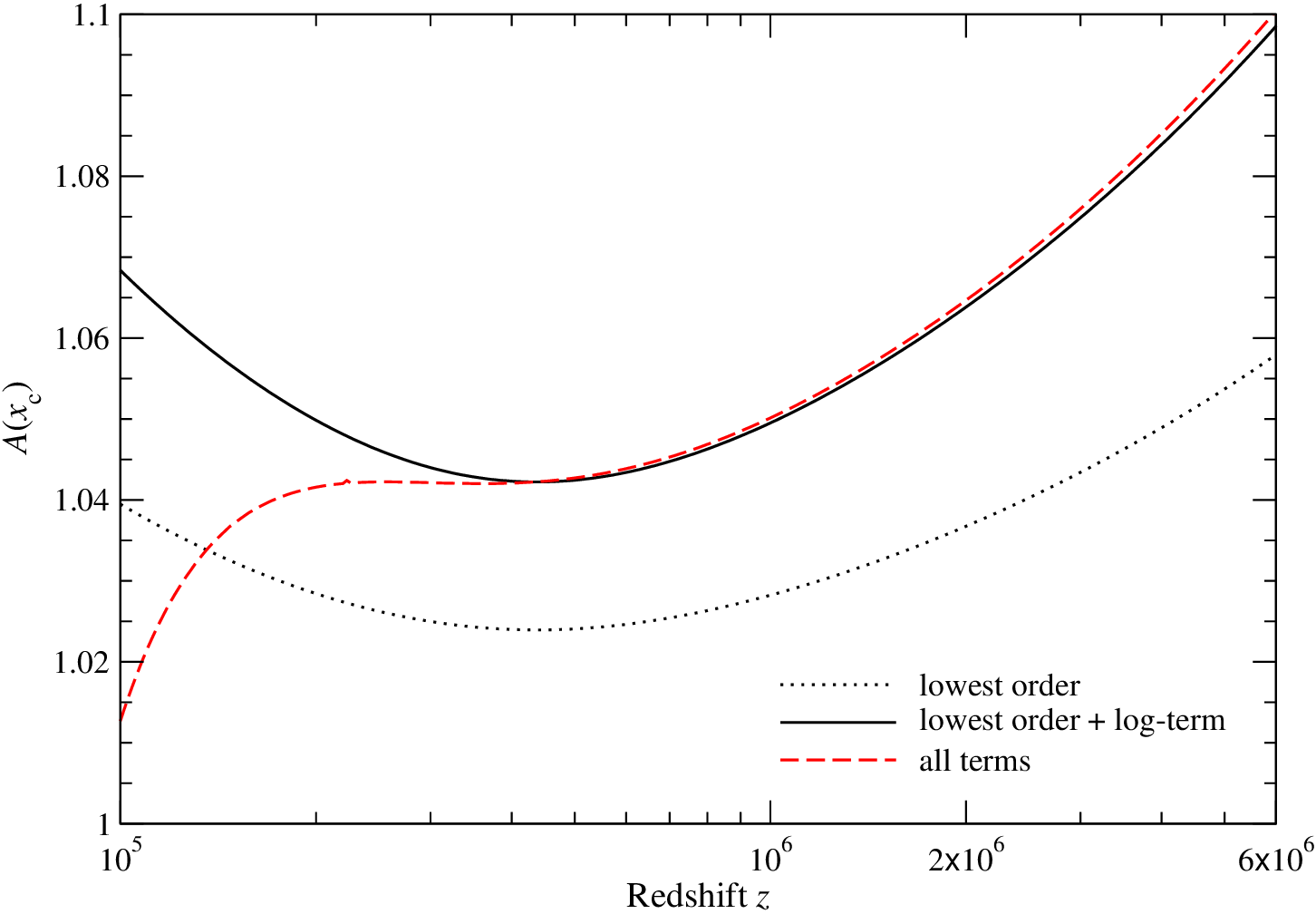}
\caption{Normalization constant $A(\xc)$ for the three cases discussed in Sect.~\ref{sec:const_C1}. The change in the normalization when adding $D_{\rm em}(x)$ is very small and can generally be neglected.}
\label{fig:Normalization}
\end{figure}

In Fig.~\ref{fig:Normalization}, we show the numerical results for $A(\xc)$ in different cases. The changes in the normalization are $\simeq 1\%-10\%$ in the shown redshift range. The $\ln(x/\xc)\,\expf{-\xc/x}$ term is clearly important at all times, and especially at late times the contribution from $D_\mu(\xc/x)$ becomes large. The correction due to $D_{\rm em} \xc$ is not as crucial and the normalization constant for the full solution is well represented by adding all terms from Eq.~\eqref{eq:kappa_zero}--\eqref{eq:kappa_zero_log_Dmu}.

\begin{figure}
\centering
\includegraphics[width=\plotwd]{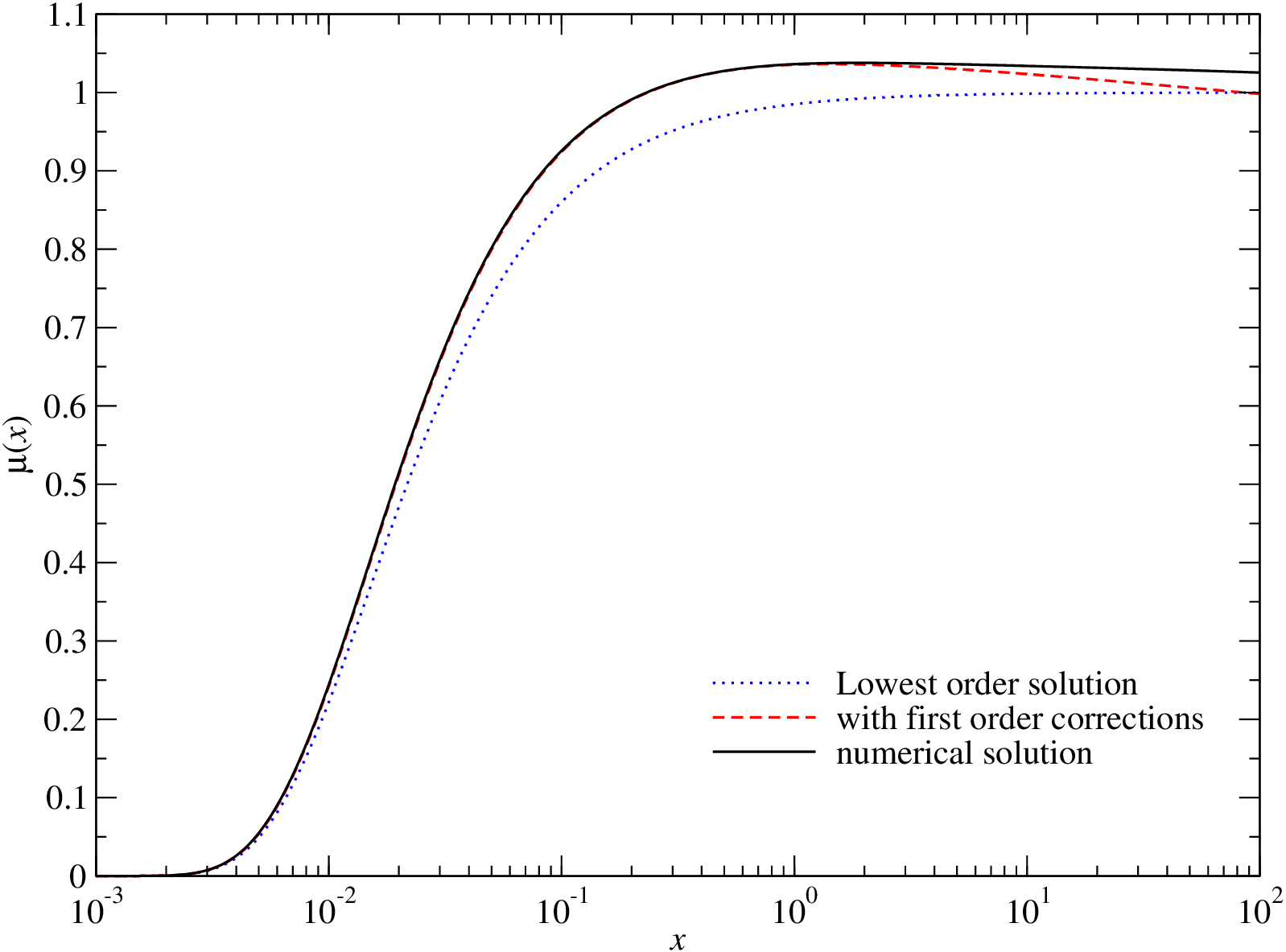}
\\[3mm]
\includegraphics[width=\plotwd]{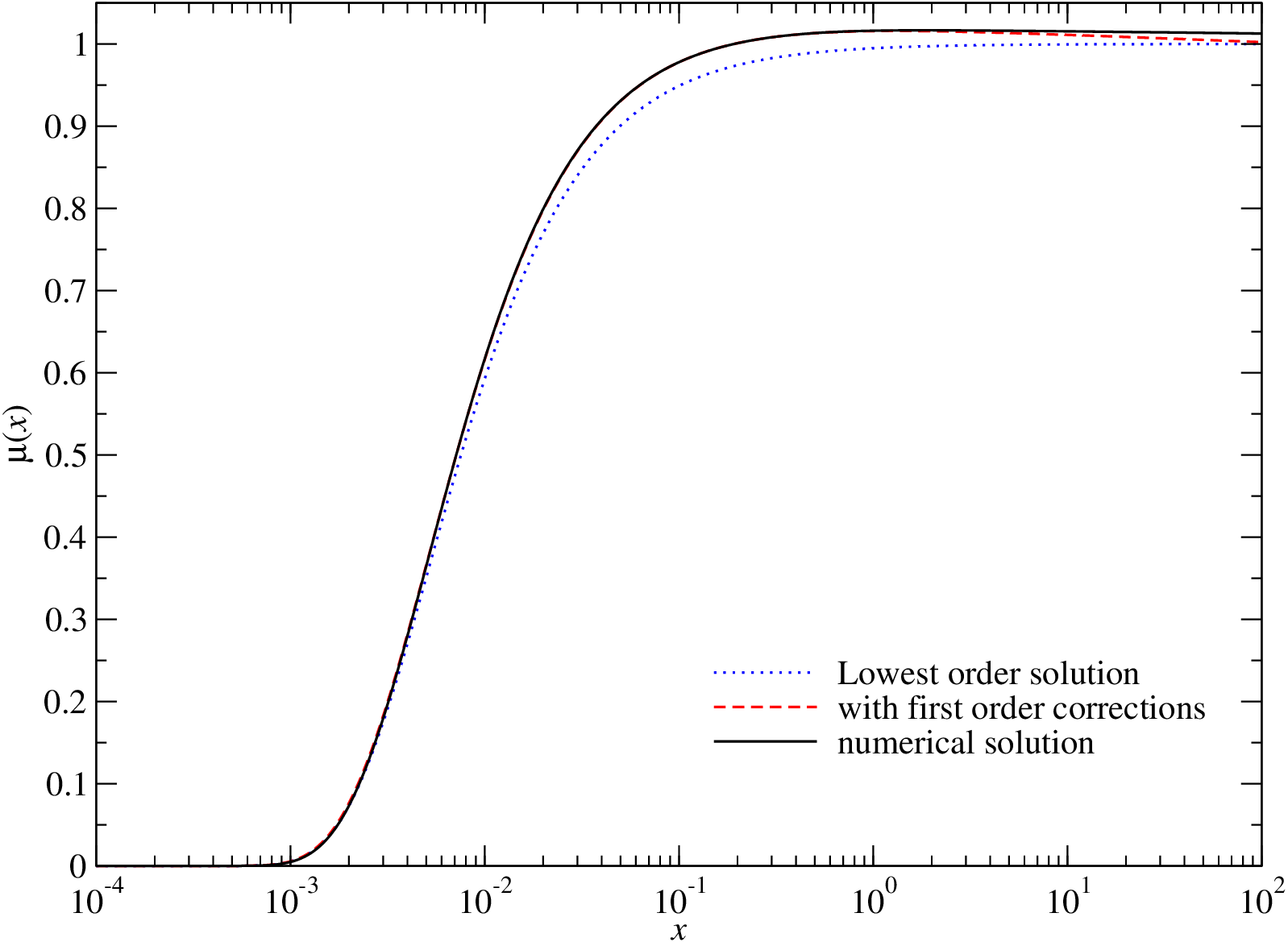}
\caption{Comparison of the lowest order solution $\hat{\mu}^{(0)}=\expf{-\xc/x}$ and Eq.~\eqref{eq:mu_sol_upto_O1} with the numerical result obtained with {\sc CosmoTherm}. We scaled the numerical solution for $\mu(t, x)$ by $\mu_\infty(t)=[3\Delta \rho_\gamma(t)/\rho_\gamma-4\Delta N_\gamma(t)/N_\gamma]/\kappa^{\rm c}_\rho$ in agreement with our normalization condition for $\hat\mu$. 
The upper panel shows the solution for $\xc=0.015$  ($z\simeq \pot{6}{6}$), while in the lower panel we have $\xc=\pot{5.3}{-3}$  ($z\simeq \pot{4.8}{5}$).
 The difference at high frequencies can be captured by matching with the high-frequency limit of the photon Boltzmann equation, giving extremely good agreement with the numerical result over the full range of frequencies (Sect.~\ref{sec:match} and Fig.~\ref{fig:Dmu_plot}).
}
\label{fig:compare_solutions}
\end{figure}

The solution for $\hat\mu$ is illustrated in Fig.~\ref{fig:compare_solutions} for two redshifts. The largest corrections are due to the renormalization factor, $A(\xc)$, and the logarithmic term in Eq.~\eqref{eq:mu_sol_upto_O1}, which are significant at both $x\simeq \xc$ and large $x$. These two terms capture the main behavior of the expression given by KS12. Notice that they normalized their solution using the condition $\mu_\infty=\mu(x=0.5)$. 
They furthermore neglected the correction caused by the $\simeq \mathcal{O}(\xc)$ emission terms leading to $D_{\rm em}$, as well as the time derivative of $\xc$, which does become significant at the later stages (see Fig.~\ref{fig:derivatives_mu_xc}). However, the difference due to this does not seem as crucial. 

In Fig.~\ref{fig:compare_solutions}, we show a comparison with the numerical result obtained with {\sc CosmoTherm}. Our approximation captures the shape of the solution very well. Only at high frequencies, which matters for the overall energetics of the solution but not as much for the photon production, the approximation Eq.~\eqref{eq:mu_sol_upto_O1} deviates noticeably from the full numerical result. This is expected since our perturbative approach is meant to work at $x\lesssim 1$ only. We can further improve the solution by matching with the high-frequency solution obtained in Sect.~\ref{sec:match}, reaching agreement at all relevant frequencies to $\simeq 0.1\%-1\%$ at $z\gtrsim \pot{\rm few}{5}$.

\subsubsection{Compton equilibrium temperature}
\label{sec:Compton_temp_first}
By construction, we should find $\Te=\TRJ$ at least to $\mathcal{O}(\xc^2\ln\xc)$, which implies the condition
\beal
\label{eq:condition_first_TRJ_Te}
0=\int x^2 \hat\mu(t, x) Y_{\rm SZ}(x) \id x=\left<\hat\mu\right>_{\rm SZ},
\end{align}
like in Sect.~\ref{sec:TRJ_Te_condition}. For this, we only need to worry about the first two terms of Eq.~\eqref{eq:mu_sol_upto_O1}, as the other are energetically much less important. We find 
\bsub
\label{eq:Integrals_SZ}
\beal
&\left<\expf{-\xc/x}\right>_{\rm SZ}\approx - 0.127 \xc
\\
&\left<\ln(x/\xc)\expf{-\xc/x}\right>_{\rm SZ}\approx - 0.277 (1+0.336 \xc^{0.598}),
\end{align}
\esub
which with $\partial_y\ln \mu^{(0)}_\infty\approx -0.7769 \xc$ implies an imbalance of order $\simeq 0.1 \xc$. This means that our solution is slightly inconsistent, but at this point we have no freedom left to `fix' this discrepancy. The situation is improved a bit once we correct the high-frequency solution using the asymptotic behavior determined in Sect.~\ref{sec:other_terms}. However, a small difference larger than $\mathcal{O}(\xc^2)$ remains. For the photon production rate, this discrepancy does not seem to matter and will be neglected below.

\begin{figure}
\centering
\includegraphics[width=\plotwd]{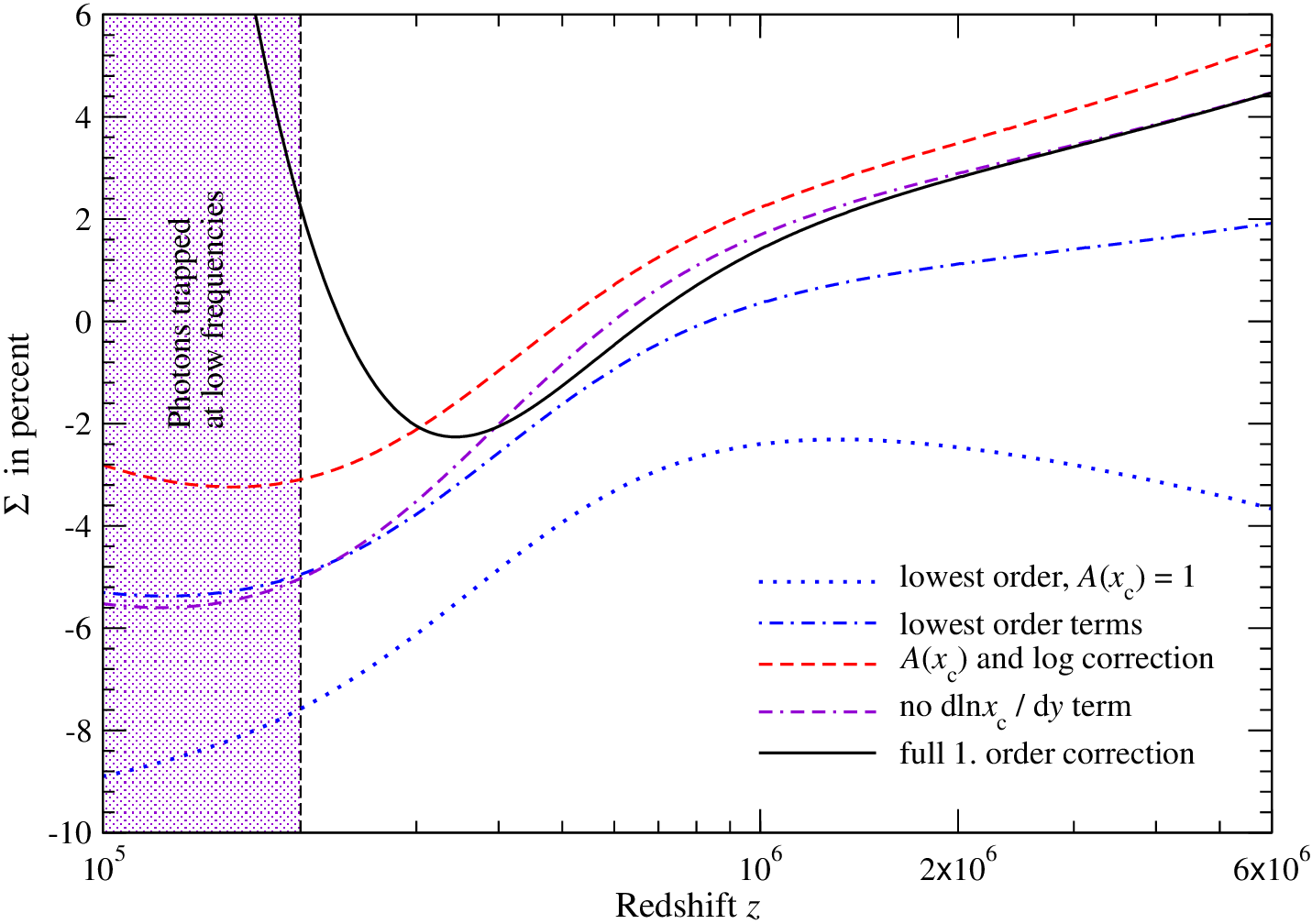}
\caption{Changes of $\Sigma= \mathcal{I}_{\hat{\mu}} -1$ for different approximations of $\hat\mu$ discussed in Sect.~\ref{sec:const_C1}. For the lowest order solution, we used $\hat\mu=A(\xc)\,\expf{-\xc/x}$ with $A(\xc)=1$ and $A(\xc)=A_0(\xc)$ defined by Eq.~\eqref{eq:kappa_zero}. For the dashed red line, we used $\hat\mu=A(\xc)\expf{-\xc/x}+\ln (x/\xc) \,{\rm e}^{-\xc/x} \partial_y\ln \mu^{(0)}_\infty$, while for the solid black line we used the full first-order expression, Eq.~\eqref{eq:mu_sol_upto_O1}, each with their corresponding normalization constants, $A(\xc)$. The double-dash dotted curve also gives the result using Eq.~\eqref{eq:mu_sol_upto_O1}, but when neglecting the contribution from $\partial_y \ln\xc$, which becomes large at low redshifts. The shaded region indicates where the high-frequency photon number freezes out.}
\label{fig:DImu_plot}
\end{figure}
\subsubsection{Change in the photon production rate}
\label{sec:corrs_O1_dN}
In Fig.~\ref{fig:DImu_plot}, we show how $\Sigma= \mathcal{I}_{\hat{\mu}} -1$ changes for different approximations. This is the relevant quantity for the optical depth correction, $\Delta \tau_\mu$, defined by Eq.~\eqref{eq:tau_mu_b}. Just using $\hat\mu=\expf{-\xc/x}$ already gives a significant correction; however, for consistency the renormalization $A(\xc)\neq 1$ has to be included. We note also that during the BR-era the logarithmic dependence of the Gaunt factor is very important, and assuming $\Lambda(x)={\rm const}$ gives incorrect results at $z\lesssim 10^6$.

If we neglect the contributions from $D_\mu$ and $D_{\rm em}$ in the approximation Eq.~\eqref{eq:mu_sol_upto_O1}, we obtain the dashed red line, where the difference to the previous case is only caused by the logarithmic term, $\propto \ln(x/\xc)\expf{-\xc/x}$.
For the violet double-dash dotted curve, we used the full first-order expression, Eq.~\eqref{eq:mu_sol_upto_O1}, for $\hat\mu$, but neglected the contribution from $\partial_y \ln\xc$, which becomes large at low redshifts. At $z\lesssim \pot{5}{5}$, neglecting the extra emission terms brings $\Sigma$ again closer to the lowest order case, while the change is much smaller at earlier times.
When also adding the contribution from $\partial_y \ln\xc$ we find a large change of $\Sigma$ during the BR-era. This difference is not as important eventually, since the total optical depth contribution becomes rather small at late times. Also, at $z\lesssim \pot{\rm few}{5}$, the approximation is not expected to be as accurate (we find that it works well until $z\simeq \pot{3}{5}$), but the overall effect on the visibility function remains small.

\begin{figure}
\centering
\includegraphics[width=0.99\columnwidth]{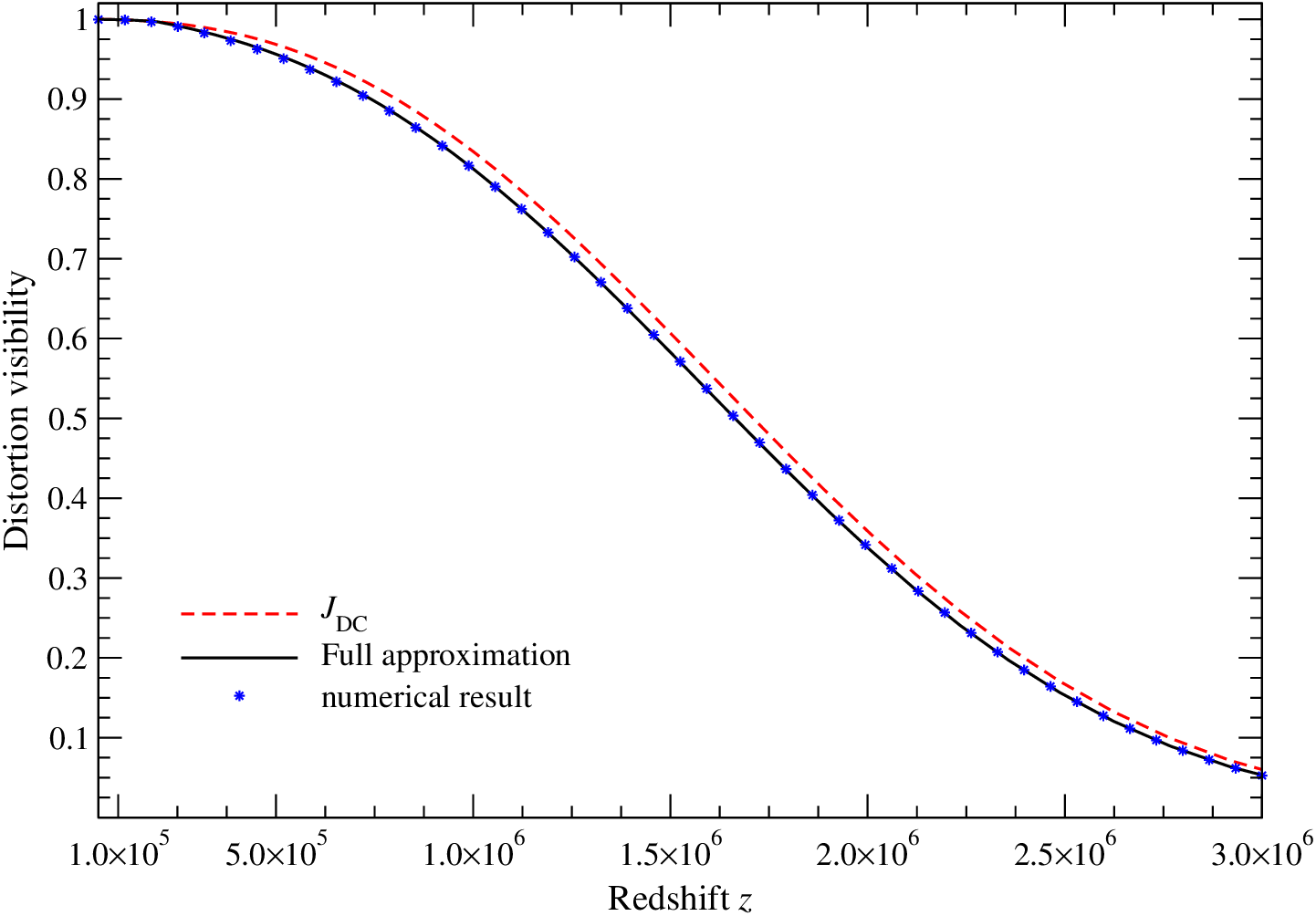}
\\[2mm]
\includegraphics[width=0.99\columnwidth]{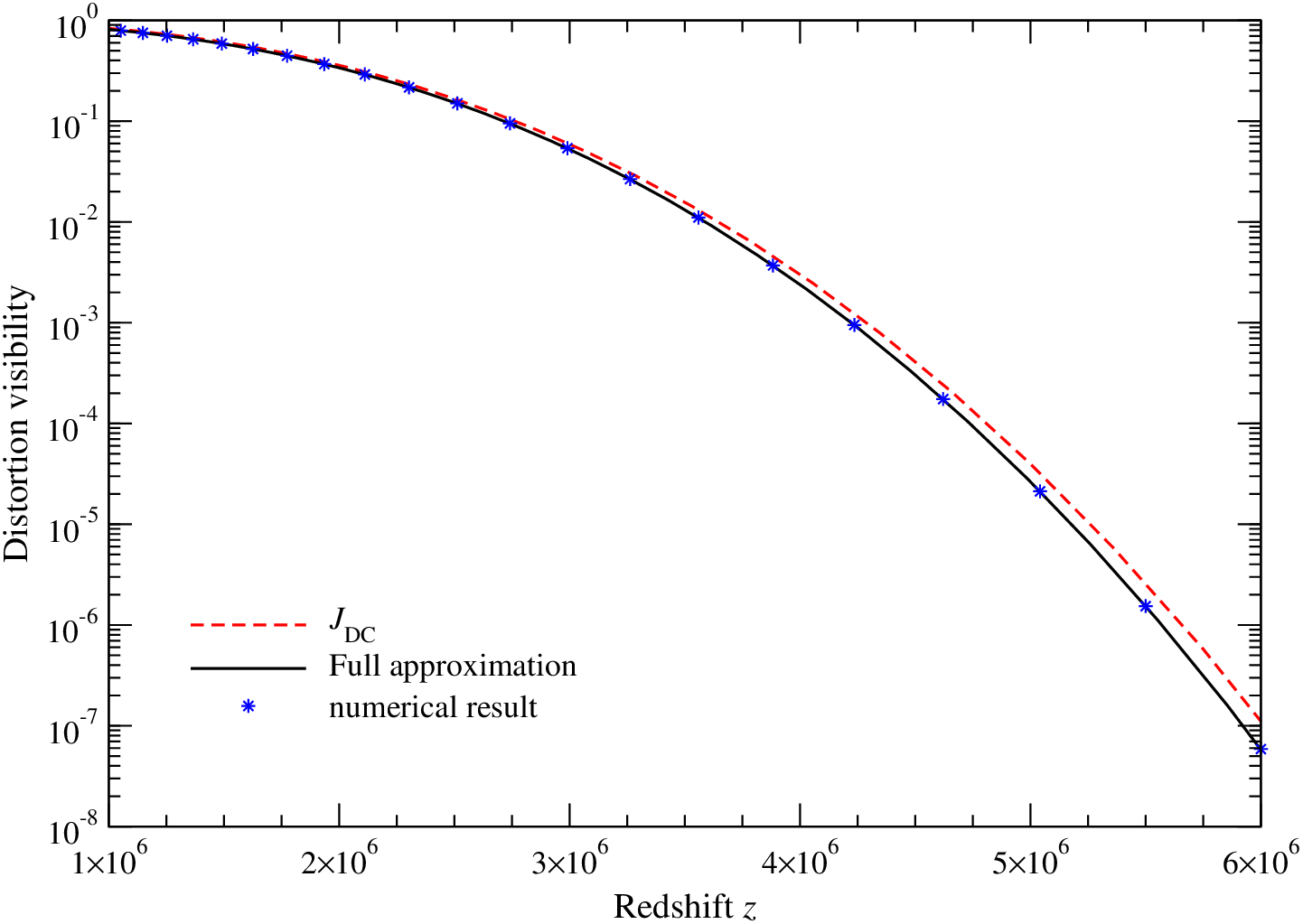}
\caption{Distortion visibility function at different redshifts. The red dashed curve shows $\mathcal{J}_{\rm DC}=\expf{-(z/\zmudc)^{5/2}}$ with $\zmudc=\pot{1.98}{6}$. The solid black line gives our approximation based on Eq.~\eqref{eq:mu_sol_upto_O1}, with all terms included. The numerical result was obtained using {\sc CosmoTherm}.}
\label{fig:Vis_plot}
\end{figure}

\begin{figure}
\centering
\includegraphics[width=\columnwidth]{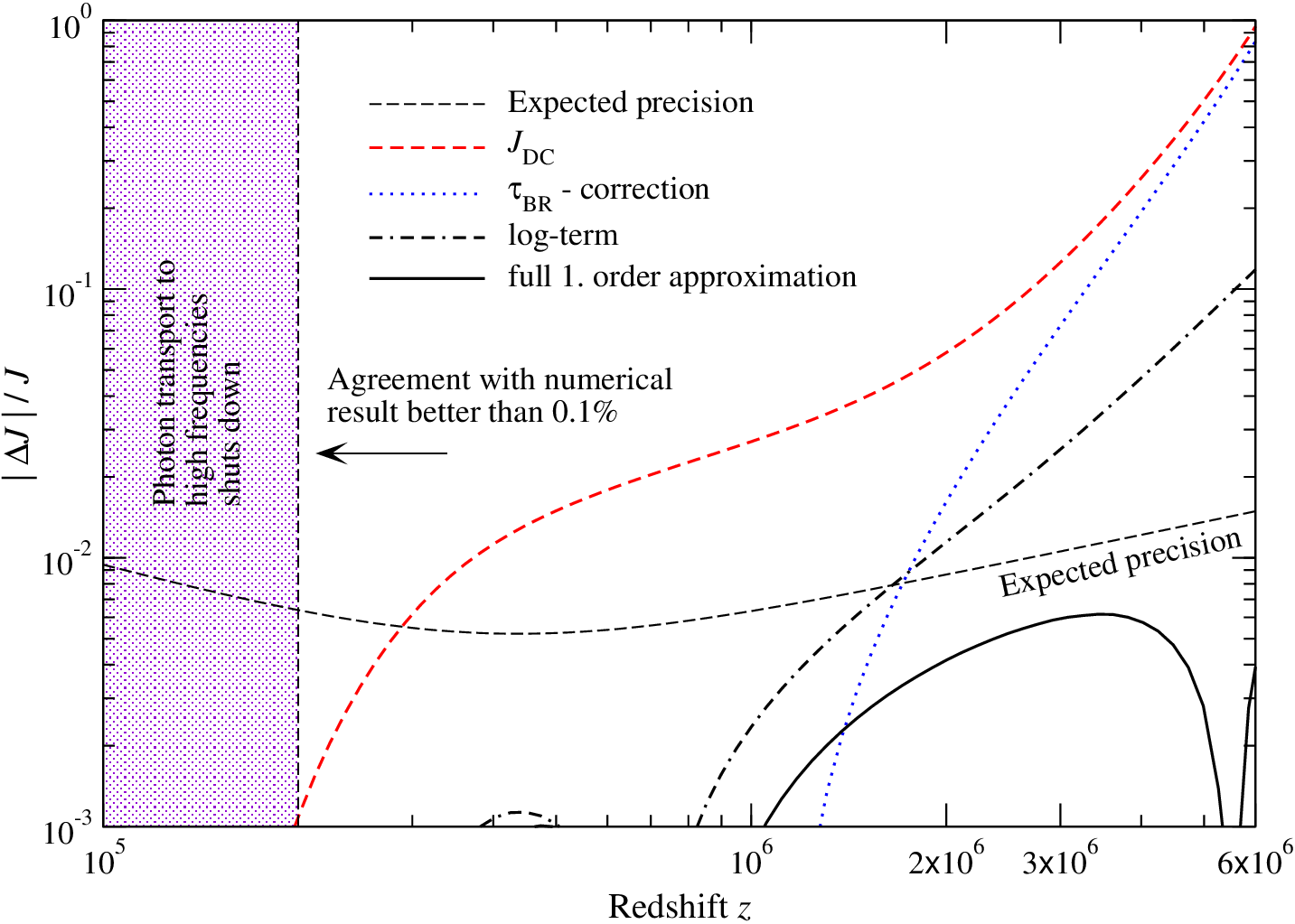}
\caption{Corrections to the distortion visibility function at different redshifts. For all curves, the numerical result obtained using {\sc CosmoTherm} was used as reference. The dashed red line shows $\mathcal{J}_{\rm DC}=\expf{-(z/\zmudc)^{5/2}}$ with $\zmudc=\pot{1.98}{6}$. When only including the BR correction to the optical depth (Sect.~\ref{sec:classical_sol}), we obtain the dotted blue line. 
Only adding the $\ln(x/\xc)\expf{-\xc/x}$ term, we improve the agreement at early times. The solid black line gives our approximation based on Eq.~\eqref{eq:mu_sol_upto_O1}, with all terms included, showing precision below the level expected in terms of perturbation order $\simeq \xc$.}
\label{fig:DVis_plot}
\end{figure}

\subsubsection{Effect on the distortion visibility function}
\label{sec:corrs_O1_DJ}
With Fig.~\ref{fig:DImu_plot}, we can now compute the visibility function for different approximation.
In Fig.~\ref{fig:Vis_plot}, we show the comparison of $\mathcal{J}_{\rm DC}$ and our approximation with the numerical result obtained with {\sc CosmoTherm}. The agreement of our approximation is extremely good, without any matching with the numerical solution being carried out. Also, evaluation of the simple integrals over the emission term and the optical depth integrals take no more than a few seconds as opposed to a couple of hours for the full numerical calculation, giving a huge improvement of the performance. We note that the full effect of the distortion visibility function and the full shape of the distortion are also captured by the efficient Green's function method introduced earlier \citep{Chluba2013Green}.

In Fig.~\ref{fig:DVis_plot}, we illustrate more clearly which terms actually matter most. The simplest approximation, $\mathcal{J}_{\rm DC}=\expf{-(z/\zmudc)^{5/2}}$, shows excellent agreement with the numerical result until $z\simeq \pot{2}{5}$, when low-frequency photons produced by BR start reaching the high-frequency domain. In particular at $z\gtrsim10^6$, the visibility is significantly lower than estimated with $\mathcal{J}_{\rm DC}$. Adding the BR correction to the optical depth, significantly improves the solution below $z\lesssim 10^6$ even to the sub-$0.1\%$ level. Clearly, by calculating the full optical depth integral and realizing that at $z\simeq \pot{2}{5}$ photon transport to high frequencies shuts down, one can improve the approximation significantly. All the physics of this correction were already included by the early treatments \citep{Sunyaev1970mu, Burigana1991, Hu1993}, but since at $z\lesssim 10^6$, also $\mathcal{J}_{\rm DC}$ already has $\lesssim 3\%$ precision, it was previously not of much interest and only added recently by KS12 in preparation for high-precision spectral distortion measurements.

Once we also add the $\ln(x/\xc)\expf{-\xc/x}$ term to the expression for $\hat\mu$, we further improve the agreement at $z\gtrsim 10^6$. The slight disagreement introduced at lower redshifts is cancelled mostly when all terms are added to the approximation. This shows, the importance of both $D_\mu$ and $D_{\rm em}$ at $z\gtrsim 10^6$; in our approach these terms need to be included to obtain an approximations below the expected level of precision which is comparable to $\simeq \xc$.

\begin{figure}
\centering
\includegraphics[width=\columnwidth]{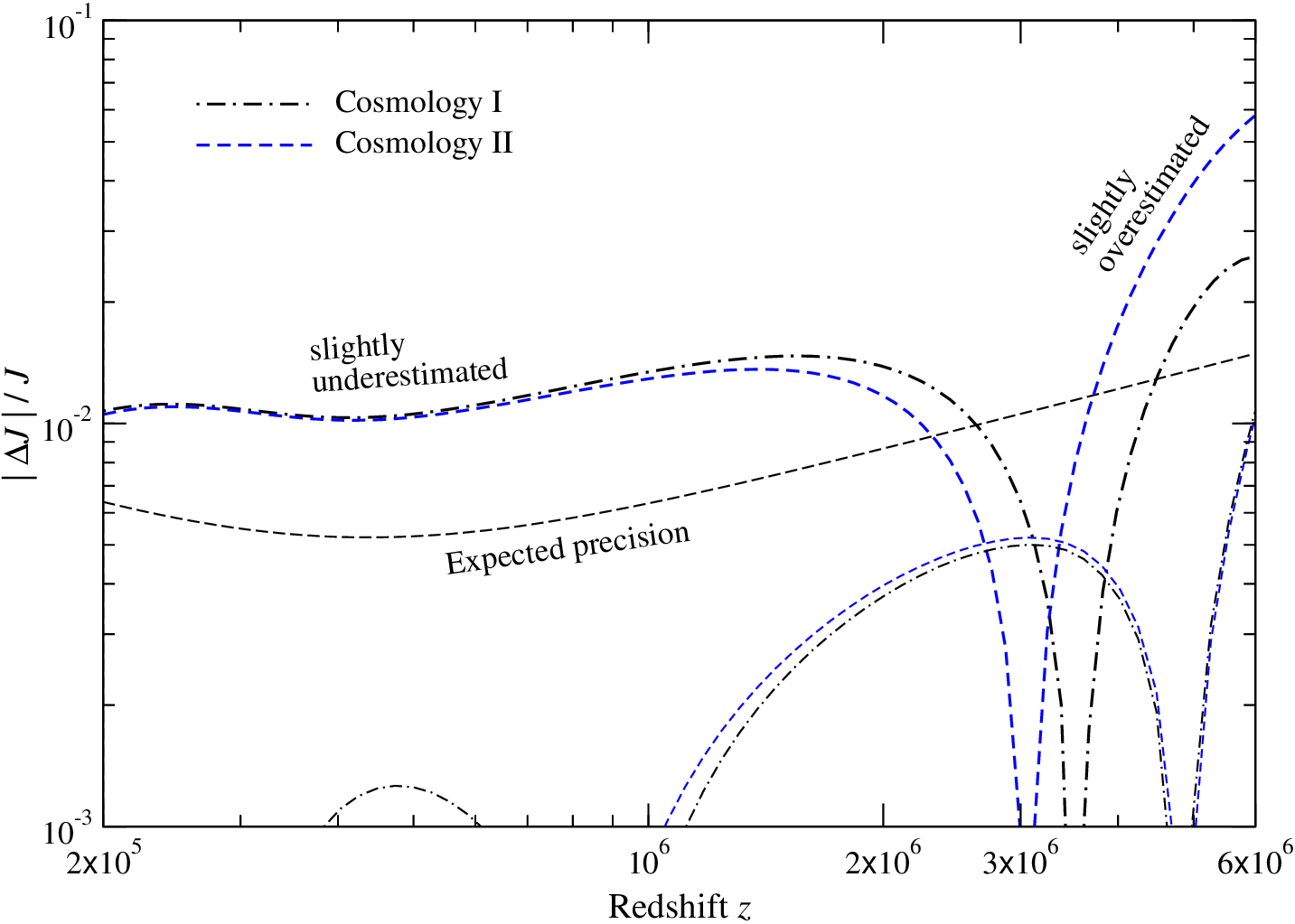}
\caption{Comparison of the approximation given by KS12 with our numerical result from {\sc CosmoTherm}. We compare for the cosmology used in KS12 (blue dashed) and the one used here (black dash-dotted). Their simple expression (heavy lines) works very well overall. Our approximation (thin lines) represents our numerical result below the expected precision $\simeq \xc$ at all redshifts and giving $\lesssim 0.1\%$ precision at $z\lesssim 10^6$. In Sect.~\ref{sec:problem}, we briefly discuss the possible explanations for difference with KS12.}
\label{fig:DVis_Khatri_plot}
\end{figure}
\subsection{Comparison with Khatri \& Sunyaev 2012}
In Fig.~\ref{fig:DVis_Khatri_plot}, we compare our numerical results directly with the approximations for the distortion visibility function given by KS12. We included DC relativistic corrections, because KS12 applied the expressions from \citet{Chluba2011therm} in their numerical computations, which included these aspects (see Sect.~\ref{sec:rel_corrs} for more discussion). 
Overall, their approximation captures the full numerical result very well. We give the comparison for two slightly different cosmologies, showing that their expression represents our result for the distortion visibility to a few percent precision. Our approximation performs a little better, representing our numerical result below the expected precision $\simeq \xc$ at all redshifts and giving much higher precision at $z\lesssim 10^6$. Also, the cosmology dependence is not as pronounced. This is reassuring, demonstrating that our perturbative approach works very well; the achieved level of precision is, however, generally very futuristic, although the computational cost is also very small.

At low redshift ($z\lesssim 10^6$), the approximation of KS12 slightly underestimates the true distortion visibility function, an effect that is also visible in their Fig.~7. This is because they did not take into account that photon transport from low to high frequencies stops below $z\simeq \pot{2}{5}$ (see Sect.~\ref{eq:transport_estimate}). Also, at those epochs, it becomes difficult even numerically to define the amplitude of $\mu_\infty$ without using energetic arguments, because the shape of the distortions starts departing from a pure $\mu$-distortion. Our approach avoids this  complication (see discussion below).

\subsubsection{Possible causes for the small differences with KS12}
\label{sec:problem}
Although pretty small, the differences between our numerical result and the approximations of KS12 are larger than the stated precision of their formulae. In particular, at $z\gtrsim 10^6$, they obtain sub-percent agreement with their numerical solution. What could be the possible causes for the differences? 

One possibility is simply the numerical treatment. This is however unlikely, since both KS12 and our approximate solutions provide an approach that is independent of the more delicate partial differential equation solving, finding excellent agreement internally. The next possibility is the included physics. Again, this seems unlikely, since they also base their physical setup on \citet{Chluba2011therm} and what went into {\sc CosmoTherm}. The only small problem could be related to the fact that KS12 did not explicitly separate the physics of DC relativistic corrections, possibly explaining some part of the cosmology dependence we find.

The most plausible cause of the differences is the normalization condition. In the derivation of Eq.~\eqref{eq:evol_mu_DT_b}, it was explicitly required that $\partial_\tau \kappa_\rho=0$. In our formulation, this is directly achieved using the normalization condition $\kappa_\rho(t)=\kappa_\rho^{\rm c}\approx 2.1419$ to fix the free integration constant in Eq.~\eqref{eq:mu_sol_upto_O1}. In contrast, KS12 just normalized their solution at one fixed frequency. This generally gives $\partial_\tau \kappa_\rho\neq 0$, so that the equivalent of Eq.~\eqref{eq:evol_mu_DT_b} reads
\beal
\label{eq:evol_mu_DT_b_mod}
\frac{\id (\hat\kappa_\rho \mu_\infty)}{\id t} 
&\approx 
\frac{3}{\kappa_\rho^{\rm c}}\frac{{\rm d} \ln a^4 \rho_\gamma}{{\rm d} t}
-\frac{4}{\kappa_\rho^{\rm c}}\frac{{\rm d} \ln a^3 N_\gamma}{{\rm d} t}.
\end{align}
This adds another small time-dependent term to the problem, which can be thought of as an equivalent of the effective heat capacity for the distorted photon field. This term was not discussed by KS12, while we absorbed it in the definition of $\mu^\ast_\infty$.
Physically, this probably implies that the distortion visibility function of KS12 does not exactly represent the fraction of energy that is stored by the distortion at a given moment. However, since the difference is small, we address this question in some future work.

\subsection{High-frequency matching and corrections due to other neglected terms}
\label{sec:other_terms}
While in terms of perturbation theory, we have already included all contributions $\mathcal{O}(\xc)$ into the analysis of Sect.~\ref{sec:corrs_O1}, it is interesting to understand the role of higher order corrections in $x$. These are expected to become relevant at intermediate frequencies $x\simeq 1$, reaching similar amplitudes as the other terms. We start by using the solution obtained in the high-frequency limit and match it with the low-frequency solution discussed in Sect.~\ref{sec:corrs_O1}. We then proceed by adding the temperature drift term and higher order Compton corrections in frequency.

\begin{figure}
\centering
\includegraphics[width=\plotwd]{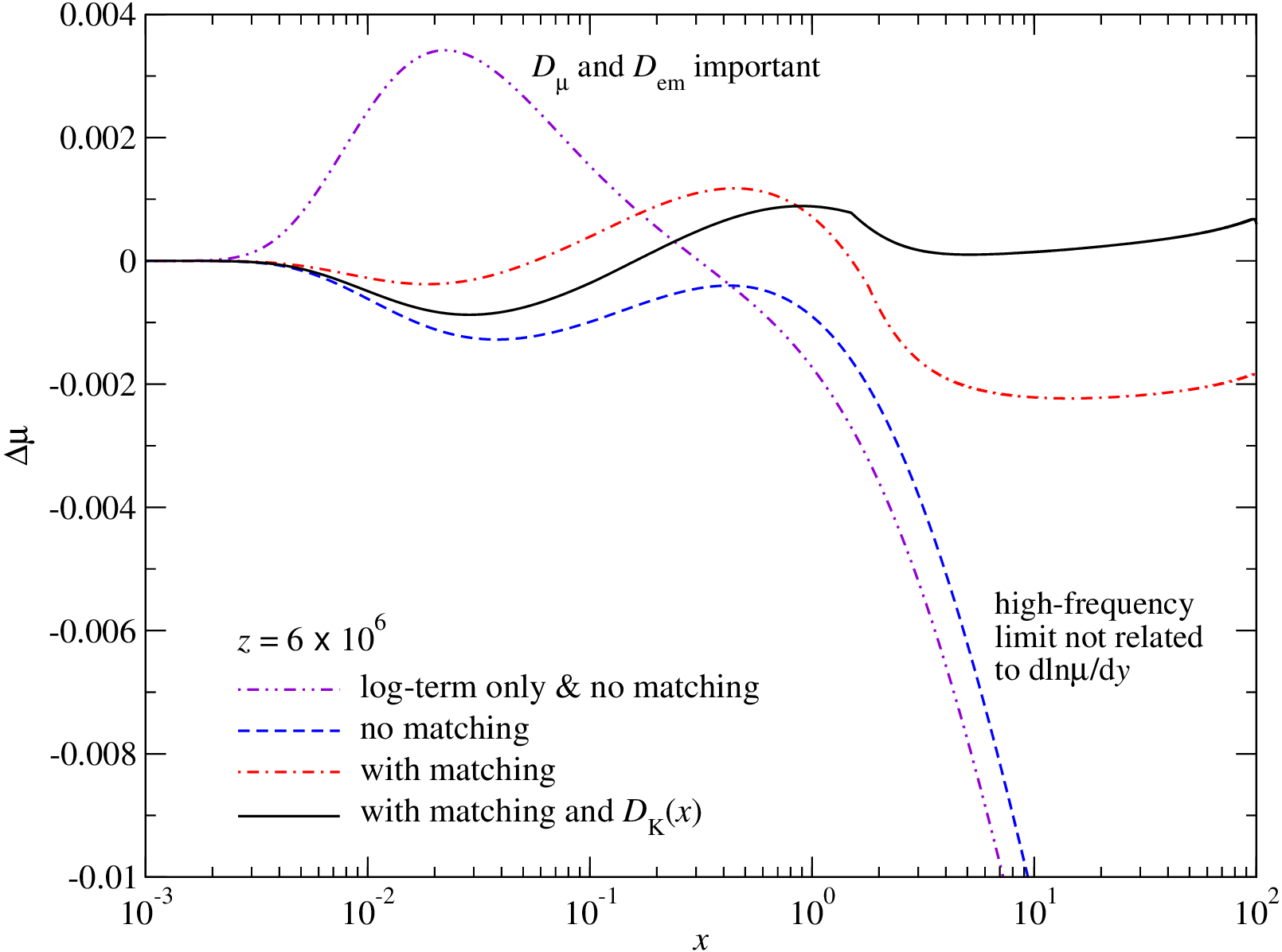}
\\[2mm]
\includegraphics[width=\plotwd]{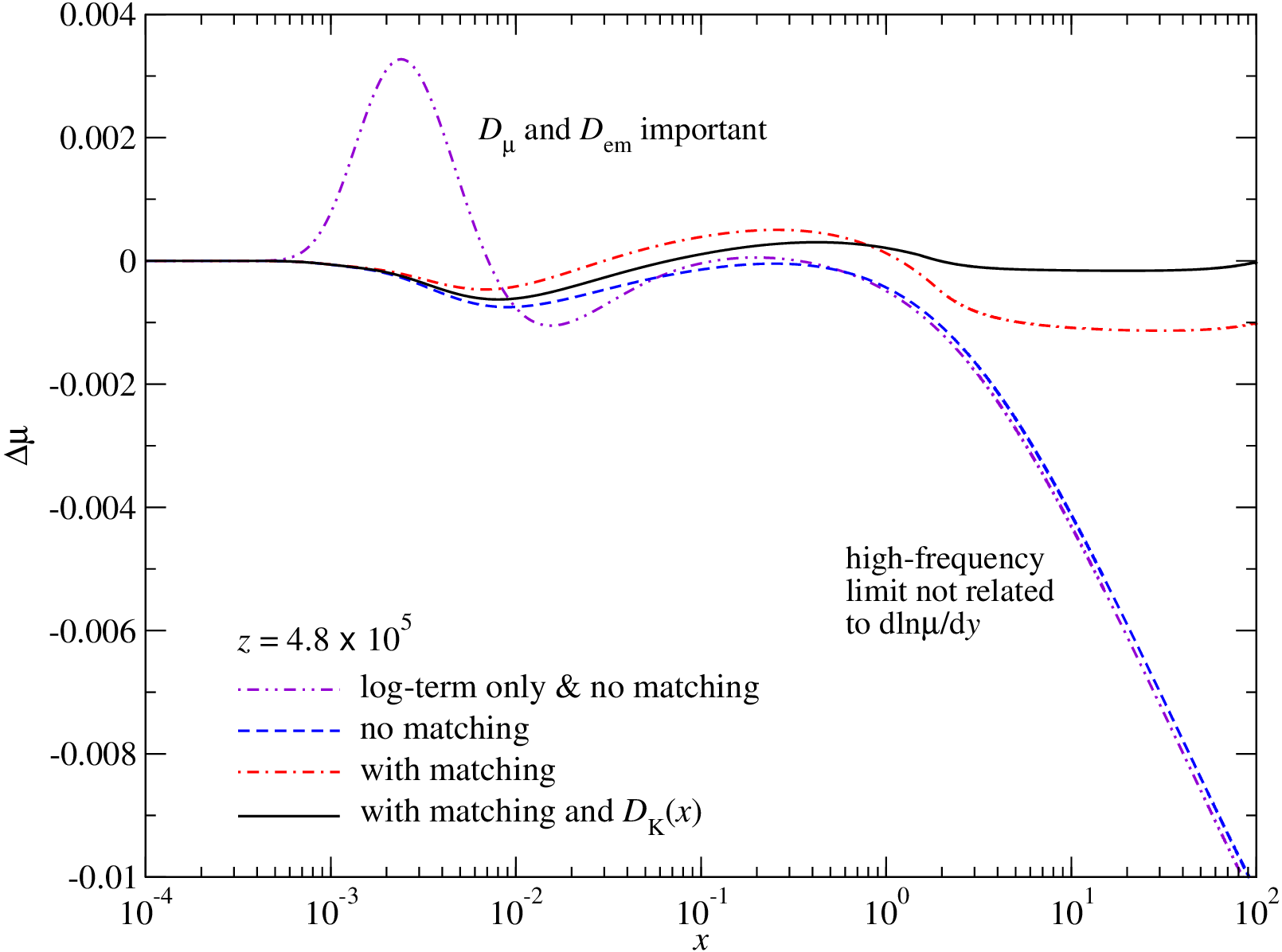}
\caption{Difference of the analytic approximation for $\hat\mu$ with respect to the numerical solution obtained from {\sc CosmoTherm} at $z=\pot{6}{6}$ and $z=\pot{4.8}{5}$. The approximation, Eq.~\eqref{eq:mu_sol_upto_O1}, shown as dashed blue line captures the behavior well at low frequencies, while at high frequencies it deviates from the numerical solution at the level of a few percent. Neglecting the contributions from $D_\mu$ and $D_{\rm em}$ (violet dash-dot-dotted line) degrades the solution at low frequencies. Matching with the high-frequency solution, Eq.~\eqref{eq:mu_high_1}, gives sub-percent agreement. Also adding the Compton scattering correction $D_{\rm K}$ to the low-frequency solution and using Eq.~\eqref{eq:mu_high_1_II} for the high-frequency part further improves the agreement.}
\label{fig:Dmu_plot}
\end{figure}

\subsubsection{High-frequency solution matching}
\label{sec:match}
Our numerical results show that the high-frequency behavior is not well represented by extrapolating the low-frequency  solution Eq.~\eqref{eq:mu_sol_upto_O1}. Instead, we should separately consider the high-frequency limit of the photon Boltzmann equation and then match the solutions at some matching point $\xmat$. 

At high frequencies, emission and absorption terms can be neglected and we only need to worry about the effect of Compton scattering. This gives the evolution equation
\beal
\label{eq:mu_equation_start_high}
\partial_y \mu-x\, \partial_y (\Te/\Tg)
&\approx
x^2 \,{\mu}'' -x^2 {\mu}'.
\end{align}
%
Assuming that the time-derivative terms can be treated as perturbations, we find the lowest order solution
$\mu^{(0)}_{\rm high}(x) \approx {\rm C}_{\rm high}$,
which is consistent with the low-frequency solution $\mu= {\rm C}_1 \expf{-\xc/x}$.
In the next iteration, we find
\beal
\label{eq:mu_high_1}
\mu_{\rm high}(x) &\approx {\rm C}_{\rm high}+\ln(x)\, \partial_y (\Te^{(0)}/\Tg),
\end{align}
which shows that energetically the time derivative of the electron temperature plays the most important role at high frequencies. Physically, this makes a lot of sense as well, since the energy exchange is dominated by the high-energy spectrum, which is mainly driven by the Compton process and hence directly related to the electron temperature and its rate of change. From Eq.~\eqref{eq:evol_mu_DT_a}, we find $\partial_y (\Te^{(0)}/\Tg)\approx (\mathcal{M}_3/4)\,\partial_y \mu^{(0)}_\infty\approx - 0.2157\, \xc  \mu^{(0)}_\infty$, which implies that the contribution of the log-term is reduced roughly $3$ times with respect to the low-frequency solution. Looking at Fig.~\ref{fig:compare_solutions} suggests that this goes into the right direction. Note, however, that according to Eq.~\eqref{eq:evol_mu_DT_a} also $(\mathcal{M}_2\mathcal{M}_3/\kappa^{\rm c}_\rho)\!\id\ln(\mathcal{M}_3/\mathcal{M}_2)/\!\id y\approx -0.25\xc(1+0.88\ln\xc)(1+1.81\xc)\!\id\ln \xc/\!\id y$ contributes to the derivative $\partial_y (\Te^{(0)}/\Tg)$ at late times.

As the next step, we should continuously match the two limiting solutions at some frequency $\xmat$. We know that the terms $D_\mu$ and $D_{\rm em}$ in Eq.~\eqref{eq:mu_sol_upto_O1} are rather small at $x\gtrsim 1$, so that for the matching condition we can use
\beal
\label{eq:matching}
A(\xc) 
+\ln (\xmat/\xc)  \partial_y\ln \mu^{(0)}_\infty
&\approx {\rm C}_{\rm high}+\ln(\xmat/\xc) \frac{\partial_y(\Te^{(0)}/\Tg)}{\mu^{(0)}_\infty} 
\end{align}
Note that we scaled the whole solution relative to $\mu^{(0)}_\infty$. To determine the best value for $\xmat$, we require smooth derivatives of the solution.
This then implies ${\rm C}_{\rm high} \approx A(\xc)-0.56\,\xc\ln(\xmat/ \xc)$.
The matching point is always close to $\xmat\approx 1.8$. 

We find that this procedure improves the agreement of the numerical and analytical solutions significantly, in particular capturing the high-frequency scaling  (cf. Fig.~\ref{fig:Dmu_plot}). However, since we leave the low-frequency solution practically unchanged, for the computation of the visibility function this modification can be omitted. In addition, a small correction to the normalization arises but it is $\simeq \mathcal{O}(\xc^2)$, and can also be neglected.

In terms of the photon spectrum, Eq.~\eqref{eq:mu_high_1} implies that the high-frequency solution for the photon occupation number is 
\beal
\label{eq:n_sol}
n(t, \xe)\simeq \alpha(t) \xe^\gamma(t)\,\expf{-\xe}, 
\end{align}
with power-law coefficient $\gamma=-\partial_y (\Te^{(0)}/\Tg)$ and $\xe=h\nu/k\Te$. This shows that due to the lack of high-frequency photons, the spectrum only slowly reaches a pure Wien-spectrum at the electron temperature, $n(t, \xe)\simeq \expf{-\xe}$. If the electron temperature changes, the shape of the spectrum is determined by the transport of photons between low and high frequencies. In particular, this indicates that the chemical potential generally remains non-zero at high frequencies. This limiting behavior is not captured by the solution given by KS12.

\subsubsection{Temperature drift term}
\begin{figure}
\centering
\includegraphics[width=\plotwd]{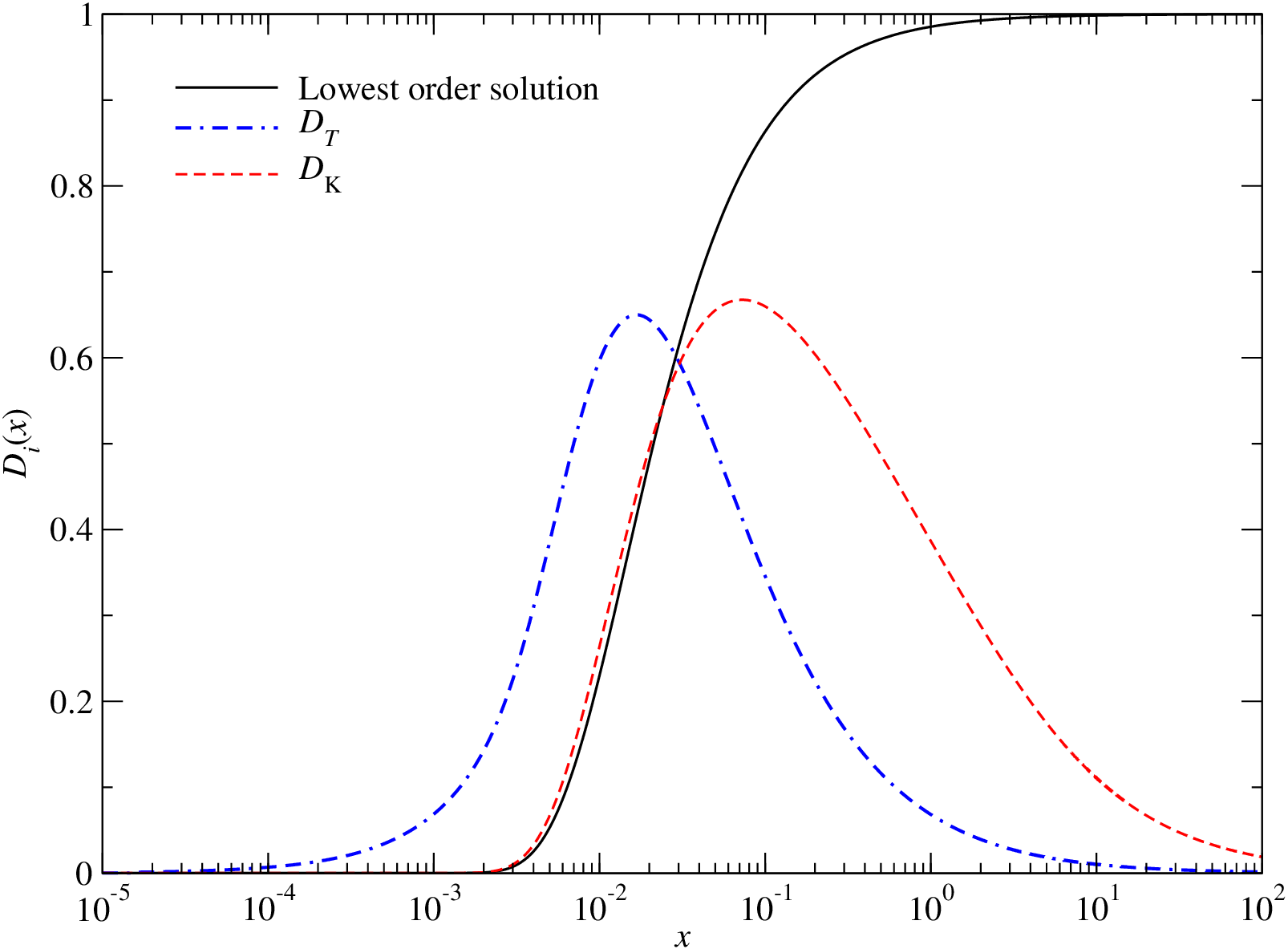}
\caption{Comparison of $\hat\mu=\expf{-\xc/x}$ with $D_T(\xc/x)$ and $D_{\rm K}(\xc, \xc/x)$ for $\xc=0.015$ ($z\simeq \pot{6}{6}$).}
\label{fig:D_T_D_K_functions}
\end{figure}

Let us consider the term $-x\,\partial_y (\Te/\Tg)$ on the left-hand side of Eq.~\eqref{eq:mu_equation_full}. As before we shall treat it as a source term in Eq.~\eqref{eq:mu1_sol}. Carrying out the integrals and absorbing any contribution $\propto \expf{\xc/x}$ (these can be absorbed by the boundary condition at $x\rightarrow 0$) and $\propto \expf{-\xc/x}$ at $x\gg 1$, we find:
\beal
\label{eq:dmu1_corrs_T}
\hat \mu_T(\tau, x)&=\left[\xc D_T(\xc/x)-x\right] \frac{\partial_y (\Te^{(0)}/\Tg)}{2\,\mu^{(0)}_\infty} 
\nonumber\\
D_T(\zeta)&=\frac{1}{2}\left[{\rm e}^{-\zeta}\,{\rm Ei}(\zeta)-{\rm e}^{\zeta}\,{\rm Ei}(-\zeta)\right].
\end{align}
The frequency dependence of $D_T$ is illustrated in Fig.~\ref{fig:D_T_D_K_functions}. While the corresponding correction is $\mathcal{O}(\xc^2)$, we find that $D_T$ peaks slightly below $\simeq 2\xc$, with a long tail towards lower frequencies, making its fractional contribution rather significant in comparison to the corrections discussed in the previous section (compare Fig.~\ref{fig:D_mu_D_em}). Still, we neglect this second-order correction, as we expect other terms to contribute at similar order.

The second correction is physically more interesting, exhibiting a $\hat\mu\propto x$ scaling. This term has to be interpreted as a shift in the electron temperature $\Delta \Te^{(1)}/\Tg\propto \frac{1}{2}\partial_y (\Te^{(0)}/\Tg)$. Since for $\hat\mu\propto x$, the photon emission integral diverges (see Sect.~\ref{sec:DNN_DEE}), this term eventually does not appear as contribution to the distortion and is absorbed as small correction to $\Te=\TRJ$, again regularizing the expression. Overall, the corrections due to the temperature drift term should be neglected at $\mathcal{O}(\xc)$ and all frequencies.

\subsubsection{Compton scattering corrections}
\label{sec:compton_x_corrs}
Earlier we argued that correction caused by $g_1(x)$ is of second-order in $\xc$. While this is true at very low frequencies, it turns out to be incorrect at $\xc\leq x \leq 1$. We can again include the effect be simply adding the associated corrections as source term to the lowest order solution. For the Compton scattering corrections, it reads
\beal
S^{(1)}_{\rm K}(x, \mu^{(0)})&=2[1-g_1(x)]x\,\partial_x \mu^{(0)}
=\mu^{(0)}\,\frac{\xc}{x}\,\left[x\,\frac{1+{\rm e}^{-x}}{1-{\rm e}^{-x}}-2\right].
\nonumber
\end{align}
Inserting this into the solution Eq.~\eqref{eq:mu1_sol}, we find
\beal
\label{eq:dmu1_corrs_K}
\hat \mu_{\rm K}(\tau, x)&=\xc \ln(x/\xc) \,\expf{-\xc/x}+ 6 \,D_{\rm K}(\xc, \xc/x) \,\xc
\nonumber
\\
D_{\rm K}(\xc, \zeta)&=\frac{1}{6}\left[F_{\rm K}(\xc, 0)\,{\rm e}^{-\zeta}- F_{\rm K}(\xc, \zeta)\right]
\\\nonumber
F_{\rm K}(\xc, \zeta)&=\ln (x) \,{\rm e}^{-\zeta} + \frac{{\rm e}^{-\zeta}}{\xc}\! \int_0^x \left[\frac{x'}{2}\,\frac{1+{\rm e}^{-x'}}{1-{\rm e}^{-x'}}-1\right]\left[{\rm e}^{2(\zeta-\zeta')}-1\right]\frac{\id  x'}{x'}.
\end{align}
The frequency dependence of $F_{\rm K}(x)$ is illustrated in Fig.~\ref{fig:D_T_D_K_functions}. It has most of its contributions at frequencies $x>2\xc$, so that in comparison with the previous correction functions it dominates in this range. In particular, the typical amplitude of the correction is {\it not} $\simeq \mathcal{O}(\xc^2)$ but rather $\mathcal{O}(\xc)$. At intermediate frequencies, this is the dominant correction we missed in our treatment above.

We furthermore see that the correction $\hat \mu_{\rm K}$ picks up a contribution $\simeq \xc \ln(x/\xc) \,\expf{-\xc/x}$, changing the extrapolated behavior of the total low-frequency solution to $\hat\mu(\tau, x)\approx A(\xc) + 0.233\xc \ln(x/\xc)$ rather than $\hat\mu(\tau, x)\approx A(\xc) -0.777\xc \ln(x/\xc)$ at high redshifts. This behavior suggests a problem with matching the low- and high-frequency solutions smoothly, since from Eq.~\eqref{eq:mu_high_1} we find a {\it negative} derivative for $\hat\mu_{\rm high}$ with respect to $x$. The problem is solved when including the next-order corrections in $x$ for the high-frequency limit, giving the evolution equation
\beal
\label{eq:mu_equation_start_high_more}
\partial_y \mu-x\, \partial_y (\Te/\Tg)
&\approx
x^2 \,{\mu}'' +(4-x)x {\mu}'.
\end{align}
and hence
\beal
\label{eq:mu_high_1_II}
\hat\mu_{\rm high}(x) &\approx {\rm C}_{\rm high}+\ln(x)\, \frac{\partial_y (\Te^{(0)}/\Tg)}{\mu^{(0)}_\infty}
\nonumber\\
&\qquad
+ \left( \partial_y\ln\mu^{(0)}_\infty-3 \frac{\partial_y (\Te^{(0)}/\Tg)}{\mu^{(0)}_\infty}\right)\frac{1+\frac{1}{x}+\frac{2}{3x}}{x}.
\end{align}
Matching with this high-frequency approximation is no problem, with the typical matching frequency $\xmat\approx 1.5$, from requiring smooth first derivatives. Including the Compton correction term in Eq.~\eqref{eq:dmu1_corrs_K}, we find agreement of the analytic solution with the numerical solution at the level $\lesssim 0.1\%$ at $0.1\xc \leq x \leq 100$ and $z\gtrsim \pot{3}{5}$ (e.g., see Fig.~\ref{fig:Dmu_plot}). That is without any direct fitting to the full numerical result, underlining the advantages of our approach.

\section{First-order relativistic corrections}
\label{sec:rel_corrs}
We finish our analysis for the early $\mu$-distortions including lowest order relativistic corrections to the DC and Compton processes. Part of the corrections were studied analytically in \citet{Chluba2005}. The DC corrections were also already included numerically by \citet{Chluba2011therm} as part of {\sc CosmoTherm}, but no more detailed discussion was given. 

\subsection{DC corrections}
\label{sec:DC_correction_Vis}
The effect of relativistic corrections caused by DC scattering is straightforward to include in our perturbation approach. The main effect is driven by a shift in the critical frequency of a few percent (see Fig.~\ref{fig:xc_fig}) and a change in the emission integral, $\mathcal{I}_{\hat\mu}$, caused by the frequency dependence of the DC Gaunt factor.
These effects can be estimated relative to the standard DC visibility function.

Neglecting the additional frequency dependence of the DC Gaunt factor, $\Lambda_{\rm DC}(x, \Thg)\approx \Lambda_{\rm DC}(0, \Thg)$, the dominant effect can be captured by re-evaluating the optical depth integral for $\tau_{\mu, 0}$ with modified $\xc$ [obtained from Eq.~\eqref{eq:xc_DC_appr_b}], giving \citep[cf.][]{Chluba2005}
\beal
\label{eq:DC_Imu_correction}
\Delta \tau_{\mu}^{\rm DC}(z, 0) \approx  
-5.06\, \Thg \,(z/\zmudc)^{5/2},
\end{align}
which at $z\simeq \pot{6}{6}$ implies a $\Delta \Jbb/\mathcal{J}_{\rm DC}\simeq 5.06\, \Thg\simeq 22\%$ visibility increase relative to $\mathcal{J}_{\rm DC}$. 
\begin{figure}
\centering
\includegraphics[width=\plotwd]{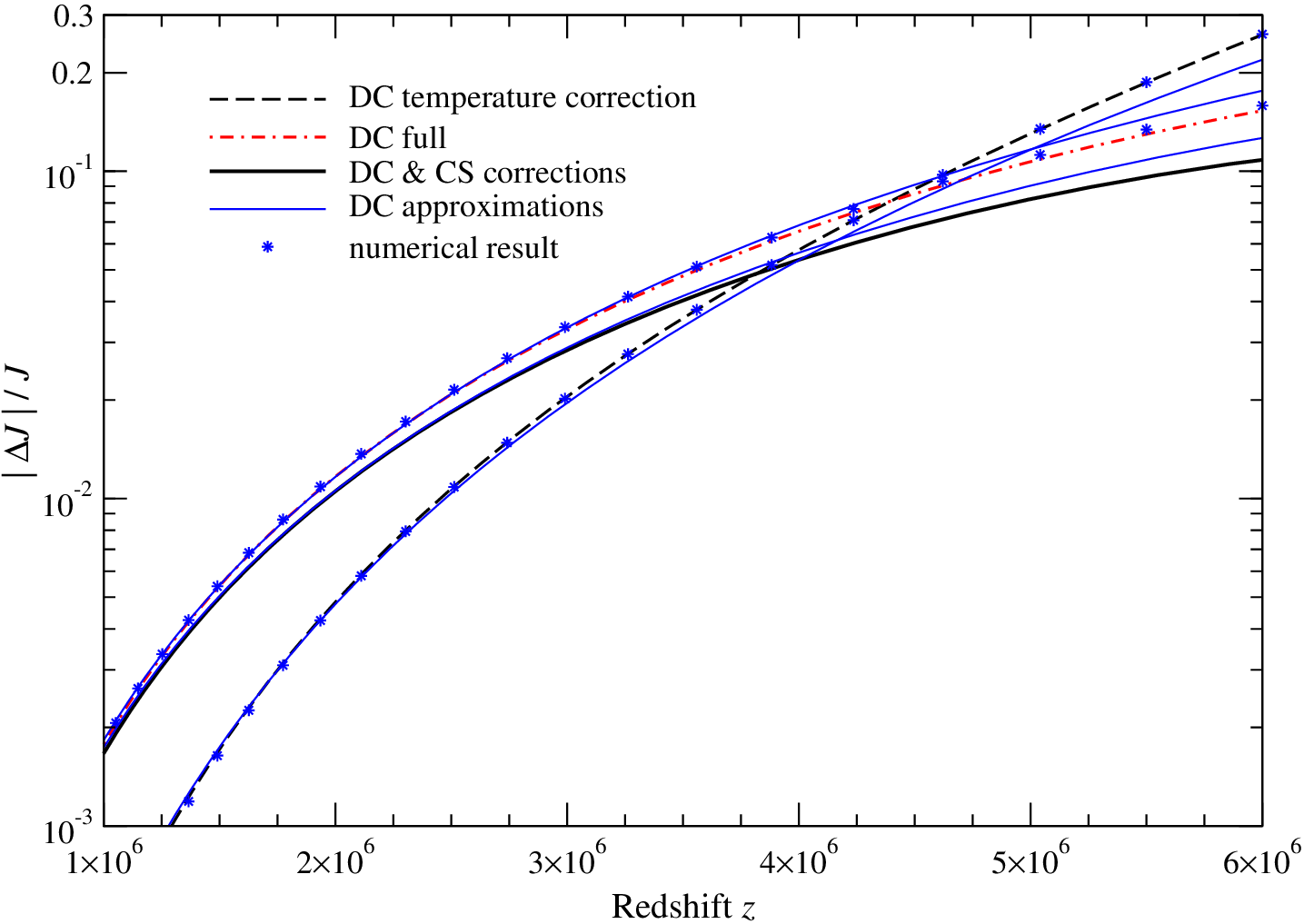}
\caption{DC and CS corrections to the distortion visibility function. The full non-relativistic result is used as a reference. The thin solid blue lines give the simple approximations discussed in Sect.~\ref{sec:rel_corrs}, while the other lines were obtained by modifying our perturbation expansion, Eq.~\eqref{eq:mu_sol_upto_O1}, accordingly. For the DC corrections, we also show the full numerical result for the correction obtained with {\sc CosmoTherm}.}
\label{fig:DVis.DC_CS}
\end{figure}
The effect is illustrated in Fig.~\ref{fig:DVis.DC_CS}, where we compared with the full non-relativistic result instead of $\mathcal{J}_{\rm DC}$. At high redshifts, the correction-to-correction is noticeable, and the simple expression, Eq.~\eqref{eq:DC_Imu_correction}, slightly underestimates the effect; however, our full perturbation approximation agrees very well with the numerical result obtained with {\sc CosmoTherm}, taking only a few seconds to evaluate rather than hours.

When also accounting for the frequency dependence of the DC Gaunt factor, both Eq.~\eqref{eq:tau_mu_a} and \eqref{eq:tau_mu_b} have to be re-evaluated, where we simply insert $\hat\mu\approx \expf{-\xc/x}$. We find $\Delta I_{\hat\mu}\approx \pot{3.70}{-3} + 1.46 \xc$ captures the correction to $I_{\hat\mu}$ pretty well, where we used $\Lambda_{\rm DC}(x, \Thg)/\Lambda_{\rm DC}(0, \Thg)\approx 1+x/2$. Evaluating the optical depth integrals then gives
\beal
\label{eq:DC_Imu_correction_xc}
\Delta \tau_{\mu}^{\rm DC}(z, 0) \approx  [\pot{3.70}{-3} + 1.43 \xc^{\rm DC, 0}] \,(z/\zmudc)^{5/2}.
\end{align}
The contribution to $\tau_{\mu, 0}(z)$, Eq.~\eqref{eq:tau_mu_a}, just coming from the shift in the critical frequency by $\Delta \xc\approx \xc/4$ [cf. Eq.~\eqref{eq:xc_DC_appr_b}] is $\simeq 0.21 \xc^{\rm DC, 0}$, while re-evaluation of Eq.~\eqref{eq:tau_mu_b} gave rise to the rest. This correction cancels the DC temperature correction, leading to a net change $\Delta \Jbb/\mathcal{J}_{\rm DC} \simeq -17\%$ at $z\simeq \pot{6}{6}$. The full result is illustrated in Fig.~\ref{fig:DVis.DC_CS}. This time, our simple approximation  overestimates the effect slightly, due to corrections-to-corrections that were not included. Our full perturbation approximation again agrees very well with the numerical result obtained with {\sc CosmoTherm}. 

We also mention, that at lowest order in $x$, the frequency modulation of the emission term, Eq.~\eqref{eq:DC_term}, caused by the factor $(1-\expf{-x})/x\approx 1-x/2$ and $\Lambda_{\rm DC}(x, \Thg)\propto 1+x/2$ cancel identically.  
Since previously we corrected for the effect of $(1-\expf{-x})/x$, inclusion of the frequency correction to $\Lambda_{\rm DC}(x, \Thg)$ reverses this correction. In our perturbation approach, this is easy to account for, and for the correction Eq.~\eqref{eq:DC_Imu_correction_xc} we also included it.

\subsection{CS temperature corrections}
\label{sec:CS_corrs}
To include CS temperature correction, we return to Eq.~\eqref{eq:mu_terms} and insert $\hat\mu^{(0)}=\expf{-\xc/x}$. By keeping only terms at lowest order in $x\ll 1$, we find the additional source term
\beal
\label{eq:S1_CS}
S^{(1)}_{\rm CS}(x)&=\Thg\left(\frac{17}{10}+\frac{14}{5}\frac{\xc}{x}-\frac{7}{10}\frac{\xc^2}{x^2}\right)\frac{\xc^2}{x^2}\,\hat\mu^{(0)}.
\end{align}
Inserting this into the integral of Eq.~\eqref{eq:mu1_sol}, we obtain the following frequency dependent correction \citep[cf.][]{Chluba2005}
\beal
\label{eq:mu1_CS}
\hat\mu_{\rm CS}(x)&\approx{\rm C}_{\rm CS}\,\expf{-\xc/x}-\Thg\,D_{\rm CS}(\xc/x)
\\
D_{\rm CS}(\zeta)
&= \frac{\zeta}{2}\left(\frac{11}{4}+\frac{21}{20}\zeta-\frac{7}{30}\zeta^2\right)\,\expf{-\zeta}.
\end{align}
We absorbed any term $\propto\expf{-\xc/x}$ into the constant and also ensured $\mu\rightarrow 0$ for small $x$. 
Since $D_{\rm CS}(\zeta)>0$ around $x\simeq 2\xc$, the main effect of CS temperature corrections is to move the critical frequency of the solution towards slightly higher values. The mean shift of the photon energy per scattering is given by $\Delta\nu/\nu\simeq 4\The[1+(5/2)\The]$ \citep[e.g.,][]{Sazonov2000}, which makes CS win the upper hand over DC at slightly higher frequencies, but this time decreasing the effective photon production rate, because according to $\Lambda_{\rm DC}/[\Thg(1+(5/2)\Thg)]=\xc^2$ the effective critical frequency decreases by $\Delta\xc\simeq -(5/4)\xc \Thg$. 
Inserting the correction function $D_{\rm CS}(\zeta)$ into the emission integral Eq.~\eqref{eq:DC_BR_I_mu}, we find
\beal
\label{eq:DA_CS}
\Delta I^{\rm CS}_{\hat\mu}&\approx -[1.72-0.82\xc]\,\Thg,
\end{align}
which is in agreement with the argument given above. The effect is slightly larger than expected from the simple estimate $\Delta I^{\rm CS}_{\hat\mu}\approx -(5/4)\Thg\xc$. This is likely due to the higher derivative terms and the precise shape of $D_{\rm CS}(\zeta)$. The final correction to the thermalization optical depth thus is
\beal
\label{eq:CS_tau_correction}
\Delta \tau_{\mu}^{\rm CS}(z, 0) \approx  - 1.23\Thg\,(z/\zmudc)^{5/2},
\end{align}
which is roughly $\simeq 4$ times smaller than the DC temperature correction, but it goes into the same direction. This is in good agreement with the estimates of \citet{Chluba2005}.

Adding the CS correction to our perturbation treatment, we obtain the thick solid black line in Fig.~\ref{fig:DVis.DC_CS}. 
Since the correction due to CS appears to be relatively small, we did not go through the trouble of implementing the effect numerically for {\sc CosmoTherm}. It is possible to iteratively include the correction using a Compton kernel approach, similar to how it was done in connection with refined helium recombination calculations \citep{Chluba2012HeRec}. However, we leave a numerical confirmation of the CS scattering correction to some future work.

For completeness, by redetermining the normalization of our full solution (see Sect.~\ref{sec:const_C1}), we find a small negative correction $\Delta A_{\rm CS}\approx \xc A_0(\xc)[0.138+1.06\ln\xc]\,\Thg$, which usually is negligible.

\section{Conclusion}
\label{sec:conclusions}
We carried out a systematic study of approximations for the distortion visibility function and early $\mu$-distortions, basing our analysis on a perturbative expansion of the solution in terms of the critical frequency, $\xc\ll 1$. Our approximations for both the distortion visibility function and the $\mu$-type distortions, over a wide range of redshifts and frequencies, agree very well with the numerical solutions obtained with {\sc CosmoTherm}. Only a few simple integrals have to be evaluated numerically, speeding the computation up from several hours\footnote{For sampling the visibility function at $\simeq 60$ redshifts.} down to seconds.

We demonstrate that the high-frequency chemical potential scales like $\mu(x)\propto {\rm const} +  \partial_y (\Te/\Tg) \ln x$ (see Sect.~\ref{sec:match}). The shape of the high-frequency spectrum is thus driven by the evolution of the electron temperature, giving rise to an $n(x)\simeq x^\gamma \expf{-x}$ dependence of the photon occupation number at large $x=h\nu/k\Tg$. The non-stationary correction caused by the time derivative of the chemical potential amplitude discussed by KS12 is noticeable at low frequencies, although a slightly larger term $\propto \ln(x/\xc) \expf{-\xc/x}$ arises due to Compton scattering frequency corrections [see Eq.~\eqref{eq:dmu1_corrs_K}]. At intermediate frequencies $x\simeq 1$, additional modifications due to the Compton process become noticeable (Sect.~\ref{sec:compton_x_corrs}). These corrections allow us to smoothly match the solutions for the $\mu$-distortion obtained in the high- and low-frequency limit, giving an accurate description of the distortion at $z\gtrsim \pot{3}{5}$, with no extra calibration of constants relative to the numerical result required (Fig.~\ref{fig:Dmu_plot}). This extends the validity for the $\mu$-distortion approximations in comparison to the approach of KS12 at practically no additional cost.

Overall our results for the distortion visibility function agree well with the approximations of KS12. Being slightly more elaborate, our approach seems to agree a bit better with our numerical result (see Fig.~\ref{fig:DVis_Khatri_plot}). We argue (see Sect.~\ref{sec:problem}) that part of the difference could be related to the precise definition of what the distortion visibility function really is, which in our case is ensured to represent the momentary fraction of energy that is stored by the distortion. We will investigate the source of the deviations in a future work.
A simple code for computing the distortion visibility function will be made available at \url{www.Chluba.de/CosmoTherm}.

Finally, in Sect.~\ref{sec:rel_corrs} we explain how DC and CS relativistic corrections affect the distortion visibility at $10^6 \lesssim z$. DC and CS temperature corrections decrease the thermalization efficiency, with the effect reaching $\Delta \Jbb/\mathcal{J}_{\rm DC} \simeq 27\%$ at $z\simeq \pot{6}{6}$. This is canceled by DC frequency-dependent corrections to the Gaunt factor, giving rise to a net $\Delta \Jbb/\mathcal{J}_{\rm DC} \simeq - 10\%$ at $z\simeq \pot{6}{6}$ (see Fig.~\ref{fig:DVis.DC_CS}). Including all corrections discussed here, at $z\simeq \pot{6}{6}$ the distortion visibility function thus is about a factor of $\simeq 2$ smaller than the simplest approximation, $\mathcal{J}_{\rm DC}(z)=\expf{-(z/\zmudc)^{2.5}}$, a modification that is important for the interpretation of future spectral distortion data (see Figs.~\ref{fig:Vis_plot} and \ref{fig:DVis_plot} for more details).

\small

\section*{Acknowledgments}
JC thanks Liang Dai, Donghui Jeong, Marc Kamionkowski and Josef Pradler for stimulating discussions of the problem. He is also grateful to Geoff Vasil for useful discussions on multi-scale perturbation theory, and Rishi Khatri and Rashid Sunyaev for their helpful feedback on the paper.
Use of the GPC supercomputer at the SciNet HPC Consortium is acknowledged. SciNet is funded by: the Canada Foundation for Innovation under the auspices of Compute Canada; the Government of Ontario; Ontario Research Fund - Research Excellence; and the University of Toronto. This work was supported by DoE SC-0008108 and NASA NNX12AE86G.

\begin{appendix}

\section{Bose-Einstein spectrum for fixed number and energy density}
\label{app:BE_spect}
\begin{figure}
\centering
\includegraphics[width=\plotwd]{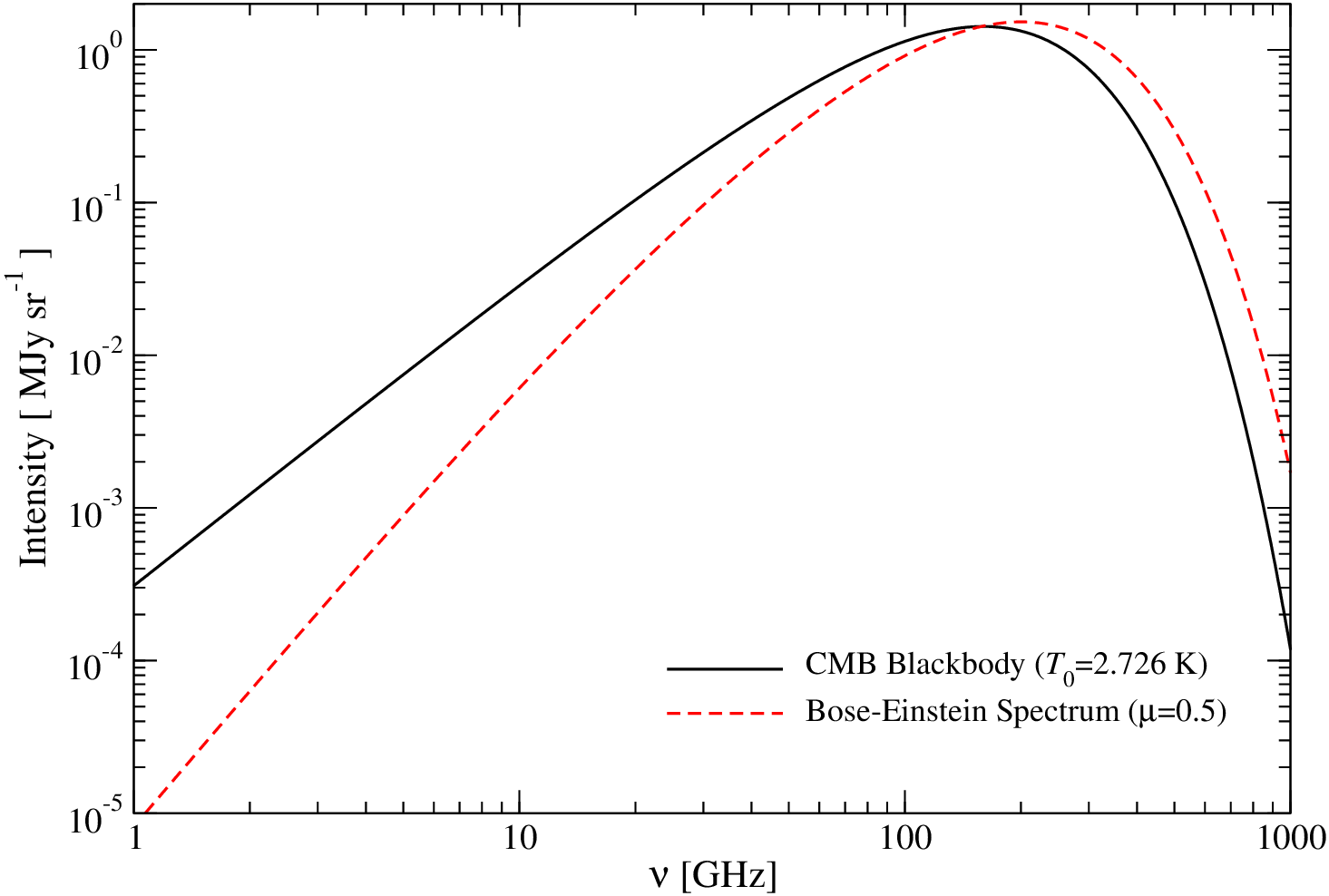}
\caption{Shape of the CMB spectrum with large chemical potential. For the considered case, the crossover frequency is at $\nu\approx 158\GHz$. Number changing processes at low frequencies were neglected, but would restore the blackbody shape at $\nu \lesssim 1\,\GHz$.}
\label{fig:BE_spec}
\end{figure}

Assuming that the photon occupation number is given by a Bose-Einstein spectrum, we can determine the precise shape from the number and energy density of the distribution. Using the ansatz, $n=1/(\expf{x\phi+\bar\mu}-1)$ [$\phi$ is needed to fix the correct number density and $\mu>0$ is constant], we can write
\beal
\label{eq:rho_def} 
\phi&=\left(\frac{2 \,{\rm Li}_3(\expf{-\mu})}{G_2^{\rm Pl}}\right)^{1/3}\approx 1- 0.4561\mu-0.137 \mu^2 \ln\mu,
\end{align}
where ${\rm Li}_n(x)$ is the polylogarithm. We assumed that the number density of the photon distribution did not change. With this solution, one can obtain the correct Bose-Einstein spectrum as a function of $x$ and $\mu$ (see Fig.~\ref{fig:BE_spec}). Fixing the energy density, we find that
\beal
\label{eq:rho_def} 
1+\frac{\Delta \rho_\gamma}{\rho_\gamma}&=\frac{6 \,{\rm Li}_4(\expf{-\mu})}{\rho^4 G_3^{\rm Pl}}
\approx 1+ 0.7140\mu +(0.815 + 0.555 \ln\mu) \mu^2
\end{align}
can be used to determine the value of $\mu$. Evidently, at lowest order one has $\mu\approx 1.401 \Delta \rho_\gamma/\rho_\gamma$, as expected.

One interesting aspect is that for larger values of $\mu$, the zero crossing of the distortion with respect to the blackbody increases. The crossover frequency is roughly given by $\nu_{\rm cr}\approx 124\GHz (1- 0.304\,\mu \ln\mu)$, so that even for very large values of $\mu\simeq 0.01$ the zero does not change dramatically.

\section{Entropy of a non-equilbrium Bose-Einstein spectrum}
\label{app:entropy}
In terms of the photon occupation number, $n=1/(\expf{x+\bar\mu}-1)$, the photon entropy density can be written as \citep{LandauStatPhys1980}
\beal\label{eq:entropy_general} 
s_\gamma &=8\pi k \left(\frac{k\Tg}{hc}\right)^3\,\int x^2 \, [(1+n)\ln(1+n)-n\ln n]\id x
\nonumber
\\
&=8\pi k\left(\frac{k\Tg}{hc}\right)^3\,\int x^2\, [\ln(1+n)+n(x+\bar\mu)]\id x
\nonumber
\\
&=\frac{4}{3}\frac{\rho_\gamma}{\Tg}-\frac{8\pi k}{3} \left(\frac{k\Tg}{hc}\right)^3\,\int x^3 \,\bar\mu \,\partial_x n \id x
\nonumber
\\
&\!\!\stackrel{\stackrel{\bar\mu\ll 1}{\downarrow}}{\approx} 
\frac{4}{3}\frac{\rho^{\rm Pl}_\gamma(\Tg)}{\Tg}\left[1+3\frac{\Delta \Te}{\Tg}\right] - \frac{\rho^{\rm Pl}_\gamma(\Tg)}{\Tg}\mu_\infty \mathcal{M}_3
\nonumber\\
&=\frac{4\mathcal{G}^{\rm Pl}_3}{3\mathcal{G}^{\rm Pl}_2} k N_\gamma +\frac{\kappa_\rho}{3}\frac{\rho^{\rm Pl}_\gamma(\Tg)}{\Tg}\,\mu_\infty \approx 3.601 k N_\gamma[1+0.5355\mu^\ast_\infty],
\end{align}
where we used $\rho^{\rm Pl}_\gamma(T)=(\mathcal{G}^{\rm Pl}_3/\mathcal{G}^{\rm Pl}_2) kT N^{\rm Pl}_\gamma(T)\approx 2.701 kT N^{\rm Pl}_\gamma(T)$ and the effective chemical potential $\mu^\ast_\infty=\hat\kappa_\rho\mu_\infty$.


\end{appendix}

\bibliographystyle{mn2e}
\bibliography{Lit}

\end{document}